\documentclass[a4paper]{article}

\usepackage{fullpage}
\usepackage{booktabs} 
\usepackage{subcaption}

\usepackage{amsmath}
\usepackage{amsfonts}
\usepackage{amssymb}
\usepackage{amsthm}
\usepackage{pxfonts}
\usepackage{mathtools}
\usepackage{proof}
\usepackage{nameref,hyperref,cleveref}
\usepackage[normalem]{ulem}

\usepackage{paralist}
\usepackage{dashbox}
\usepackage{diagbox}

\usepackage{color}
\usepackage{tikz}
\usetikzlibrary{positioning}
\usetikzlibrary{decorations.markings}
\usetikzlibrary{cd}
\usetikzlibrary{calc}

\newcommand\blfootnote[1]{%
  \begingroup
  \renewcommand\thefootnote{}\footnote{#1}%
  \addtocounter{footnote}{-1}%
  \endgroup
}

\definecolor{lightblue}{RGB}{173,216,230}
\definecolor{indianred}{RGB}{205,92,92}
\definecolor{darkgreen}{RGB}{20,200,10}
\definecolor{purplerain}{RGB}{155,30,220}
\definecolor{yellowish}{RGB}{220,230,130}

\definecolor{darkred}{RGB}{155,42,42}
\definecolor{darkblue}{RGB}{62,142,205}
\definecolor{axwhite}{RGB}{255,240,240}
\definecolor{cutgray}{RGB}{80,120,120}

\tikzstyle{pt}=[circle,fill=black,draw=black,scale=0.2]
\tikzstyle{root}=[circle,fill=white,draw=black,scale=0.2]
\tikzstyle{fuse}=[circle,fill=black,draw=black,scale=0.5]
\tikzstyle{glue}=[circle,fill=white,draw=black,scale=0.2]
\tikzstyle{bigglue}=[circle,fill=white,draw=black,scale=0.5]
\tikzstyle{axiom}=[circle,fill=axwhite,draw=black,scale=0.5]
\tikzstyle{cut}=[circle,fill=cutgray,draw=black,scale=0.5]

\tikzstyle{lam}=[circle,fill=indianred,draw=black,scale=0.5]
\tikzstyle{app}=[circle,fill=lightblue,draw=black,scale=0.5]
\tikzstyle{ptlam}=[circle,fill=indianred,draw=black,scale=0.2]
\tikzstyle{ptapp}=[circle,fill=lightblue,draw=black,scale=0.2]
\tikzstyle{counit}=[circle,fill=lightblue,draw=black,scale=0.5]
\tikzstyle{unit}=[circle,fill=indianred,draw=black,scale=0.5]

\tikzset{->-/.style={decoration={
  markings,
  mark=at position .5 with {\arrow{>}}},postaction={decorate}}}
\tikzset{-->-/.style={decoration={
  markings,
  mark=at position .75 with {\arrow{>}}},postaction={decorate}}}
\tikzset{->--/.style={decoration={
  markings,
  mark=at position .25 with {\arrow{>}}},postaction={decorate}}}
\tikzset{-<-/.style={decoration={
  markings,
  mark=at position .5 with {\arrow{<}}},postaction={decorate}}}
\tikzset{--<-/.style={decoration={
  markings,
  mark=at position .75 with {\arrow{<}}},postaction={decorate}}}
\tikzset{-<--/.style={decoration={
  markings,
  mark=at position .25 with {\arrow{<}}},postaction={decorate}}}
\tikzset{->>-/.style={decoration={
  markings,
  mark=at position .25 with {\arrow{>}},mark=at position .75 with {\arrow{>}}},
  postaction={decorate}}}
\tikzset{->>>-/.style={decoration={
  markings,
  mark=at position .166 with {\arrow{>}},mark=at position .5 with {\arrow{>}},mark=at position .833 with {\arrow{>}}},
  postaction={decorate}}}
\tikzset{!->-/.style={decoration={
  markings,
  mark=at position #1 with {\arrow{>}}},postaction={decorate}}}

\tikzset{->-neg/.style={color=darkblue,ultra thick,postaction={ultra thin,decoration={markings,mark=at position .5 with {\arrow{>}}},decorate}}}
\tikzset{->-pos/.style={color=darkred,ultra thick,postaction={ultra thin,decoration={markings,mark=at position .5 with {\arrow{>}}},decorate}}}

\tikzset{->-negpos/.style={color=darkred,ultra thick,decoration={
  markings,
  mark=at position .5 with {\arrow{>}},
  mark=at position .5 with {\pgfsetfillcolor{blue};\pgfusepath{fill};}},
  postaction={decorate}}}

\newtheorem{theorem}{Theorem}[section]

\crefname{theorem}{theorem}{theorems}
\crefname{proposition}{prop.}{propositions}
\crefname{definition}{defn.}{definitions}
\crefname{defprop}{Def.-Prop.}{definition-propositions}
\newtheorem{definition}[theorem]{Definition}
\newtheorem{proposition}[theorem]{Proposition}
\newtheorem{corollary}[theorem]{Corollary}
\newtheorem{lemma}[theorem]{Lemma}
\newtheorem{example}[theorem]{Example}
\newtheorem{defprop}[theorem]{Definition-Proposition}
\newtheorem{observation}[theorem]{Observation}
\newtheorem{remark}[theorem]{Remark}

\newtheorem*{notation}{Notation}

\newif\ifwebversion
\webversiontrue

\def\comment#1{}

\newcommand\imp{\multimap}
\newcommand\impcon{\multimap}

\newcommand\rimp[2]{#1 \imp #2}
\newcommand\mul{\cdot}
\newcommand\bmul{\bullet}

\def\hide#1{}

\newcommand\oeis[1]{\href{https://oeis.org/#1}{#1}}

\newcommand\defn[1]{{\bf #1}}

\newcommand\LRA{\Longrightarrow}
\newcommand\LLA{\Longleftarrow}
\newcommand\LLRA{\Longleftrightarrow}

\newcommand\compose{\tau}
\newcommand\unit{\delta}
\newcommand\init{\iota}
\newcommand\assoc{\rho}

\newcommand\op{{\mathrm{op}}}

\newcommand\defeq{:=}

\newcommand\set[1]{\{\,#1\,\}}

\newcommand\Down[1]{{#1}^{\downarrow}}
\newcommand\Up[1]{{#1}^{\uparrow}}

\newcommand\UpDown[1]{{#1}^{\uparrow\downarrow}}

\newcommand\FunImp[1]{P[#1]}
\newcommand\FunAbImp[1]{\tilde{P}[#1]}

\newcommand\refs{\sqsubset}
\newcommand\refsup{\mathbin{\sqsubset^\uparrow}}

\newcommand\refsdown{\mathbin{\sqsubset^\downarrow}}

\newcommand\irel[1]{\mathrel{\#_{#1}}}
\newcommand\quot[2]{#1/#2}

\newcommand\Z{\mathbb{Z}}
\newcommand\V{\mathbb{V}}

\DeclareMathOperator{\nhdi}{in}
\DeclareMathOperator{\nhdo}{out}

\DeclareMathOperator\orbit{orbit}
\DeclareMathOperator\orbits{orbits}

\DeclarePairedDelimiter\card{\lvert}{\rvert}

\DeclareMathOperator{\pull}{pull}

\newcommand\vpos[3]{[#1,#2,#3]^+}
\newcommand\vneg[3]{[#1,#2,#3]^-}

\newcommand\mimp[2]{#1 \multimapdot #2}

\begin{document}

\title{A theory of linear typings as flows on 3-valent graphs}
\author{Noam Zeilberger \\ {\small University of Birmingham}}

\maketitle

\begin{abstract}
Building on recently established enumerative connections between lambda calculus and the theory of embedded graphs (or ``maps''), this paper develops an analogy between typing (of lambda terms) and coloring (of maps).
Our starting point is the classical notion of an abelian group-valued ``flow'' on an abstract graph (Tutte, 1954).
Typing a linear lambda term may be naturally seen as constructing a flow (on an embedded 3-valent graph with boundary) valued in a more general algebraic structure consisting of a preordered set equipped with an ``implication'' operation and unit satisfying composition, identity, and unit laws.
Interesting questions and results from the theory of flows (such as the existence of nowhere-zero flows) may then be re-examined from the standpoint of lambda calculus and logic.
For example, we give a characterization of when the local flow relations (across vertices) may be categorically lifted to a global flow relation (across the boundary), proving that this holds just in case the underlying map has the orientation of a lambda term.
We also develop a basic theory of rewriting of flows that suggests topological meanings for classical completeness results in combinatory logic, and introduce a polarized notion of flow, which draws connections to the theory of proof-nets in linear logic and to bidirectional typing.
\end{abstract}

\section{Introduction}
\label{sec:intro}

The study of graphs embedded on surfaces, or \emph{maps}, has a long history, much of it linked with the rich history of the Four Color Problem (now the Four Color Theorem, or 4CT) \cite{Thomas1998,Gonthier2008}.
Formally, 4CT is a statement about maps, namely that \emph{every bridgeless planar map has a proper face-4-coloring.}
\blfootnote{This is the preliminary arXiv version of a paper to be presented at LICS 2018. The content is the same as the conference version, with the addition of two appendices containing proofs of results (\Cref{sec:appendix:proofs}) and an extended example (\Cref{sec:appendix:polflows}).}

A mathematician who made great contributions to the study of maps and colorings was Bill Tutte, including his observation of a duality between \emph{chromatic polynomials} and \emph{flow polynomials} \cite{Tutte1954}, which turns on the very natural notion of an \emph{abelian group-valued flow} over a graph.
By a well-known reduction going back to Tait \cite{Tait1880}, 4CT is equivalent to the statement that \emph{every bridgeless planar 3-valent map has a proper edge-3-coloring,}
\begin{center}\includegraphics[width=8cm]{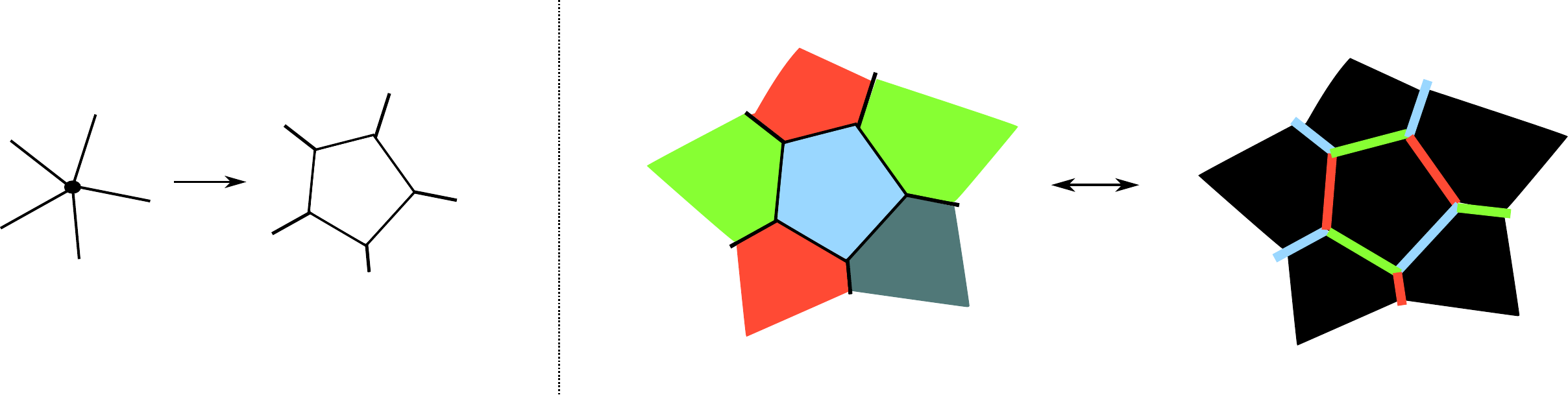}\end{center}
and within Tutte's theory this may be reformulated as the statement that every such map has a \emph{nowhere-zero $\V$-flow}, where 
$\V \cong \Z_2\times\Z_2$ stands for the Klein Four Group \cite{Tutte1966}.

A separate line of work that Tutte began in the 1960s (also originally motivated by 4CT, see \cite[Ch.~10]{Tutte1998}) was the \emph{enumerative} study of maps, establishing some remarkably simple formulas for the number of (rooted) planar maps of a given size satisfying varying constraints.
Though Tutte's approach to 4CT was ultimately side-stepped by the Appel-Haken proof \cite{AppelHaken1977},
enumeration of maps remains a very active area of combinatorics, with links to wide-ranging domains such as algebraic geometry, knot theory, and mathematical physics.

\begin{table}
\begin{center}
\small
\begin{tabular}{l|l|l}
{\bf family of lambda terms} & {\bf family of rooted maps} & {\bf OEIS}\\
\hline\hline
linear & 3-valent (of genus $g\ge 0$) & \oeis{A062980} \\
planar & planar 3-valent & \oeis{A002005} \\
unitless linear & bridgeless 3-valent ($g\ge 0$) & \oeis{A267827}\\
unitless planar & bridgeless planar 3-valent & \oeis{A000309} \\
\hline
$\beta$-normal linear/$\sim$  & (all maps of genus $g\ge 0$)  & \oeis{A000698} \\
$\beta$-normal planar & planar & \oeis{A000168}  \\
$\beta$-normal unitless linear/$\sim$ & bridgeless ($g \ge 0$)  & \oeis{A000699} \\
$\beta$-normal unitless planar  & bridgeless planar & \oeis{A000260}
\end{tabular}
\end{center}
\caption{Known correspondences \cite{BoGaJa2013,ZG2015corr,Z2015counting,Z2016trivalent,Z2017assoc,CYZ2017chordmaps} between families of lambda terms and rooted maps, as combinatorial classes.
Here ``$\sim$'' stands for an equivalence relation defined in \cite{Z2015counting}, and the rest of the terminology is explained in \Cref{sec:imploid-flows,sec:flow-rewriting:background}.
(Indices on the right refer to the Online Encyclopedia of Integer Sequences \cite{OEIS}.)
}
\label{table:lambda}
\end{table}

The ideas developed in this paper sprang from the recent discovery of a host of surprising links between map enumeration and lambda calculus: see \Cref{table:lambda}.
Bridges between lambda calculus and graph theory trace back to the pioneering works of Statman \cite{Statman1974thesis} and Girard \cite{Girard1987}, but these combinatorial connections go even further in suggesting that very concrete objects of study are actually shared: for example, it turns out that many of the sequences first computed by Tutte also count natural families of lambda terms!

Although the correspondences listed in the lower half of \Cref{table:lambda} for the moment rest mainly at the level of enumeration, there is a simple bijection between linear lambda terms and rooted 3-valent maps (originally described in \cite{BoGaJa2013}) that may be restricted to account for the entire upper half of the table.
This was used in \cite{Z2016trivalent} to state a lambda calculus reformulation of 4CT, essentially by turning the existence of a nowhere-zero $\V$-flow into a \emph{typing} problem for a certain family of terms.
In turn, that motivated the development of a more general theory of linear typings-as-flows on 3-valent maps, and this paper represents a preliminary sketch of such a theory.

For example, rather than limiting types to abelian groups, flows over linear lambda terms are naturally valued in a more general algebraic structure consisting of a preordered set equipped with an ``implication'' operation and unit element satisfying composition, identity, and unit laws: what we call an \emph{imploid} (and is otherwise known as a ``thin skew-closed category'').
Considering imploid-valued flows over arbitrary well-oriented 3-valent maps with boundary, a natural question is when can the local flow relations (across vertices) be lifted to a global flow relation (across the boundary).
We will see that this holds just in case the map is equipped with the canonical orientation of a lambda term, giving a new perspective on the interplay between linear typing and compositionality.

The rest of the paper is structured as follows.
In \Cref{sec:imploids} we establish some elementary properties of imploids, leading up through the construction of quotients.
In \Cref{sec:imploid-flows}, after a review of topics in graph theory, we introduce the basic definition of imploid-valued flows and nowhere-unit flows over well-oriented 3-valent maps.
We recall the bijection between rooted 3-valent maps and linear lambda terms based on the \emph{topological orientation} of a map with boundary, and use it to prove the above-mentioned characterization of the global flow condition.
We also briefly explain how the definitions may be recast algebraically in terms of a certain quotient imploid we call the \emph{fundamental imploid} of a well-oriented 3-valent map (analogous to the fundamental quandle of a knot).
In \Cref{sec:flow-rewriting}, we discuss rewriting of flows across operations such as $\beta$-reduction and $\eta$-expansion, and prove a \emph{topological completeness} theorem which is closely related to classical completeness results in combinatory logic.
Finally, in \Cref{sec:polarized-flows} we briefly describe a polarized notion of flow, pointing out connections with linear logic proof-nets, as well as with bidirectional typing.

{\bf Acknowledgments.}
The ideas described in this paper have been in gestation for a while and have benefited from interaction with numerous individuals and seminar audiences, as well as from the friendly and supportive atmosphere within the Theory Group at Birmingham.
I am especially grateful to Jason Reed for a long series of email exchanges on map coloring and lambda calculus that (in addition to being highly enjoyable) really brought these ideas into focus.

\section{The elementary theory of imploids}
\label{sec:imploids}

\subsection{Preliminary definitions and examples}
\label{sec:imploids:prelim}

We recall some standard terminology and conventions. 
A \emph{preorder} on a set $P$ is a binary relation which is reflexive and transitive, typically indicated $\le$ or $\le_P$.
We write $\equiv$ for the induced equivalence relation $a\equiv b \defeq (a \le b) \wedge (b \le a)$, and say that the preorder is a \emph{partial order} when $a \equiv b$ implies $a = b$.
A \emph{preordered set} is a set equipped with a preorder $(P,\le)$, and we may also sometimes write $P$ to stand for the preordered set with $\le$ left implicit.
To any preordered set $P$ is associated an \emph{opposite} $P^\op$ with the same underlying set of elements but the opposite preorder, $a \le^\op b$ iff $b \le a$.
To any pair of preordered sets $P$ and $Q$ is associated a \emph{product} $P \times Q$ with underlying set of elements given by the cartesian product and with the componentwise ordering $(a_1,b_1) \le_{P\times Q} (a_2,b_2)$ iff $a_1 \le_P a_2$ and $b_1 \le_Q b_2$.
Finally, we write $f : P \to Q$ to indicate that $f$ is an order-preserving (a.k.a.~\emph{monotonic}) function from $P$ to $Q$ (i.e., that $a \le_P b$ implies $f(a) \le_Q f(b)$ for all $a,b \in P$), and we say that $f$ is \emph{left adjoint} to $g : Q \to P$ (or $g$ is \emph{right adjoint} to $f$)
just in case $f(a) \le_Q b$ iff $a \le_P g(b)$ for all $a\in P$, $b \in Q$.
\begin{definition}
An \defn{imploid} is a preordered set $P$ equipped with an operation $\impcon$ (called \emph{implication}) which is contravariant in its left argument and covariant in its right argument,
\begin{equation}\label{clopset:imp}
\infer{a_1\imp b_1 \le a_2\imp b_2}{a_2 \le a_1 & b_1 \le b_2} \tag{imp}
\end{equation}
together with a distinguished element $I \in P$,
satisfying \emph{composition}, \emph{identity}, and \emph{unit} laws:
\begin{align}
\label{clopset:comp}
\rimp bc &\le \rimp{(\rimp ab)}{(\rimp ac)} \tag{comp}
\\ \label{clopset:id}
I &\le \rimp aa \tag{id}
\\ \label{clopset:unit}
\rimp Ia &\le a \tag{unit}
\end{align}
A \defn{non-unital imploid} is given by the same except that we do not require the element $I$ and omit axioms \eqref{clopset:id} and \eqref{clopset:unit}.
\end{definition}
\noindent
It is worth emphasizing that what we call an imploid is simply the preorder restriction of what Street has recently referred to as a \emph{skew-closed category} \cite{Street2013skew}, corresponding to a slight relaxation of Eilenberg and Kelly's original definition of a (non-monoidal) closed category \cite{EilenbergKelly1966}.
Although the development we give here is limited to preordered sets (which include our motivating examples), it is likely that many of these constructions could be lifted with care to the more general context of skew-closed categories.
\begin{definition}[cf.~\cite{Street2013skew}]
In any imploid $P$ we have that $a \le b$ entails $I \le \rimp ab$, by \eqref{clopset:id} and \eqref{clopset:imp}.
We say that $P$ is \defn{left normal} if this entailment is reversible, that is, if $I \le \rimp ab$ entails $a \le b$.
\end{definition}
\begin{example}
\label{ex:quantale-imploid}
Any Heyting algebra 
defines a left normal imploid with $\rimp ab \defeq a \supset b$ and $I \defeq \top$.
More generally, any (unital) \emph{quantale} \cite{Yetter1990} gives an example of a (left normal, unital) imploid.
\end{example}
\begin{example}
\label{ex:group-imploid}
Any group can be seen as a left normal imploid under the discrete order ($a\le b$ iff $a = b$), where $I$ is the unit element of the group and implication can be defined by either right division $\rimp ab \defeq b \mul a^{-1}$ or left division $\rimp ab \defeq a^{-1}\mul b$.
\end{example}
\begin{definition}[cf.~\cite{EilenbergKelly1966,BourkeLack2017braidedskew}]
We say that a (unital or non-unital) imploid is \defn{symmetric} if it satisfies the law of \emph{exchange:}
\begin{equation}
\rimp a{(\rimp bc)} \le \rimp b{(\rimp ac)} \tag{exch} \label{clopset:exch}
\end{equation}
\end{definition}
\begin{proposition}\label{prop:exchange-from-dni}
The law of \emph{double-negation introduction}
\begin{equation}\label{clopset:dni}
a \le \rimp{(\rimp ab)}b \tag{dni}
\end{equation}
implies the law of exchange.
Conversely, \eqref{clopset:exch} implies \eqref{clopset:dni} under assumption of left normality.
\end{proposition}
\begin{example}
In the above examples, the imploid associated to a group/quantale is symmetric whenever the underlying multiplication operation is commutative.
\end{example}

\subsection{Skew monoids, upsets and downsets}
\label{sec:imploids:dmonoids}

One of the motivations for the recent study of skew-closed categories (the category-theoretic version of imploids) is their close connection to \emph{skew-monoidal categories} \cite{Szlachanyi2012}.
We will refer to the order-theoretic versions of the latter as \emph{skew monoids}.
\begin{definition}
A (left) \defn{skew monoid} is a preordered set $M$ equipped with an operation $\bmul$ which is covariant in both arguments,
\begin{equation}\label{monset:mul}
\infer{a_1\bmul b_1 \le a_2\bmul b_2}{a_1 \le a_2 & b_1 \le b_2} \tag{mul}
\end{equation}
as well as a distinguished element $I \in M$, satisfying \emph{semi-associativity}, \emph{left unit}, and \emph{right unit} laws:
\begin{align}
\label{mopset:assocr}
(a \bmul b)\bmul c &\le a \bmul (b\bmul c) \tag{assocr}
\\ \label{mopset:lunit}
I \bmul a &\le a \tag{lunit}
\\ \label{mopset:runit}
a &\le a \bmul I \tag{runit}
\end{align}
A \defn{non-unital skew monoid} is given by the same except that we do not require $I$ and omit \eqref{mopset:lunit} and \eqref{mopset:runit}.
\end{definition}
\noindent
One simple relationship between imploids and skew monoids is via adjunction:
any family of right adjoints to the partially instantiated multiplication operations $-\bmul b : M \to M$ of a skew monoid $M$ can be extended to a an operation $\impcon : M^\op \times M \to M$
satisfying the imploid laws, and dually, any family of left adjoints to the partially instantiated implication operations $\rimp{b}{-} : P \to P$ of an imploid $P$ can be extended to an operation $\bmul : P \times P \to P$
satisfying the skew monoid laws.\footnote{A categorical version of this fact is proved by Street \cite{Street2013skew}, who mentions that it resolves a nagging asymmetry in the traditional setting (cf.~\cite{DayLaplaza1978}), where the existence of left adjoints is not enough to ensure that an Eilenberg-Kelly closed category can be given the structure of an ordinary monoidal category.}
We will make use of a related duality between the \emph{downsets} of a skew-monoid and the \emph{upsets} of an imploid.
\begin{definition}
A subset $R$ of a preordered set $A$ is said to be \defn{downwards closed} (or a \defn{downset}), written $R \refsdown A$, if $b \in R$  and $a \le b$ implies $a \in R$.
Dually, it is said to be \defn{upwards closed} (or an \defn{upset}), written $R \refsup A$, if $a \in R$ and $a\le b$ implies $b \in R$.
(We sometimes use mirror notation ``$R \ni a$'' to denote the elementhood relation in an upset.
Also, we sometimes write $R \refs A$ to indicate that $R$ is a subset but emphasize that it is not necessarily closed with respect to the order on $A$.)
\end{definition}
\noindent 
Recall that every element $x \in A$ of a preordered set induces both a \emph{principal downset} $\Down{x} \defeq \set{a \mid a \le x}$ and a \emph{principal upset} $\Up{x} \defeq \set{a \mid x \le a}$, and that these define a pair of faithful embeddings $\Down{(-)} : A \to \hat A$ and $\Up{(-)} : A \to \check A$, where $\hat{A}$ denotes the set of downsets of $A$ partially ordered by inclusion, and $\check{A}$ the set of upsets partially ordered by reverse inclusion (an order-preserving function $f : P \to Q$ is said to be \emph{faithful} when $f(a) \le_Q f(b)$ implies $a \le_P b$).
The following constructions amount to a ``skew'' variation of the well-known \emph{Day construction} (cf.~\cite{DayLaplaza1978,Street2013skew}).
\begin{proposition}\label{prop:imp2mon}
If $P = (P,\le,\impcon,I)$ is an imploid, then $\check{P}$ can be given the structure of a skew monoid as follows (for all $R,S \refsup P$):
\begin{align*}
R \bmul S \ni p &\iff \exists q.\ \  R \ni \rimp qp \ \  \wedge \ \  S\ni q \\
I \defeq \Up{I} \ni p &\iff I \le p
\end{align*}
\end{proposition}
\begin{observation}[cf.~\cite{BourkeLack2017braidedskew}]\label{prop:impcommpsh}
Let $P$ be an imploid.
Then
\begin{enumerate}
\item $a \le (a \imp b) \imp b$ for all $a,b \in P$ iff $R \bmul S \supseteq S \bmul R$ for all $R,S \refsup P$; and
\item $a \imp (b \imp c) \le b \imp (a \imp c)$ for all $a,b,c \in P$ iff $(R \bmul S)\bmul T \supseteq (R \bmul T)\bmul S$ for all $R,S,T \refsup P$.
\end{enumerate}
\end{observation}
\begin{proposition}\label{prop:mon2imp}
If $M = (M,\le,\bmul,I)$ is a skew monoid, then $\hat{M}$ can be given the structure of an imploid as follows (for all $K,L \refsdown M$):
\begin{align*}
m \in \rimp KL &\iff \forall n.\ \  n\in K \ \  \Rightarrow\ \  m\bmul n \in L \\
m \in I \defeq \Down{I} &\iff m \le I
\end{align*}
\end{proposition}
\begin{definition}
An order-preserving function $f : P \to Q$ between two imploids is said to be a \defn{(lax) homomorphism} if it weakly preserves the imploid structure in the sense that $I \le_Q f(I)$ and $f(\rimp ab) \le_Q \rimp{f(a)}{f(b)}$ for all $a,b \in P$.
It is said to be \defn{strong} if it preserves this structure up to equivalence, $I \equiv_Q f(I)$ and $f(\rimp ab) \equiv_Q \rimp{f(a)}{f(b)}$.
\end{definition}
\begin{proposition}\label{prop:double-yoneda-embedding}
The composite $\UpDown{(-)} : P \to \hat{\check{P}}$
is a strong homomorphism of imploids.
\end{proposition}

\subsection{Deductive closure, dni  and imploid quotients}
\label{sec:imploids:quotients}

Given that any group can be seen as an imploid (\Cref{ex:group-imploid}), it is natural to wonder what is the imploid analogue for \emph{subgroups}.
In fact, there are at least two different natural substructures of an imploid that could be considered as generalizations of the group-theoretic concept, one starting from the view of a subgroup as the image of an injective homomorphism, the other from the view of a (normal!) subgroup as the kernel of a surjective homomorphism.
\begin{definition}
\label{defn:subimploid}
A subset $R \refs P$ of an imploid $P$ is said to be a \defn{subimploid} if 
\begin{inparaenum}[1)]
\item $I \in R$, and
\item $a \in R$ and $b \in R$ implies $\rimp ab \in R$.
\end{inparaenum}
\end{definition}
\begin{definition}
\label{defn:dedupset}
An upset $R \refsup P$ of an imploid $P$ is said to be \defn{deductively closed} (or a \defn{dedupset}) if
\begin{inparaenum}[1)]
\item $I \in R$, and
\item $a \in R$ and $\rimp ab \in R$ implies $b \in R$.
\end{inparaenum}
\end{definition}
\noindent
It is easy to check that for any group $G$ viewed as a discrete imploid, a subset $H \subseteq G$ is a subimploid iff it is a dedupset iff it is a subgroup.
However, in general these two notions are quite different.
Since imploid quotients will play an important role in this paper, we spend the rest of the section on elaborating the second definition, beginning with the following simple observation.
\begin{observation}
A dedupset of $P$ is the same thing as a comonoid in $\check{P}$ (relative to the skew monoid structure defined in \Cref{prop:imp2mon}), i.e., an upset $R \refsup P$ such that $R \supseteq I$ and $R \supseteq R \bmul R$.
\end{observation}
\begin{corollary}\label{prop:deductive-closure}
The \defn{deductive closure} $!R$ of an upset $R$ is given by the formula $!R \defeq \bigwedge_{n\ge 0} R^{\bmul n}$, where $\bigwedge$ denotes the meet in $\check{P}$ (corresponding to union of subsets), and where $R^{\bmul n}$ denotes the left-associated product $R^{\bmul 0} = I$, $R^{\bmul n+1} = R^{\bmul n} \bmul R$.
\end{corollary}
\begin{definition}
The \defn{induced relation} of an upset $R \refsup P$ is a binary relation $\irel{R}$ on the elements of $P$ defined by $a \irel{R} b$ iff $\rimp ab \in R$.
\end{definition}
\begin{proposition}\label{prop:leRpreorder}
If $R$ is a dedupset, 
$\irel{R}$ is a preorder extending $\le$.
\end{proposition}
\noindent
Given an imploid $P$ and a dedupset $R \refsup P$, a natural candidate for the \emph{quotient imploid} is given by $\quot PR = (P,\irel{R},\impcon,I)$ (i.e., by the same underlying set and operations considered relative to a coarser order)\ldots but the problem is that this does not necessarily define an imploid.
Although the three imploid axioms obviously remain valid (since $\irel{R}$ is an extension of $\le$), and one can even verify that the implication $\rimp ab$ is monotone in $b$ relative to $\irel{R}$, nothing guarantees that it is also antitone in $a$.
To ensure that implication restricts to an operation of type $(\quot PR)^\op \times \quot PR \to \quot PR$ we impose a further condition on dedupsets, which is the precise analogue of the restriction to \emph{normal} subgroups in the construction of group-theoretic quotients.
\begin{definition}
We say that an upset $R \refsup P$ is \defn{dni-closed} if $a \in R$ implies $\rimp{(\rimp ab)}b \in R$ for all $b \in P$.
\end{definition}
\begin{proposition}\label{prop:dni-closure-exists}
Any upset $R$ has a \defn{dni-closure} (i.e., a dni-closed upset containing $R$ and maximal wrt.~$\supseteq$).
\end{proposition}
\begin{proposition}\label{prop:norm:tfae}
Let $R \refsup P$. The following are equivalent:
\begin{enumerate}
\item \label{prop:norm:tfae:case1} $R$ is dni-closed.
\item \label{prop:norm:tfae:case2} $R\bmul S \supseteq S\bmul R$ for all $S \refsup P$.
\item \label{prop:norm:tfae:case3} $R$ satisfies the following closure conditions:
 \begin{enumerate}
 \item \label{prop:norm:tfae:case3i} $a \in R \Rightarrow \rimp Ia \in R$
 \item \label{prop:norm:tfae:case3ii} $\rimp ab \in R \Rightarrow \forall c \in P, \rimp{(\rimp bc)}{(\rimp ac)} \in R$
 \end{enumerate}
\item \label{prop:norm:tfae:case4} The induced relation $\irel{R}$ satisfies the following rules:
$$\infer{I \irel{R} a}{a \in R} \qquad \infer{\rimp{a_1}{b_1} \irel{R} \rimp{a_2}{b_2}}{a_2 \irel{R} a_1 & b_1 \irel{R} b_2}$$ 
\end{enumerate}
\end{proposition}
\begin{corollary}\label{prop:tensor-preserves-dni}
\begin{inparaenum}[1)]
\item
If $R$ and $S$ are dni-closed then so is $R \bmul S$;
\item
if $R$ is dni-closed then so is $!R$.
\end{inparaenum}
\end{corollary}
\noindent
\begin{proposition}\label{prop:quot:imploidLN}
If $R\refsup P$ is deductively closed and dni-closed, then $\quot PR \defeq (P,\irel{R},\impcon,I)$ is a left normal imploid.
\end{proposition}
\noindent
\begin{proposition}\label{prop:pullback}
Let $f : P \to Q$ be any order-preserving function, and $S \refsup Q$ any upset.
\begin{enumerate}
\item The inverse image of $S$ along $f$ is an upset $f^{-1}(S) \refsup P$.
\item If $f$ is a homomorphism of imploids and $S$ is deductively closed, then $f^{-1}(S)$ is deductively closed.
\item If $f$ is a strong homomorphism and $S$ is dni-closed, then $f^{-1}(S)$ is dni-closed.
\end{enumerate}
\end{proposition}
\begin{proposition}\label{prop:Idnidedup}
The upset $\Up{I}$ is deductively and dni-closed.
\end{proposition}
\begin{definition}
Let $f : P \to Q$ be a homomorphism of imploids.
The \defn{kernel} of $f$ is the upset $\ker f \refsup P$ given by the inverse image of the unit, $\ker f \defeq f^{-1}(I_Q)$.
\end{definition}
\begin{proposition}\label{prop:kerclosure}
Let $f : P \to Q$ be a homomorphism. Then $\ker f \refsup P$ is deductively closed, and dni-closed if $f$ is strong.
\end{proposition}
\begin{proposition}\label{prop:kerfaithful}
Let $f : P \to Q$ be a strong homomorphism of left normal imploids. Then $f$ is faithful iff $\ker f = I$.
\end{proposition}
\begin{proposition}[Universal property of the quotient]
\label{prop:quotient-is-quotient}
For any imploid $P$ and dni-closed dedupset $R \refsup P$, the function acting as the identity on elements defines a strong homomorphism of imploids $[-] : P \to \quot PR$ whose kernel is $R$.
Moreover, for any other left normal imploid $Q$ and lax (respectively, strong) homomorphism $f : P \to Q$ such that $R \subseteq \ker{f}$, there exists a unique lax (respectively, strong) imploid homomorphism $\bar{f} : \quot PR \to Q$ such that $f = \bar{f} \circ [-]$.
\end{proposition}
\begin{corollary}
For any collection $C$ of ordered pairs of elements of an imploid $P$, we can define the \defn{quotient of $P$ modulo the relations} $[a \le b]_{(a,b) \in C}$ as $\quot P{\tilde{R}_C}$, where $\tilde{R}_C$ is the deductive closure of the dni-closure of the upwards closure of the set $\set{\rimp ab \mid (a,b) \in C}$.
\end{corollary}
\noindent
Note that although our construction of the imploid quotient only defines a coarser preorder on the existing elements of $P$, we can always obtain a partially ordered set by considering the image of $\quot PR$ under the $\UpDown{(-)}$ embedding (\Cref{prop:double-yoneda-embedding}).
In the case of a group $G$ seen as a discrete imploid, quotienting by a dni-closed dedupset (= normal subgroup) $H \lhd G$ corresponds to introducing an equivalence relation on the elements of $G$, namely $a \equiv_H b$ iff $b\mul a^{-1} \in H$.
Applying $\UpDown{(-)}$ then corresponds to taking equivalence classes, and what results is just the usual construction of the group-theoretic quotient as the group of cosets of a normal subgroup.

\section{Imploid-valued flows on 3-valent maps}
\label{sec:imploid-flows}

\subsection{Background: graphs, orientations, flows, maps}
\label{sec:imploid-flows:background}

In this paper we take \defn{graph} to mean finite, undirected graph with loops and/or parallel edges \cite{Serre1980,Tutte1984book}.
Formally, such a graph can be considered as a diagram 
$
\begin{tikzcd}[cramped]
A \arrow[loop left,"e"]\arrow[r,shift left=1.75,"s"] \arrow[r,shift right=1.75,"t"'] & V
\end{tikzcd}
$
where $e$ is a fixpoint-free involution such that $t = s \circ e$.
Elements of the set $V$ are called \emph{vertices} and elements of $A$ are called \emph{arcs}, while the functions $s$ and $t$ return the \emph{source} and \emph{target} vertex of an arc.
The involution $e$ matches each arc $x$ with an \emph{opposite arc} $-x \defeq e(x)$, and the unordered pair $\orbit(e,x) = \set{x,-x}$ is called an \emph{edge}.
We write $E \defeq \orbits(e)$ for the set of edges.
The \emph{degree} of a vertex $v$ is the cardinality of the set $\set{x \mid s(x) = v}$, or equivalently of the set $\set{x \mid t(x) = v}$.
A graph is said to be \emph{trivalent} (or \emph{3-valent} or \emph{cubic}) if every vertex has degree three.
A graph is \emph{connected} if it is neither empty nor the sum of two smaller graphs.
A \emph{bridge} in a connected graph is an edge whose removal disconnects the graph.
More generally, given a non-empty subset of vertices $V^+ \subseteq V$, the set of edges $C(V^+) = \set{\orbit(e,x) \mid s(x) \in V^+ \wedge t(x) \notin V^+}$ with one end in $V^+$ and the other outside is called a \emph{cut} (a bridge is a cut containing a single edge).
A connected graph is \emph{bridgeless} (or \emph{2-edge-connected}) if it has no bridges; contrarily, it is a \emph{tree} if every edge is a bridge.

An \defn{orientation} of a graph corresponds to the selection of one arc from every edge, or in other words to the choice of a subset $A^+ \subseteq A$ of arcs such that $A = A^+ \uplus -A^+$.
Given an orientation $A^+$, we define the \emph{inputs} of a vertex as the set $\nhdi(v) \defeq t^{-1}(v) \cap A^+$ and the \emph{outputs} as the set $\nhdo(v) \defeq s^{-1}(v) \cap A^+$.
We say that a trivalent graph is \emph{well-oriented} (cf.~\cite{Lebed2015}) if it is oriented so that every vertex either has two inputs and one output (we refer to this as a \emph{negative} vertex), or one input and two outputs (we refer to this as a \emph{positive} vertex).
Note that a connected trivalent graph can always be well-oriented by considering any \emph{spanning tree} (we describe this more systematically in \Cref{sec:imploid-flows:canonical-orientation}).

Let $\Gamma = (V,A,s,t,e)$ be a connected graph equipped with an orientation $A^+\subseteq A$, and let $G$ be any abelian group.
A \defn{group-valued flow} (or $G$-flow) \cite{Tutte1954,Jaeger79} on $\Gamma$ (relative to $A^+$) is a function $\phi : E \to G$ satisfying the equation 
\begin{equation}\label{eq:kirchoff-law} \tag{Kirchhoff's law}
\sum_{x \in \nhdi v} \!\!\phi(x)\ \ = \sum_{x \in \nhdo v} \!\!\phi(x)
\end{equation}
at every vertex $v$.
(Observe that the commutativity condition on $G$ is necessary for the equation to be well-defined.)
A \emph{nowhere-zero flow} is a flow $\phi$ such that $\phi(x) \ne 0$ for all $x \in E$.
For example, below on the left we show an orientation of the complete graph $K_4$ with a nowhere-zero $\Z$-flow, and on the right another graph with a $\Z_3$-flow which is \emph{not} nowhere-zero:
$$
\vcenter{\hbox{\scalebox{1.2}{\begin{tikzpicture}[scale=1.8]
\draw (-0.433,-0.25) node (ly) [style=pt] {};
\draw (0,0.5) node (lz) [style=pt] {};
\draw (0.433,-0.25) node (ax) [style=pt] {};
\draw (0,0) node (ay) [style=pt] {};
\draw [->-] (lz) to node [left] {$\tiny 3$} (ly);
\draw [->-] (ly) to node [below,yshift=1pt] {$\tiny 2$} (ax);
\draw [->-] (ly) to node [right] {$\tiny 1$} (ay);
\draw [->-] (lz) to node [right,xshift=-2pt] {$\tiny 2$} (ay);
\draw [->-] (ax) to node [right] {$\tiny 5$} (lz);
\draw [->-] (ay) to node [left] {$\tiny 3$} (ax);
\end{tikzpicture}}}}
\qquad\qquad
\vcenter{\hbox{\scalebox{1.2}{\begin{tikzpicture}[scale=1]
\draw (0.5,1) node (a) [style=pt] {};
\draw (0,0.5) node (b) [style=pt] {};
\draw (0.5,0) node (c) [style=pt] {};
\draw (1,0.5) node (d) [style=pt] {};
\draw (1.5,0.5) node (e) [style=pt] {};
\draw (2,0) node (f) [style=pt] {};
\draw (2.5,0.5) node (g) [style=pt] {};
\draw (2,1) node (h) [style=pt] {};
\draw [->-] (a) to node [above,yshift=-2pt,xshift=-2pt] {$\tiny 2$} (b);
\draw [->-] (b) to node [below,yshift=2pt,xshift=-1pt] {$\tiny 1$} (c);
\draw [->-] (c) to node [right,yshift=-2pt,xshift=-2pt] {$\tiny 1$} (d);
\draw [->-] (d) to node [above,xshift=2pt,yshift=-2pt] {$\tiny 2$} (a);
\draw [->-] (d) to node [above,yshift=-1pt] {$\tiny 2$} (b);
\draw [->-] (d) to node [above] {$\tiny 0$} (e);
\draw [->-] (e) to node [below,yshift=2pt,xshift=-1pt] {$\tiny 2$} (f);
\draw [->-] (e) to node [above,yshift=-1pt,xshift=1pt] {$\tiny 2$} (g);
\draw [->-] (f) to node [right,yshift=-2pt,xshift=-2pt] {$\tiny 2$} (g);
\draw [->-] (g) to node [above,xshift=2pt,yshift=-2pt] {$\tiny 1$} (h);
\draw [->-] (h) to node [above,xshift=-2pt,yshift=-2pt] {$\tiny 1$} (e);
\end{tikzpicture}}}}
$$
Although the notion of flow is defined relative to a given orientation, a (nowhere-zero) flow for one orientation can be transformed into a (nowhere-zero) flow for any other simply by negating the values assigned to some edges.
Also, it is easy to prove that a graph cannot admit a nowhere-zero flow unless it is bridgeless, as a corollary of the more general fact that the net flow across any cut is always zero.
Finally, it is worth mentioning that many questions about flows on general graphs can be reduced to questions about flows on trivalent graphs
(cf.~\cite{Jaeger79}).

In this paper we take \defn{map} to mean cellular embedding of a connected graph $\Gamma$ into a connected, compact oriented surface \cite{LZgraphs,Eynard2016}.
Formally, such an embedding is determined up to orientation-preserving homeomorphism of the underlying surface by the purely combinatorial data of an additional permutation $v : A \to A$ on the arcs of $\Gamma$, assuming that $V \cong \orbits(v)$, and that $s$ factors via the function $x \mapsto \orbit(v,x)$ sending an arc to its $v$-orbit.
The \emph{faces} of a map are then defined as the orbits of the permutation $f \defeq (e \circ v)^{-1}$, and the \emph{genus} $g$ of the underlying surface can be determined from its \emph{Euler characteristic}
$\chi \defeq \card{\orbits(v)} - \card{\orbits(e)} + \card{\orbits(f)} = 2 - 2g.$
(A \emph{planar map} is a map of genus $g=0$.)
The triple of permutations $(v,e,f)$ (or equivalently the pair $(v,e)$), which up to isomorphism determines the graph, the surface, and the embedding, is sometimes referred to as a ``combinatorial'' map.
Every combinatorial map also has a \emph{dual map}
$
(v,e,f)^* \defeq (f^{-1},e,v^{-1})
$
in which the role of vertices and faces is reversed.
For example, any \emph{trivalent map} (i.e., a map whose underlying graph is trivalent) on a given surface induces a dual \emph{triangulation} of the same surface, and vice versa.
One of the reasons trivalent maps in particular arise as natural objects of study is that they have close connections to the \emph{modular group} $\mathrm{PSL}(2,\mathbb{Z}) \cong \langle v,e \mid v^3 = e^2 = I\rangle$ (cf.~\cite{Serre1980,JonesSingerman1994,Vidal2010phd}).

\begin{figure}[t]
$$
\vcenter{\hbox{\begin{tikzpicture}[scale=1]
\draw[style=dotted] (0,0) circle [radius=1];
\draw (0,0) circle [radius=0.5];
\draw (0,-1) node (root) [style=root] {};
\draw (0,-0.5) node (lx) [style=pt] {};
\draw (-0.433,-0.25) node (ly) [style=pt] {};
\draw (0,0.5) node (lz) [style=pt] {};
\draw (0.433,-0.25) node (ax) [style=pt] {};
\draw (0,0) node (ay) [style=pt] {};
\draw (lx) to (root.center);
\draw (ly) to (ay);
\draw (lz) to (ay);
\draw (ay) to (ax);
\end{tikzpicture}}}
\quad
\vcenter{\hbox{\begin{tikzpicture}[scale=1]
\draw[style=dotted] (0,0) circle [radius=1];
\draw (0,0) circle [radius=0.5];
\draw (0,-1) node (root) [style=root] {};
\draw (0,-0.5) node (lx) [style=pt] {};
\draw (-0.354,-0.354) node (ly) [style=pt] {};
\draw (-0.354,0.354) node (lz) [style=pt] {};
\draw (0.354,0.354) node (argy) [style=pt] {};
\draw (0.354,-0.354) node (argz) [style=pt] {};
\draw (lx) to (root.center);
\draw (ly) to (argy);
\draw (lz) to (argz);
\end{tikzpicture}}}
\quad
\vcenter{\hbox{\begin{tikzpicture}[scale=1]
\draw[style=dotted] (0,0) circle [radius=1];
\draw (0,-1) node (root) [style=root] {};
\draw (0,1) node (var) {};
\draw (1,0) node (var2) {};
\draw (0,-0.33) node (appx) [style=pt] {};
\draw (0.5,-0.33) node (ly) [style=pt] {};
\draw (0,0.33) node (appz) [style=pt] {};
\draw (var.center) to (appz);
\draw (appz) to (appx);
\draw (appx) to (root.center);
\draw (ly) to (appx);
\path (ly) edge [loop right,looseness=5,min distance=1.5em,in=0,out=60] (ly);
\draw (var2.center) to (appz);
\end{tikzpicture}}}
\ \ 
\vcenter{\hbox{\begin{tikzpicture}[scale=1]
\draw[style=dotted] (0,0) circle [radius=1];
\draw (0,1) node (root) {};
\draw (0,-1) node (var) [style=root] {};
\draw (-1,0) node (var2) {};
\draw (0,0.33) node (appx) [style=pt] {};
\draw (-0.5,0.33) node (ly) [style=pt] {};
\draw (0,-0.33) node (appz) [style=pt] {};
\draw (appz) to (var.center);
\draw (appz) to (appx);
\draw (appx) to (root.center);
\draw (ly) to (appx);
\path (ly) edge [loop right,looseness=5,min distance=1.5em,in=180,out=240] (ly);
\draw (var2.center) to (appz);
\end{tikzpicture}}}
$$
\caption{Some small examples of rooted 3-valent maps.}
\label{fig:rtm-examples}
\end{figure}
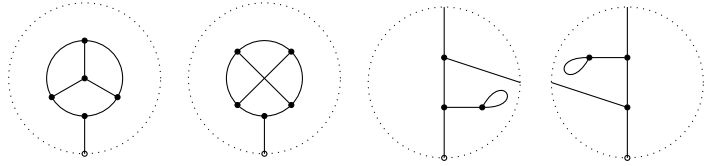

A \defn{rooted map} is a map equipped with a distinguished root arc $x_0 \in A$, considered up to root-preserving isomorphism.
The study of rooted maps was initiated by Tutte in a series of papers on the combinatorics of planar maps \cite{Tutte1962planartriangulations,Tutte1962hamiltonian,Tutte1963planarmaps}, taking advantage of the fact that rooted maps have no symmetries and so are easier to count.
A rooted map can also be seen as a \emph{map with marked boundary}.
While the classical theory of combinatorial maps \cite{JonesSingerman1978,Tutte1984book} is formulated in terms of surfaces without boundary (such as the sphere or torus), it is possible to consider boundaries as distinguished faces representing ``holes'' in the surface \cite{Eynard2016}.
After removing these faces what is left is an \emph{open graph} in the sense that some edges have only one end attached to a vertex and the other attached to the boundary: see \Cref{fig:rtm-examples} for such depictions of \defn{rooted trivalent maps} as trivalent maps with a marked boundary. 
Observe that the first two diagrams in \Cref{fig:rtm-examples} represent \emph{two different embeddings} of the same underlying graph: the first into a surface of genus 0 (the open disc), the second into a surface of genus 1 (the open disc with a handle attached -- the crossing in the diagram should be thought of as ``virtual'', arising from the projection of this higher genus surface down to the page).
In contrast, the second pair of diagrams represent \emph{two different rootings} of the same underlying map: if we forget the marking of the root, then a $180^\circ$ rotation of the disc witnesses an isomorphism between the two diagrams.

It's not unreasonable to think of the boundary of a rooted trivalent map as a single ``external vertex'' of arbitrary positive degree\footnote{This is dual to Tutte's original treatment of rooted planar triangulations \cite{Tutte1962planartriangulations}, which included an external face of unbounded degree.}, or at least as a \emph{cut} across which values can flow between its interior and its exterior.
In fact, the extra structure of the vertex permutation that comes with a combinatorial map naturally enables a more general notion of flow valued in arbitrary (not necessarily abelian) groups, but one important point of divergence with the theory of flows on abstract graphs (see \cite{GKRV2017tuttepolymaps} for a discussion) is that in the case of a non-planar map, the local condition on vertices does not automatically extend to arbitrary cuts.
After formulating the appropriate definitions, our main results in this section characterize when an \emph{imploid-valued flow} is guaranteed to have such a global extension property, relating the flow at each trivalent vertex to the flow across the boundary.

\subsection{Imploid-valued flows}
\label{sec:imploid-flows:defn}

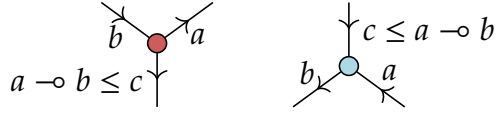
\begin{figure}[t]
$$
\vcenter{\hbox{\scalebox{1.5}{\begin{tikzpicture}
  \node (lam) [style=lam] {};
  \node (root) [below=1em of lam] {};
  \node (var) [above right=0.5em and 0.866em of lam] {};
  \node (body) [above left=0.5em and 0.866em of lam] {};
  \path
     (lam) edge [->-] node [left] {\footnotesize$\rimp{a}{b}\le c$} (root.center)
     (lam) edge [->-] node [near start,right,yshift=-2pt] {\footnotesize$a$} (var.center)
     (body.center) edge [->-] node [near end,left,yshift=-2pt] {\footnotesize$b$} (lam);  
\end{tikzpicture}}}}
\qquad
\vcenter{\hbox{\scalebox{1.5}{\begin{tikzpicture}
  \node (app) [style=app] {};
  \node (fn) [above=1em of app] {};
  \node (cont) [below left=0.5em and 0.866em of app] {};
  \node (arg) [below right=0.5em and 0.866em of app] {};
  \path
    (fn.center) edge [->-] node [right] {\footnotesize$c \le \rimp{a}{b}$} (app)
    (app) edge [->-] node [near start,left,yshift=2pt] {\footnotesize$b$} (cont.center)
    (arg.center) edge [->-] node [near end,right,yshift=2pt] {\footnotesize$a$} (app);  
\end{tikzpicture}}}}
$$
\caption{Defining relations for imploid-valued flows on well-oriented 3-valent maps.}
\label{fig:localflow}
\end{figure}

\begin{notation}
Suppose given a well-oriented 3-valent map $T$.
We write $\vpos xyz \in T$ to indicate that $T$ contains a positive vertex with output $x$, input $y$, and output $z$ as listed in \uline{counterclockwise} order.
Dually, we write $\vneg xyz \in T$ to indicate that $T$ contains a negative vertex with input $x$, output $y$, and input $z$ as listed in \uline{clockwise} order.
\end{notation}
\begin{definition}\label{defn:imploid-localflow}
An \defn{imploid-valued flow} on 
a well-oriented 3-valent map $T$
is a function $\phi : E \to P$ assigning each edge a value in some left normal imploid $P$, such that the relation
\begin{equation} \label{3-flow+} \tag{3-flow$^+$}
\phi(x) \imp \phi(y) \le \phi(z)
\end{equation}
holds at every positive vertex $\vpos xyz \in T$, and the relation
\begin{equation} \label{3-flow-} \tag{3-flow$^-$}
\phi(z) \le \phi(x) \imp \phi(y)
\end{equation}
holds at every every negative vertex $\vneg xyz \in T$.
(These relations are summarized visually in \Cref{fig:localflow}.)
\end{definition}
\noindent
Notice that in \Cref{fig:localflow} we have used colors to help visually distinguish positive vertices (blue) from negative vertices (red), as we will continue doing throughout the paper.
\begin{definition}
A flow $\phi : E \to P$ is said to be \defn{nowhere-unit}  if $I \not\le \phi(x)$ for all $x \in E$, where $I$ is the unit of $P$.
\end{definition}
\begin{proposition}\label{prop:flowpush}
A (nowhere-unit) flow $\phi : E \to P$ may be pushed forward along any (faithful) strong homomorphism $f : P \to Q$ to obtain a (nowhere-unit) flow $f \phi : E \to Q$ defined by post-composition.
\end{proposition}
\begin{example}\label{ex:bubble-nzflows}
The oriented ``bubble'' 
$
\vcenter{\hbox{\begin{tikzpicture}[scale=0.75]
\draw[style=dotted] (0,0) circle [radius=1];
\draw (0,-1) node (root) {};
\draw (0,1) node (var) {};
\draw (0,0.33) node (app) [style=ptapp] {};
\draw (0,-0.33) node (lam) [style=ptlam] {};
\draw [->-] (lam) to (root.center);
\draw [->-] (var.center) to (app);
\draw [->-,bend right=60] (lam) to (app);
\draw [->-,bend right=60] (app) to (lam);
\end{tikzpicture}}}
$
admits a nowhere-unit $P$-flow just in case $P$ contains a pair of elements $a$ and $b$ such that $I \not\le a$ and $I \not\le b$ and $a \not\le b$.
\end{example}

\begin{definition}
Let $T$ be a 3-valent map with boundary $\partial{T}$.
We say that an arc is an \defn{input of $T$} (respectively, \defn{output of $T$}) if its source (resp., target) lies in $\partial{T}$ (otherwise, the arc is \emph{internal} to $T$).
\end{definition}
\noindent
\begin{definition}
Let $T$ be a well-oriented rooted 3-valent map. 
We say that $T$ is \defn{globally well-oriented} if its orientation contains exactly one output of $T$, whose target is the root.
\end{definition}
\noindent
For example, here are two different global well-orientations of the third rooted map in \Cref{fig:rtm-examples}:
$$
\vcenter{\hbox{\begin{tikzpicture}[scale=0.8]
\draw[style=dotted] (0,0) circle [radius=1];
\draw (0,-1) node (root) {};
\draw (0,1) node (var) {};
\draw (1,0) node (var2) {};
\draw (0,-0.33) node (appx) [style=ptapp] {};
\draw (0.5,-0.33) node (ly) [style=ptlam] {};
\draw (0,0.33) node (appz) [style=ptapp] {};
\draw [->-] (var.center) to (appz);
\draw [->-] (appz) to (appx);
\draw [->-] (appx) to (root.center);
\draw [->-] (ly) to (appx);
\path (ly) edge [-->-,loop left,looseness=5,min distance=1.5em,in=60,out=0] (ly);
\draw [->-] (var2.center) to (appz);
\end{tikzpicture}}}
\qquad
\vcenter{\hbox{\begin{tikzpicture}[scale=0.8]
\draw[style=dotted] (0,0) circle [radius=1];
\draw (0,-1) node (root) {};
\draw (0,1) node (var) {};
\draw (1,0) node (var2) {};
\draw (0,-0.33) node (appx) [style=ptlam] {};
\draw (0.5,-0.33) node (ly) [style=ptapp] {};
\draw (0,0.33) node (appz) [style=ptapp] {};
\draw [->-] (var.center) to (appz);
\draw [->-] (appz) to (appx);
\draw [->-] (appx) to (root.center);
\draw [->-] (appx) to (ly);
\path (ly) edge [-->-,loop left,looseness=5,min distance=1.5em,in=60,out=0] (ly);
\draw [->-] (var2.center) to (appz);
\end{tikzpicture}}}
$$
\begin{notation}
We write $\partial T = [x_0;x_1,\dots,x_n]$ to indicate that $x_0$ is the unique output of a globally well-oriented map $T$, followed by inputs $x_1,\dots,x_n$ in clockwise order around the boundary.
\end{notation}
\begin{notation}
Let $P$ be any imploid, $\vec{a} = a_1,\dots,a_n \in P$ a list of elements, and $b \in P$ a distinguished element.
We write $\mimp{\vec{a}}{b}$ for the right-associated implication defined inductively by $\mimp{\cdot}{b} \defeq b$, $\mimp{(\vec{a},a)}{b} \defeq \mimp{\vec{a}}{(a \imp b)}$.
\end{notation}
\begin{definition}
Let $T$ be a globally well-oriented 3-valent map, with $\partial T = [x_0;x_1,\dots,x_n]$.
We say that a flow $\phi$ on $T$ satisfies the \defn{global flow condition} if the following relation holds:
\begin{equation}\tag{global flow}
I \le \mimp{(\phi(x_1),\dots,\phi(x_n))}{\phi(x_0)}
\end{equation}
\end{definition}
\noindent
Before we move on to study the notion of imploid-valued flow, let's record the following easy observation, which relates it to the classical notion.
\begin{proposition}\label{prop:classicalflow}
Let $T$ be a globally well-oriented 3-valent map, and let $G$ be an abelian group, seen as a discrete symmetric imploid.
Then a (nowhere-unit) flow $\phi : E \to G$ on $T$ is the same thing as a group-valued (nowhere-zero) flow on the underlying graph of $T$.
As a consequence, any flow $\phi : E \to G$ necessarily satisfies the global flow condition.
\end{proposition}
\begin{remark}\label{remark:nowhereunit-flow-np-complete}
Since deciding the existence of a proper edge-3-coloring for an abstract cubic graph is NP-complete \cite{Holyer1981}, the problem of deciding for an arbitrary well-oriented 3-valent map $T$ and finite imploid $P$ whether $T$ admits a nowhere-unit $P$-flow is likewise NP-complete, taking $P = \V$.
(Of course the problem might be easier in the case of particular imploids or classes of maps.
For example, every bridgeless planar 3-valent map admits a nowhere-zero $\V$-flow \cite{AppelHaken1977}.)
\end{remark}

\subsection{Topological orientations as linear lambda terms}
\label{sec:imploid-flows:canonical-orientation}

As mentioned in \Cref{sec:imploid-flows:background}, every connected trivalent graph can be well-oriented.
Indeed, given the extra data of an embedding and a rooting, there is a canonical way of reconstructing such an orientation, which we refer to as the \emph{topological orientation} of a rooted 3-valent map.
The definition rests on the fact that any rooted 3-valent map $T$ must have one of the following three schematic forms:
\begin{center}
\begin{inparaenum}[1)]
\item 
$
\vcenter{\hbox{\begin{tikzpicture}[scale=1]
\draw[style=dotted] (0,0) circle [radius=1];
\draw (0,-1) node (root) [style=root] {};
\draw (0,-0.5) node (app) [style=pt] {};
\draw (-0.433,0) node (fn) [shape=circle,draw,dotted,fill=lightgray] {$T_1$};
\draw (0.433,0) node (arg) [shape=circle,draw,dotted,fill=lightgray] {$T_2$};
\draw (-0.924,0.383) node (var1) {};
\draw (-0.383,0.924) node (var2) {};
\draw (0.383,0.924) node (var3) {};
\draw (0.924,0.383) node (var4) {};
\draw (app) to (root.center);
\draw (fn) to (app);
\draw (arg) to (app);
\draw (var1.center) to (fn);
\draw (var2.center) to (fn);
\draw (var3.center) to (arg);
\draw (var4.center) to (arg);
\end{tikzpicture}}}
$;
or 
\item 
$
\vcenter{\hbox{\begin{tikzpicture}[scale=1]
\draw[style=dotted] (0,0) circle [radius=1];
\draw (0,-1) node (root) [style=root] {};
\draw (0,-0.5) node (lam) [style=pt] {};
\draw (-0.433,0) node (body) [shape=circle,draw,dotted,fill=lightgray] {$T_1$};
\draw (-0.924,0.383) node (var1) {};
\draw (0.283,0.524) node (var3) {};
\draw (0.924,0.383) node (var4) {};
\draw (lam) to (root.center);
\draw (body) to (lam);
\draw (var1.center) to (body);
\draw (lam) to [bend right=60] (var3) to [bend right=60] (body);
\draw (var4.center) to (body);
\end{tikzpicture}}}
$;
or
\item
$
\vcenter{\hbox{\begin{tikzpicture}[scale=0.5]
\draw[style=dotted] (0,0) circle [radius=1];
\draw (0,-1) node (root) [style=root] {};
\draw (0,1) node (var) {};
\draw (var.center) to (root.center);
\end{tikzpicture}}}
$ .
\end{inparaenum}
\end{center}
In other words, if we remove the vertex incident to the root, either
\begin{inparaenum}[1)]
\item the map becomes disconnected; or 
\item it stays connected; or
\item there was no such vertex in the first place.
\end{inparaenum}
\begin{defprop}\label{prop:topological-orientation}
Every rooted 3-valent map may be globally well-oriented by its \defn{topological orientation}, defined as follows by induction on the number of vertices:
\begin{center}
\begin{inparaenum}[1)]
\item 
$
\vcenter{\hbox{\begin{tikzpicture}[scale=1]
\draw[style=dotted] (0,0) circle [radius=1];
\draw (0,-1) node (root) {};
\draw (0,-0.5) node (app) [style=ptapp] {};
\draw (-0.433,0) node (fn) [shape=circle,draw,dotted,fill=lightgray] {$T_1$};
\draw (0.433,0) node (arg) [shape=circle,draw,dotted,fill=lightgray] {$T_2$};
\draw (-0.924,0.383) node (var1) {};
\draw (-0.383,0.924) node (var2) {};
\draw (0.383,0.924) node (var3) {};
\draw (0.924,0.383) node (var4) {};
\draw [->-] (app) to (root.center);
\draw [->-] (fn) to (app);
\draw [->-] (arg) to (app);
\draw [->-] (var1.center) to (fn);
\draw [->-] (var2.center) to (fn);
\draw [->-] (var3.center) to (arg);
\draw [->-] (var4.center) to (arg);
\end{tikzpicture}}}
$;
\item 
$
\vcenter{\hbox{\begin{tikzpicture}[scale=1]
\draw[style=dotted] (0,0) circle [radius=1];
\draw (0,-1) node (root) {};
\draw (0,-0.5) node (lam) [style=ptlam] {};
\draw (-0.433,0) node (body) [shape=circle,draw,dotted,fill=lightgray] {$T_1$};
\draw (-0.924,0.383) node (var1) {};
\draw (0.283,0.524) node (var3) {};
\draw (0.924,0.383) node (var4) {};
\draw [->-] (lam) to (root.center);
\draw [->-] (body) to (lam);
\draw [->-] (var1.center) to (body);
\draw [-->-] (lam) to [bend right=60] (var3) to [bend right=60] (body);
\draw [-->-] (var4.center) to (body);
\end{tikzpicture}}}
$;
\item
$
\vcenter{\hbox{\begin{tikzpicture}[scale=0.5]
\draw[style=dotted] (0,0) circle [radius=1];
\draw (0,-1) node (root) {};
\draw (0,1) node (var) {};
\draw[->-] (var.center) to (root.center);
\end{tikzpicture}}}
$ .
\end{inparaenum}
\end{center}
\end{defprop}
\noindent
As some examples, here are the topological orientations of the four rooted 3-valent maps displayed in \Cref{fig:rtm-examples}:
$$
\vcenter{\hbox{\begin{tikzpicture}[scale=0.8]
\draw[style=dotted] (0,0) circle [radius=1];
\draw [decoration={markings, mark=at position 0.1 with {\arrow{>}}, mark=at position 0.45 with {\arrow{>}}, mark=at position 0.7 with {\arrow{>}}, mark=at position 0.85 with {\arrow{>}}},postaction={decorate}] (0,0) circle [radius=0.5];
\draw (0,-1) node (root) {};
\draw (0,-0.5) node (lx) [style=ptlam] {};
\draw (-0.433,-0.25) node (ly) [style=ptlam] {};
\draw (0,0.5) node (lz) [style=ptlam] {};
\draw (0.433,-0.25) node (ax) [style=ptapp] {};
\draw (0,0) node (ay) [style=ptapp] {};
\draw [->-] (lx) to (root.center);
\draw [->-] (ly) to (ay);
\draw [->-] (lz) to (ay);
\draw [->-] (ay) to (ax);
\end{tikzpicture}}}
\quad
\vcenter{\hbox{\begin{tikzpicture}[scale=0.8]
\draw[style=dotted] (0,0) circle [radius=1];
\draw [decoration={markings, mark=at position 0 with {\arrow{>}}, mark=at position 0.25 with {\arrow{>}}, mark=at position 0.5 with {\arrow{>}}, mark=at position 0.7 with {\arrow{>}}, mark=at position 0.83 with {\arrow{>}}},postaction={decorate}] (0,0) circle [radius=0.5];
\draw (0,-1) node (root) {};
\draw (0,-0.5) node (lx) [style=ptlam] {};
\draw (-0.354,-0.354) node (ly) [style=ptlam] {};
\draw (-0.354,0.354) node (lz) [style=ptlam] {};
\draw (0.354,0.354) node (argy) [style=ptapp] {};
\draw (0.354,-0.354) node (argz) [style=ptapp] {};
\draw [->-] (lx) to (root.center);
\draw [-->-] (ly) to (argy);
\draw [-->-] (lz) to (argz);
\end{tikzpicture}}}
\quad
\vcenter{\hbox{\begin{tikzpicture}[scale=0.8]
\draw[style=dotted] (0,0) circle [radius=1];
\draw (0,-1) node (root) {};
\draw (0,1) node (var) {};
\draw (1,0) node (var2) {};
\draw (0,-0.33) node (appx) [style=ptapp] {};
\draw (0.5,-0.33) node (ly) [style=ptlam] {};
\draw (0,0.33) node (appz) [style=ptapp] {};
\draw [->-] (var.center) to (appz);
\draw [->-] (appz) to (appx);
\draw [->-] (appx) to (root.center);
\draw [->-] (ly) to (appx);
\path (ly) edge [-->-,loop left,looseness=5,min distance=1.5em,in=60,out=0] (ly);
\draw [->-] (var2.center) to (appz);
\end{tikzpicture}}}
\ \ 
\vcenter{\hbox{\begin{tikzpicture}[scale=0.8]
\draw[style=dotted] (0,0) circle [radius=1];
\draw (0,1) node (root) {};
\draw (0,-1) node (var) {};
\draw (-1,0) node (var2) {};
\draw (0,0.33) node (appx) [style=ptapp] {};
\draw (-0.5,0.33) node (ly) [style=ptlam] {};
\draw (0,-0.33) node (appz) [style=ptapp] {};
\draw [->-] (appz) to (var.center);
\draw [->-] (appx) to (appz);
\draw [->-] (root.center) to (appx);
\draw [->-] (ly) to (appx);
\path (ly) edge [-<--,loop right,looseness=5,min distance=1.5em,in=180,out=240] (ly);
\draw [->-] (var2.center) to (appz);
\end{tikzpicture}}}
$$
To be a bit more formal, let $\Theta(n)$ denote the set of isomorphism classes of rooted 3-valent maps with $n$ non-root arcs incident to the boundary.
One way of constructing a rooted 3-valent map is to glue a pair of maps $T_1$ and $T_2$ by their roots onto a fresh trivalent vertex (case 1), corresponding to a natural family of operations
$
@:\Theta(n_1) \times \Theta(n_2) \to \Theta(n_1+n_2).
$
Another way is to pick one of the non-root arcs on the boundary of $T_1$ and glue it together with $T_1$'s root onto a fresh trivalent vertex (case 2), corresponding to a natural family of operations
$
\lambda_i : \Theta(n+1)  \to \Theta(n) 
$
for every $1 \le i \le n+1$.
Together, these two operations generate \emph{all} rooted 3-valent maps starting from the trivial rooted map (case 3), and since they naturally extend to operations 
on globally well-oriented maps (where $@$ introduces a negative vertex and $\lambda_i$ a positive vertex) this explains the definition of the topological orientation.

Now, we can also observe that $\Theta(n)$ has the structure of a \emph{symmetric operad} \cite{MarklSchniderStasheff}, meaning that there is a natural family of \emph{composition} operations
$
\circ_i : \Theta(m+1) \times \Theta(n) \to \Theta(m+n)
$ ($1 \le i \le m+1$) together with an action of the symmetric group $S_n$ on $\Theta(n)$, satisfying appropriate axioms of associativity, unitality, and equivariance.
Composition corresponds to grafting the root of one map onto a boundary arc of another,
while the (free) action of the symmetric group corresponds to permuting the boundary arcs.
\hide{
$$
\vcenter{\hbox{\begin{tikzpicture}[scale=1]
\draw[style=dotted] (0,0) circle [radius=1];
\draw (0,-1) node (root) [style=root] {};
\draw (0,-0.3) node (fn) [shape=circle,draw,dotted,fill=lightgray] {$T_1$};
\draw (0,0.55) node (arg) [shape=circle,draw,dotted,fill=lightgray,scale=0.7] {$T_2$};
\draw (-0.924,0.383) node (var1) {};
\draw (-0.383,0.924) node (var2) {};
\draw (0.383,0.924) node (var3) {};
\draw (0.924,0.383) node (var4) {};
\draw (fn) to (root.center);
\draw (arg) to (fn.north);
\draw (var1.center) to (fn);
\draw (var2.center) to (arg);
\draw (var3.center) to (arg);
\draw (var4.center) to (fn);
\end{tikzpicture}}}
\qquad
\vcenter{\hbox{\begin{tikzpicture}[scale=1]
\draw[style=dotted] (0,0) circle [radius=1];
\draw (0,-1) node (root) [style=root] {};
\draw (0,0) node (app) [shape=circle,draw,dotted,fill=lightgray] {$T$};
\draw (-0.924,0.383) node (var1) {};
\draw (-0.383,0.924) node (var2) {};
\draw (0.383,0.924) node (var3) {};
\draw (0.924,0.383) node (var4) {};
\draw (app) to (root.center);
\draw (var1.center) to (app);
\draw (var2.center) to (app.north east);
\draw (var3.center) to (app.north west);
\draw (var4.center) to (app);
\end{tikzpicture}}}
$$}

At this point, the reader with a background in lambda calculus may recognize that our description of $\Theta(n)$ as a symmetric operad exactly mirrors the syntactic structure of \defn{linear lambda terms} \emph{with $n$ free variables}, 
reading $@$ as \emph{application} $T_1(T_2)$ and $\lambda_i$ as \emph{abstraction in the $i^\text{th}$ variable} $\lambda x_i.T_1$, and interpreting grafting by \emph{substitution} and the symmetric action by \emph{variable exchange}.
Indeed, this operadic perspective (cf.~\cite{Hyland2017moderndress}) is one way of understanding the one-to-one correspondence between (isomorphism classes of) rooted 3-valent maps and ($\alpha$-equivalence classes of) linear lambda terms: the latter may be understood as \emph{complete invariants} of rooted 3-valent maps, corresponding to their topological orientations \cite{BoGaJa2013,Z2016trivalent}.
For instance, under this correspondence, the first two of the four examples above (the topological orientations of the rooted planar tetrahedron and rooted toric tetrahedron) correspond to the \emph{B combinator} $\lambda x.\lambda y.\lambda z.x(yz)$ and \emph{C combinator} $\lambda x.\lambda y.\lambda z.(xz)y$, respectively, in the sense of classical combinatory logic \cite{Hindley1997} (see \cite[Example 1]{Z2016trivalent}).

Finally, let us draw attention to the fact that $\Theta(n)$ also has several significant \emph{suboperads,} corresponding to natural subfamilies of maps and subsystems of lambda calculus.
By restricting to maps constructed using $@$ and the operation $\lambda_{n+1}$
we obtain the (non-symmetric) operad $\Theta_0(n)$ of \defn{planar} (i.e., genus 0) rooted 3-valent maps.
Note this corresponds to the restriction on linear lambda terms that variables are used in the order they are abstracted (i.e., the forbidding of exchange).
By restricting the domain and codomain of the operations $@$ and $\lambda_i$ to maps with at least one non-root boundary arc, we obtain the operad $\Theta^{2}(n)$ of bridgeless (i.e., 2-edge-connected) rooted 3-valent maps.
This corresponds to the restriction on linear lambda terms that they have no closed subterms, which we refer to as being \defn{unitless.}\footnote{
In other words, a unitless term is one that can be constructed in the absence of the empty (unit) context.
This property (with a very minor technical variation) was called being ``indecomposable'' in \cite{Z2016trivalent}.
}

\subsection{Topological flows are global}
\label{sec:imploid-flows:topological-flows}

The connection to lambda calculus suggests another way of understanding the global flow condition.
In the case of a topological orientation, the problem of building a flow on a rooted 3-valent map may be recast as one of constructing a \emph{linear typing derivation} for the corresponding lambda term.
From this the global flow condition follows by an easy proof theory-style argument (we also give a more conceptual explanation in \Cref{sec:flow-rewriting}).
\begin{proposition}\label{prop:flows-as-derivations}
Let $T$ be a topologically oriented rooted 3-valent map, and let $\partial T = [x_0;x_1,\dots,x_n]$.
Then $T$ has a flow $\phi$ such that
$\phi(x_1) = a_1, \dots, \phi(x_n) = a_n$ and $\phi(x_0) = b$
iff the judgment
$x_1:a_1,\dots,x_n:a_n \vdash T:b$
is derivable in the following type system, where the boxed rule is only needed in the non-planar case: 
$$
\small
\infer[(c \le a \imp b)]{\Gamma,\Delta \vdash T_1(T_2) : b}{\Gamma \vdash T_1 : c & \Delta \vdash T_2 : a}
\qquad
\infer[(a \imp b \le c)]{\Gamma \vdash \lambda x.T_1 : c}{\Gamma,x:a \vdash T_1 : b}
$$
$$
\small
\infer{x : a \vdash x : a}{}
\quad\qquad
\dbox{\infer{\Gamma,x:a,y:b,\Delta \vdash T : c}{\Gamma,y:b,x:a,\Delta \vdash T : c}}
$$
\end{proposition}
\noindent
\begin{lemma}\label{lemma:top-flow-is-global}
Let $T$ be a (non-planar) rooted 3-valent map equipped with its topological orientation.
Then any flow $\phi$ on $T$ valued in an arbitrary (symmetric) imploid $P$ satisfies the global flow condition.
\end{lemma}
\begin{corollary}\label{corr:nowhere-unit-bridgeless}
Let $T$ be a (non-planar) rooted 3-valent map equipped with its topological orientation.
If $T$ has a nowhere-unit flow $\phi$ valued in a (symmetric) imploid $P$, then $T$ is bridgeless.
\end{corollary}

\subsection{Non-topological orientations can violate global flow}
\label{sec:imploid-flows:non-topological-flows}

Conversely, there is no such guarantee for non-topological orientations of rooted 3-valent maps, and indeed, for any such orientation we can always exhibit an explicit \emph{counterexample} in the form of an assignment $\phi : E \to P$ valued in a specific (symmetric left normal) imploid $P$, such that $\phi$ satisfies the local relations \eqref{3-flow+} and \eqref{3-flow-} but violates the global flow condition.
For this purpose, consider the imploid $P = \hat{2}$ consisting of three linearly ordered elements $0 < 1 < 2$ with the implication $a \imp b$ defined as follows:
\begin{center}
\begin{tabular}{c|ccc}
\diagbox[height=1.25\line]{$a$}{$b$} & $0$ & $1$ & $2$ \\
\hline
$0$ & $2$ & $2$ & $2$ \\
$1$ & $0$ & $1$ & $2$ \\
$2$ & $0$ & $0$ & $2$
\end{tabular}
\end{center}
Observe that $\hat{2}$ is isomorphic to the imploid of downsets associated to the unique idempotent skew monoid with two elements $2 = (\{1,2\},\le,\max, 1)$.
\begin{lemma}\label{lemma:nontop-flow-violation}
Let $T$ be a rooted 3-valent map equipped with a well-orientation $A^+$ containing the root $x_0$ as an output, and the remaining boundary edges $x_1,\dots,x_n$ as either inputs or outputs.
If $A^+$ is non-topological then there is a $\hat{2}$-flow $\phi$ such that $\phi(x_0) = 0$, and $\phi(x_i) = 1$ or $2$ for all $1\le i\le n$.
\end{lemma}
\noindent
As a corollary, we obtain the following characterization:
\begin{theorem}\label{thm:nontop-flow-is-nonglobal}
Let $T$ be a globally well-oriented 3-valent map. The following are equivalent:
\begin{enumerate}
\item $T$ is topologically oriented, i.e., has the orientation of a linear lambda term.
\item Every $\hat{2}$-flow on $T$ satisfies the global flow condition.
\item Every $P$-flow on $T$ satisfies the global flow condition, for any symmetric imploid $P$.
\end{enumerate}
\end{theorem}
\begin{example}\label{ex:nonglobal-flow}
A pair of non-global $\hat{2}$-flows on non-lambda terms:
$$
\begin{tikzpicture}[scale=1.2]
\draw[style=dotted] (0,0) circle [radius=1];
\draw (0,-1) node (root) {};
\draw (0,1) node (var) {};
\draw (1,0) node (var2) {};
\draw (0,-0.33) node (appx) [style=lam] {};
\draw (0.5,-0.33) node (ly) [style=app] {};
\draw (0,0.33) node (appz) [style=app] {};
\draw [->-] (var.center) to node [left] {\footnotesize$1$} (appz);
\draw [->-] (appz) to node [left] {\footnotesize$1$} (appx);
\draw [->-] (appx) to node [left] {\footnotesize$0$} (root.center);
\draw [->-] (appx) to node [below,yshift=1pt] {\footnotesize$2$} (ly);
\path (ly) edge [-->-,loop left,looseness=5,min distance=1.5em,in=60,out=0] node [below] {\footnotesize$0$} (ly);
\draw [->-] (var2.center) to node [above] {\footnotesize$1$} (appz);
\end{tikzpicture}
\qquad
\begin{tikzpicture}
  \draw (0,-1.5) node (root1) {};
  \draw (0,-1) node (app1) [style=app] {};
  \draw (0.5,-0.5) node (lam2) [style=lam] {};
  \draw (-0.5,-0.5) node (lam3) [style=lam] {};
  \draw (0,0) node (app4) [style=app] {};
  \draw (0,0.5) node (lam5) [style=lam] {};
  \draw (-0.6,0.8) node (lam6) [style=lam] {};
  \draw (0.6,0.8) node (app7) [style=app] {};
  \path (lam3) edge [->-,bend right] node [left] {\footnotesize$2$} (app1);
  \path (lam2) edge [->-,bend left] node [right] {\footnotesize$0$} (app1);
  \path (app1) edge [->-] node [right] {\footnotesize$0$} (root1.center);
  \path (app4) edge [->-,bend left] node [left,yshift=-2pt,xshift=1pt] {\footnotesize$0$} (lam2);
  \path (lam3) edge [->-,bend left] node [right,xshift=-1pt,yshift=-1pt] {\footnotesize$2$} (app4);
  \path (lam5) edge [->-] node [right] {\footnotesize$0$} (app4);
  \path (lam6) edge [->-,bend right] node [left] {\footnotesize$2$} (lam3);
  \path (lam2) edge [->-,bend right] node [right] {\footnotesize$2$} (app7);
  \path (app7) edge [->-,bend right] node [above] {\footnotesize$2$} (lam6);
  \path (lam6) edge [->-,bend right] node [below,yshift=1pt] {\footnotesize$0$} (lam5);
  \path (lam5) edge [->-,bend right] node [below] {\footnotesize$2$} (app7);
\end{tikzpicture}
$$
\end{example}
\noindent
\Cref{thm:nontop-flow-is-nonglobal} says in a sense that the global flow condition acts as a ``correctness criterion'' in the terminology of linear logic \cite{Girard1987}.
(Indeed it has similarities with de~Groote's algebraic criterion for intuitionistic proof-nets \cite{DeGroote1999}. We elaborate on the relationship with proof-nets a bit more in \Cref{sec:polarized-flows}.)
It may be surprising that such a small imploid is powerful enough to distinguish topological orientations from non-topological ones, although this is consistent with the fact that the topological orientation of a rooted 3-valent map can be computed efficiently, in a single depth-first traversal \cite{BoGaJa2013}.

\subsection{Fundamental imploids and universal flows}
\label{sec:imploid-flows:funimps}

Following a familiar pattern of abstract nonsense, it is possible to bundle the notion of an imploid-valued flow into that of the \emph{fundamental imploid} of a well-oriented 3-valent map.
\begin{definition}\label{defn:funimp}
The \defn{fundamental imploid} of a well-oriented 3-valent map $T$ is the left normal imploid $\FunImp{T}$ freely generated from the edges of $T$ modulo the relations in \Cref{fig:localflow}.
\end{definition}
\noindent
The function $[-] : E \to \FunImp{T}$ sending each edge to the corresponding generator of the fundamental imploid tautologically defines a flow, and by the universal property of the quotient, any other flow $\phi : E \to P$ uniquely extends to a strong homomorphism of imploids $\bar\phi : \FunImp{T} \to P$ such that $\phi = \bar\phi[-]$.
Moreover, $\phi$ is nowhere-unit just in case $\ker\bar\phi$ does not contain a generator.
The \emph{fundamental symmetric imploid} $\FunAbImp{T}$ can be defined similarly, with analogous properties for flows valued in symmetric left normal imploids.

Although this abstract definition of the fundamental imploid (reminiscent of the \emph{fundamental quandle} of a knot \cite{Joyce1982})
allows us to express flow concepts in a more uniform language, it doesn't immediately provide us much help in understanding the space of possible flows over a given 3-valent map.
To get a more concrete handle on this space, in the rest of the paper we develop a computational perspective on imploids and flows that is inspired by their connections to combinatory logic and type theory.

\section{Rewriting and pullback of flows}
\label{sec:flow-rewriting}

\subsection{Background: beta reduction and eta expansion}
\label{sec:flow-rewriting:background}

All of the computational power of the lambda calculus lies in the rule of \emph{$\beta$-reduction} $(\lambda x.T_1)(T_2) \to T_1[T_2/x]$, and as Church originally showed, the problem of determining if a general lambda term has a 
$\beta$-normal form is undecidable \cite{Church1936}.
On the other hand, if one imposes linearity the rule becomes much more tractable: every linear lambda term has a $\beta$-normal form, and the problem of computing it is complete for polynomial time \cite{Mairson2004}.
Graphically, $\beta$-reduction corresponds to the operation of ``unzipping'' a pair of trivalent vertices of opposite polarity:
$
\vcenter{\hbox{\scalebox{0.6}{
\begin{tikzpicture}
  \node [style=lam] (lam) {};
  \node (var) [above right=0.5em and 0.866em of lam.center] {};
  \node (body) [above left=0.5em and 0.866em of lam.center] {};
  \node [style=app] (app) [below=2em of lam] {};
  \node (cont) [below left=0.5em and 0.866em of app.center] {};
  \node (arg) [below right=0.5em and 0.866em of app.center] {};
  \draw[->-] (body.center) to (lam);
  \draw[->-] (lam) to (var.center);
  \path (lam) edge [->-] (app);  
  \draw[->-] (app) to (cont.center);  
  \draw[->-] (arg.center) to (app);
\end{tikzpicture}}}}
\overset{\beta}\LRA
\vcenter{\hbox{\scalebox{0.6}{
\begin{tikzpicture}
  \node (in1) {};
  \node (in2) [right=1.732em of in1] {};
  \node (out1) [below=3.232em of in1] {};
  \node (out2) [below=3.232em of in2] {};
  \path (in1.center) edge [->-] (out1.center);
  \path (out2.center) edge [->-] (in2.center);
\end{tikzpicture}}}}
$;
we refer to the matching pair of vertices as a \emph{$\beta$-redex}.
This rule can in principle be applied whenever such a configuration appears in a well-oriented 3-valent map, but the fact that it corresponds to $\beta$-reduction of lambda terms means that it \emph{preserves topological orientation}.
The graphical rule also manifestly preserves planarity, and $\beta$-reduction correspondingly restricts to an operation on planar terms.\footnote{The precise computational complexity of $\beta$-normalization for planar terms is a natural question, which is open as far as I am aware.
(Mairson's proof of PTIME-hardness for linear lambda calculus \cite{Mairson2004} is based on an encoding of boolean circuits that uses non-planarity in an essential way.)
}
Finally, that $\beta$-reduction restricts to an operation on unitless terms implies it preserves 2-edge-connectedness when restricted to topological orientations (although it can lead to disconnected maps when applied to non-topological orientations).
Dual to $\beta$-reduction is the less computationally interesting (but still logically important) rule of \emph{$\eta$-expansion} $T \to \lambda x.T(x)$.
Graphically, this rule corresponds to ``bubbling'' an oriented edge,
$
\vcenter{\hbox{\scalebox{0.6}{
\begin{tikzpicture}
  \node (in) {};
  \node (out) [below=3.732em of in] {};
  \path (in.center) edge [->-] (out.center);
\end{tikzpicture}}}}
\overset{\eta}\LRA
\vcenter{\hbox{\scalebox{0.6}{
\begin{tikzpicture}
  \node (fn) {};
  \node [style=app] (app) [below=0.866em of fn.center] {};
  \node [style=lam] (lam) [below=2em of app.center] {};
  \node (root) [below=0.866em of lam.center]{};
  \draw[->-] (fn.center) to (app);
  \path (app) edge [->-,bend right=60] (lam);  
  \path (lam) edge [->-,bend right=60] (app);  
  \draw[->-] (lam) to (root.center);
\end{tikzpicture}}}}
$, 
an operation which manifestly preserves both planarity and 2-edge-connectedness.

\subsection{Pullback of flows}
\label{sec:flow-rewriting:pullback}

\begin{figure}[t]
$$
\vcenter{\hbox{\scalebox{1.5}{\begin{tikzpicture}
  \node (out) {};
  \node [style=unit] (unit) [above=1em of out] {};
  \path 
    (unit) edge [->-] node [left=1pt] {\footnotesize$I \le a$} (out.center);
\end{tikzpicture}}}}
\quad\qquad
\vcenter{\hbox{\scalebox{1.5}{\begin{tikzpicture}
  \node (in) {};
  \node [style=counit] (counit) [below=1em of in] {};
  \node (pad) [below=0.1em of counit] {};
  \path 
    (in.center) edge [->-] node [right=1pt] {\footnotesize$a\le I$} (counit);
\end{tikzpicture}}}}
\qquad\vrule\qquad
\vcenter{\hbox{\scalebox{1.5}{\begin{tikzpicture}
  \node (in) {};
  \node (edge) [style=bigglue,below=1em of in] {};
  \node (le) [right=0.5em of edge] {\footnotesize$(a \le b)$};
  \node (out) [below=1em of edge] {};
  \path
    (in.center) edge [->-] node [right] {\footnotesize$a$} (edge)
    (edge) edge [->-] node [right] {\footnotesize$b$} (out.center);
\end{tikzpicture}}}}
$$
\caption{Flow relations for 1-valent and 2-valent vertices.}
\label{fig:localflow-2/1-valent}
\end{figure}
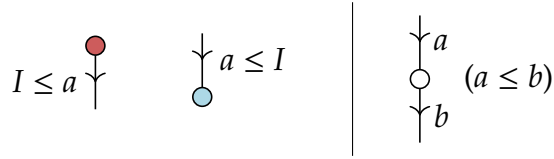

It is not hard to check that an imploid-valued flow can always be \emph{pulled back} along a $\beta$-reduction or an $\eta$-expansion, in a suitable sense.
To make this statement 
more precise, it is useful to first liberalize the notion of flow on a trivalent map to allow for arbitrary subdivision of edges by 2-valent vertices: we assume these to be well-oriented (one input, one output) and to satisfy the natural flow relation shown at the right in \Cref{fig:localflow-2/1-valent}.
Edge subdivision provides us an additional degree of flexibility 
when relating one flow to another, but it is always possible to recover a flow on a strictly 3-valent map by choosing any of the component values along a subdivided edge.

Consider again the rule of $\beta$-reduction, now written in reverse:
$$
\vcenter{\hbox{\scalebox{1.5}{
\begin{tikzpicture}
  \node [style=lam] (lam) {};
  \node (var) [above right=0.5em and 0.866em of lam] {};
  \node (body) [above left=0.5em and 0.866em of lam] {};
  \node [style=app] (app) [below=2.5em of lam] {};
  \node (cont) [below left=0.5em and 0.866em of app] {};
  \node (arg) [below right=0.5em and 0.866em of app] {};
  \draw[->-] (body.center) to node [left] {\footnotesize$b_1$} (lam);
  \draw[->-] (lam) to node [right] {\footnotesize$a_1$} (var.center);
  \path (lam) edge [->>-] node [pos=0.25,right] {\footnotesize$a_1 \imp b_1$} node [pos=0.5,style=glue] {} node [pos=0.75,right] {\footnotesize$a_2 \imp b_2$} (app);  
  \draw[->-] (app) to node [left] {\footnotesize$b_2$} (cont.center);  
  \draw[->-] (arg.center) to node [right] {\footnotesize$a_2$} (app);
\end{tikzpicture}}}}
\qquad\overset{\pull[\beta]}\LLA\qquad
\vcenter{\hbox{\scalebox{1.5}{
\begin{tikzpicture}
  \node (in1) {};
  \node (in2) [right=1.732em of in1] {};
  \node (out1) [below=4em of in1] {};
  \node (out2) [below=4em of in2] {};
  \path (in1.center) edge [->>-] node [pos=0.25,left] {\footnotesize$b_1$} node [pos=0.5,style=glue] {} node [pos=0.75,left] {\footnotesize$b_2$} (out1.center);
  \path (out2.center) edge [->>-] node [pos=0.25,right] {\footnotesize$a_2$} node [pos=0.5,style=glue] {} node [pos=0.75,right] {\footnotesize$a_1$} (in2.center);
\end{tikzpicture}}}}
$$
Here we have annotated the rule as it acts in the backwards direction on flows, as a \emph{$\beta$-expansion}
taking a pair of subdivided edges and ``rezipping'' them into a $\beta$-redex.
That this is a well-defined operation on flows reduces to the \emph{totality} of the implication $a \imp b$ and its monotonicity properties \eqref{clopset:imp}.
Dually, pullback along $\eta$-expansion 
$$
\vcenter{\hbox{\scalebox{1.5}{
\begin{tikzpicture}
  \node (in) {};
  \node (out) [below=4em of in] {};
  \path (in.center) edge [->>-] node [pos=0.25,right] {\footnotesize$c_1$} node [pos=0.75,right] {\footnotesize$c_2$} node [pos=0.5,style=glue] {} (out.center);
\end{tikzpicture}}}}
\qquad\overset{\pull[\eta]}\LLA\qquad
\vcenter{\hbox{\scalebox{1.5}{
\begin{tikzpicture}
  \node (fn) {};
  \node [style=app] (app) [below=1em of fn] {};
  \node [style=lam] (lam) [below=2.5em of app] {};
  \node (root) [below=1em of lam]{};
  \draw[->-] (fn.center) to node [right] {\footnotesize$c_1$} (app);
  \path (app) edge [->>-,bend right=60] node [pos=0.25,left] {\footnotesize$b_1$} node [pos=0.5,style=glue] {} node [pos=0.75,left] {\footnotesize$b_2$}  (lam);  
  \path (lam) edge [->>-,bend right=60] node [pos=0.25,right] {\footnotesize$a_2$} node [pos=0.5,style=glue] {} node [pos=0.75,right] {\footnotesize$a_1$} (app);  
  \draw[->-] (lam) to node [right] {\footnotesize$c_2$} (root.center);
\end{tikzpicture}}}}
$$
may be justified by \eqref{clopset:imp} and \emph{uniqueness} of implication.

Formally, these pullback operations on flows may be analyzed in terms of fundamental imploids as follows.
The rules of $\beta$-reduction and $\eta$-expansion both lift to \emph{strong homomorphisms} $\FunImp{T_L} \to \FunImp{T_R}$ from the fundamental imploid of the map on the left-hand side to that of the right-hand side.
By the universal properties of $\FunImp{T_L}$ and $\FunImp{T_R}$, pulling a flow $\phi : E_R \to P$ back along these operations reduces to pre-composing $\bar\phi : \FunImp{T_R} \to P$ with the corresponding homomorphism $\FunImp{T_L} \to \FunImp{T_R}$.
Moreover, these homomorphisms are \emph{boundary-preserving} in the sense that they fix all of the generators in $\partial T_L = \partial T_R$, which implies that the corresponding transformations 
can be applied locally 
anywhere inside a larger flow.
On the other hand, observe that nothing guarantees we can push a flow $\phi : E_L \to P$ \emph{forward} along $\FunImp{T_L} \to \FunImp{T_R}$, and it is easy to come up with counterexamples to such a principle for $\beta$-reduction (e.g., taking $P = \Z_2$, $a_1 = b_2 = 1$, $b_1 = a_2 = 0$ in the first diagram above).

\subsection{Imploid moves and topological completeness}
\label{sec:flow-rewriting:imploid-moves}

All this suggests a more powerful theory of rewriting for well-oriented maps, where each rule $T_L \Rightarrow T_R$ corresponds to a principle for pulling back flows on $T_R$ to flows on $T_L$.
To gain the full benefits of such a theory, it is natural to further generalize the notion of flow to allow for 1-valent vertices, with the relations shown in \Cref{fig:localflow-2/1-valent};
we refer to maps containing only vertices of positive degree $\le 3$ as \emph{essentially trivalent maps}.
\begin{definition}
A transformation $T_L \Rightarrow T_R$ between a pair of well-oriented essentially 3-valent maps with the same boundary $\partial \defeq \partial T_L = \partial T_R$
is called a \defn{(symmetric) imploid move} if it is realizable by a
strong homomorphism $\FunImp{T_L} \to \FunImp{T_R}$ (respectively, $\FunAbImp{T_L} \to \FunAbImp{T_R}$) fixing every element in $\partial$.
\end{definition}
\begin{proposition}\label{prop:imploid-moves}
The following are imploid moves:
$$
\vcenter{\hbox{\scalebox{0.8}{
\begin{tikzpicture}
  \node (in) {};
  \node (out) [below=2em of in] {};
  \path (in.center) edge [->>-] node [pos=0.5,style=glue] {} (out.center);
\end{tikzpicture}}}}
\LRA
\vcenter{\hbox{\scalebox{0.8}{
\begin{tikzpicture}
  \node (in) {};
  \node (out) [below=2em of in] {};
  \path (in.center) edge [->-] (out.center);
\end{tikzpicture}}}}
\qquad
\vcenter{\hbox{\scalebox{0.8}{
\begin{tikzpicture}
  \node (in) {};
  \node (out) [below=2em of in] {};
  \path (in.center) edge [->>-] node [pos=0.5,style=glue] {} (out.center);
\end{tikzpicture}}}}
\LRA
\vcenter{\hbox{\scalebox{0.8}{
\begin{tikzpicture}
  \node (in) {};
  \node (out) [below=2em of in] {};
  \path (in.center) edge [->>>-] node [pos=0.33,style=glue] {} node [pos=0.66,style=glue] {} (out.center);
\end{tikzpicture}}}}
\qquad\ \ 
\vcenter{\hbox{\scalebox{0.8}{\begin{tikzpicture}
  \node (root) {};
  \node [style=lam] (lam) [above=1em of root] {};
  \node (body) [above left=0.5em and 0.866em of lam] {};
  \node (var) [above right=0.5em and 0.866em of lam] {};
  \path
    (lam) edge [->-] (var.center)
    (body.center) edge [->-] (lam)
    (lam) edge [->-] (root.center);
\end{tikzpicture}}}}
\LRA
\vcenter{\hbox{\scalebox{0.8}{\begin{tikzpicture}
  \node (root) {};
  \node [style=lam] (lam) [above=1em of root] {};
  \node (body) [above left=0.5em and 0.866em of lam] {};
  \node (var) [above right=0.5em and 0.866em of lam] {};
  \path
    (lam) edge [->>-] node [pos=0.5,style=glue] {} (var.center)
    (body.center) edge [->>-] node [pos=0.5,style=glue] {} (lam)
    (lam) edge [->>-] node [pos=0.5,style=glue] {} (root.center);
\end{tikzpicture}}}}
\ \ 
\vcenter{\hbox{\scalebox{0.8}{\begin{tikzpicture}
  \node (fn) {};
  \node [style=app] (app) [below=1em of fn] {};
  \node (cont) [below left=0.5em and 0.866em of app] {};
  \node (arg) [below right=0.5em and 0.866em of app] {};
  \path
    (arg.center) edge [->-] (app)
    (app) edge [->-] (cont.center)
    (fn.center) edge [->-] (app);
\end{tikzpicture}}}}
\LRA
\vcenter{\hbox{\scalebox{0.8}{\begin{tikzpicture}
  \node (fn) {};
  \node [style=app] (app) [below=1em of fn] {};
  \node (cont) [below left=0.5em and 0.866em of app] {};
  \node (arg) [below right=0.5em and 0.866em of app] {};
  \path
    (arg.center) edge [->>-] node [pos=0.5,style=glue] {} (app)
    (app) edge [->>-] node [pos=0.5,style=glue] {} (cont.center)
    (fn.center) edge [->>-] node [pos=0.5,style=glue] {} (app);
\end{tikzpicture}}}}
$$
$$
\vcenter{\hbox{\scalebox{0.8}{
\begin{tikzpicture}
  \node [style=lam] (lam) {};
  \node (var) [above right=0.5em and 0.866em of lam] {};
  \node (body) [above left=0.5em and 0.866em of lam] {};
  \node [style=app] (app) [below=2.5em of lam] {};
  \node (cont) [below left=0.5em and 0.866em of app] {};
  \node (arg) [below right=0.5em and 0.866em of app] {};
  \draw[->-] (body.center) to (lam);
  \draw[->-] (lam) to (var.center);
  \path (lam) edge [->>-] node [pos=0.5,style=glue] {} (app);  
  \draw[->-] (app) to (cont.center);  
  \draw[->-] (arg.center) to (app);
\end{tikzpicture}}}}
\overset{\beta}\LRA
\vcenter{\hbox{\scalebox{0.8}{
\begin{tikzpicture}
  \node (in1) {};
  \node (in2) [right=1.732em of in1] {};
  \node (out1) [below=4em of in1] {};
  \node (out2) [below=4em of in2] {};
  \path (in1.center) edge [->>-] node [pos=0.5,style=glue] {} (out1.center);
  \path (out2.center) edge [->>-] node [pos=0.5,style=glue] {} (in2.center);
\end{tikzpicture}}}}
\qquad\qquad
\vcenter{\hbox{\scalebox{0.8}{
\begin{tikzpicture}
  \node (in) {};
  \node (out) [below=4em of in] {};
  \path (in.center) edge [->>-] node [pos=0.5,style=glue] {} (out.center);
\end{tikzpicture}}}}
\ \overset{\eta}\LRA
\vcenter{\hbox{\scalebox{0.8}{
\begin{tikzpicture}
  \node (fn) {};
  \node [style=app] (app) [below=1em of fn] {};
  \node [style=lam] (lam) [below=2.5em of app] {};
  \node (root) [below=1em of lam]{};
  \draw[->-] (fn.center) to (app);
  \path (app) edge [->>-,bend right=60] node [pos=0.5,style=glue] {} (lam);  
  \path (lam) edge [->>-,bend right=60] node [pos=0.5,style=glue] {} (app);  
  \draw[->-] (lam) to (root.center);
\end{tikzpicture}}}}
$$
$$
\vcenter{\hbox{\scalebox{0.8}{\begin{tikzpicture}
  \node (root) {};
  \node [style=lam] (lam) [above=1em of root] {};
  \node (body) [above left=0.5em and 0.866em of lam] {};
  \node (var) [above right=0.5em and 0.866em of lam] {};
  \path
    (lam) edge [->-] (var.center)
    (body.center) edge [->-] (lam)
    (lam) edge [->-] (root.center);
\end{tikzpicture}}}}
 \overset\compose\LRA
\vcenter{\hbox{\scalebox{0.8}{\begin{tikzpicture}
  \node (root) {};
  \node [style=lam] (lam1) [above=1em of root] {};
  \node [style=lam] (lam2) [above left=1.5em and 0.866em of lam1] {};
  \node [style=app] (app) [above right=1.5em and 0.866em of lam1] {};
  \node (C) [above left=0.5em and 0.866em of lam2] {};
  \node (B) [above right=0.5em and 0.866em of app] {};
  \path
    (lam1) edge [->-] (root.center)
    (lam1) edge [->-] (app)
    (lam2) edge [->-] (lam1)
    (lam2) edge [->-] (app)
    (C.center) edge [->-] (lam2)
    (app) edge [->-] (B.center);
\end{tikzpicture}}}}
\quad\qquad
\vcenter{\hbox{\scalebox{0.8}{\begin{tikzpicture}
  \node (app) [style=app] {};
  \node (fn) [style=unit,above=1em of app] {};
  \node (cont) [below left=0.5em and 0.866em of app] {};
  \node (arg) [below right=0.5em and 0.866em of app] {};
  \path
    (fn) edge [->-] (app)
    (app) edge [->-] (cont.center)
    (arg.center) edge [->-] (app);  
\end{tikzpicture}}}}
\overset\init\LLRA\ 
\vcenter{\hbox{\scalebox{0.8}{
\begin{tikzpicture}
  \node (in) {};
  \node (out) [left=1.5em of in] {};
  \draw[->>-,bend right=90,min distance=1em] (in.center) to node [pos=0.5,style=glue] {} (out.center);
\end{tikzpicture}}}}
\quad\qquad
\vcenter{\hbox{\scalebox{0.8}{
\begin{tikzpicture}
  \node (in) {};
  \node (out) [below left=1.5em and 0.5em of in] {};
  \draw[->>-,bend left=30] (in.center) to node [pos=0.5,style=glue] {} (out.center);
\end{tikzpicture}}}}
 \ \ \overset\unit\LRA
\vcenter{\hbox{\scalebox{0.8}{\begin{tikzpicture}
  \node (app) [style=app] {};
  \node (fn) [above=1em of app] {};
  \node (cont) [below left=0.5em and 0.866em of app] {};
  \node (arg) [style=unit,below right=0.5em and 0.866em of app] {};
  \path
    (fn.center) edge [->-] (app)
    (app) edge [->-] (cont.center)
    (arg) edge [->-] (app);  
\end{tikzpicture}}}}
$$
$$
\vcenter{\hbox{\scalebox{0.8}{\begin{tikzpicture}
  \node (root) {};
  \node [style=app] (lam) [below=1em of root] {};
  \node (body) [below left=0.5em and 0.866em of lam] {};
  \node (var) [below right=0.5em and 0.866em of lam] {};
  \path
    (lam) edge [-<-] (var.center)
    (body.center) edge [-<-] (lam)
    (lam) edge [-<-] (root.center);
\end{tikzpicture}}}}
\overset{\bar\compose}\LRA
\vcenter{\hbox{\scalebox{0.8}{\begin{tikzpicture}
  \node (root) {};
  \node [style=app] (lam1) [below=1em of root] {};
  \node [style=lam] (lam2) [below left=1.5em and 0.866em of lam1] {};
  \node [style=app] (app) [below right=1.5em and 0.866em of lam1] {};
  \node (C) [below left=0.5em and 0.866em of lam2] {};
  \node (B) [below right=0.5em and 0.866em of app] {};
  \path
    (lam1) edge [-<-] (root.center)
    (lam1) edge [-<-] (app)
    (lam2) edge [-<-] (lam1)
    (lam2) edge [->-] (app)
    (C.center) edge [-<-] (lam2)
    (app) edge [-<-] (B.center);
\end{tikzpicture}}}}
\quad\qquad
\vcenter{\hbox{\scalebox{0.8}{\begin{tikzpicture}
  \node (out) {};
  \node [style=unit] (unit) [above=1em of out] {};
  \path 
    (unit) edge [->-] (out.center);
\end{tikzpicture}}}}
 \ \ \overset{\mathrm{B}}\LRA\ \ 
\vcenter{\hbox{\scalebox{0.8}{\begin{tikzpicture}
\draw [decoration={markings, mark=at position 0.1 with {\arrow{>}}, mark=at position 0.45 with {\arrow{>}}, mark=at position 0.7 with {\arrow{>}}, mark=at position 0.85 with {\arrow{>}}},postaction={decorate}] (0,0) circle [radius=0.5];
\draw (0,-1) node (root) {};
\draw (0,-0.5) node (lx) [style=lam] {};
\draw (-0.433,-0.25) node (ly) [style=lam] {};
\draw (0,0.5) node (lz) [style=lam] {};
\draw (0.433,-0.25) node (ax) [style=app] {};
\draw (0,0) node (ay) [style=app] {};
\draw [->-] (lx) to (root.center);
\draw [->-] (ly) to (ay);
\draw [->-] (lz) to (ay);
\draw [->-] (ay) to (ax);
\end{tikzpicture}}}}
\quad\qquad
\vcenter{\hbox{\scalebox{0.8}{\begin{tikzpicture}
  \node (out) {};
  \node [style=unit] (unit) [above=1em of out] {};
  \path 
    (unit) edge [->-] (out.center);
\end{tikzpicture}}}}
 \ \ \overset{\mathrm{I}}\LRA
\vcenter{\hbox{\scalebox{0.8}{\begin{tikzpicture}
  \node (root) {};
  \node [style=lam] (lam) [above=1em of root] {};
  \path (lam) edge [->-,loop,min distance=2em,in=135,out=45] (lam);
  \path (lam) edge [->-] (root.center);
\end{tikzpicture}}}}
$$
$$
\vcenter{\hbox{\scalebox{0.8}{
\begin{tikzpicture}
  \node [style=lam] (lam) {};
  \node (var) [above right=0.5em and 0.866em of lam.center] {};
  \node (body) [above left=0.5em and 0.866em of lam.center] {};
  \node [style=app] (app) [below=2em of lam] {};
  \node (cont) [below left=0.5em and 0.866em of app.center] {};
  \node (arg) [below right=0.5em and 0.866em of app.center] {};
  \draw[->-] (body.center) to (lam);
  \draw[->-] (lam) to (var.center);
  \path (lam) edge [->-] (app);  
  \draw[->-] (app) to (cont.center);  
  \draw[->-] (arg.center) to (app);
\end{tikzpicture}}}}
\overset{IH}\LRA
\vcenter{\hbox{\scalebox{0.8}{
\begin{tikzpicture}
  \node [style=lam] (lam) {};
  \node (body) [above left=0.866em and 0.5em of lam.center] {};
  \node (cont) [below left=0.866em and 0.5em of lam.center] {};
  \node [style=app] (app) [right=2em of lam] {};
  \node (var) [above right=0.866em and 0.5em of app.center] {};
  \node (arg) [below right=0.866em and 0.5em of app.center] {};
  \draw[->-] (body.center) to (lam);
  \draw[->-] (lam) to (cont.center);
  \path (lam) edge [->-] (app);  
  \draw[->-] (app) to (var.center);  
  \draw[->-] (arg.center) to (app);
\end{tikzpicture}}}}
\quad\qquad
\vcenter{\hbox{\scalebox{0.8}{\begin{tikzpicture}
  \node (app) [style=app] {};
  \node (fn) [above=1em of app] {};
  \node (app2) [style=app,below left=0.866em and 0.5em of app] {};
  \node (arg2) [below right=0.866em and 0.5em of app2.center] {};
  \node (cont) [below left=0.866em and 0.5em of app2.center] {};
  \node (arg1) [right=1em of arg2.center] {};
  \path
    (fn.center) edge [->-] (app)
    (app) edge [->-] (app2)
    (app2) edge [->-] (cont.center);  
  \path
    (arg1.center) edge [->-] (app)
    (arg2.center) edge [->-] (app2);
\end{tikzpicture}}}}
\overset\assoc\LRA
\vcenter{\hbox{\scalebox{0.8}{\begin{tikzpicture}
  \node (app) [style=app] {};
  \node (fn) [above=1em of app] {};
  \node (app2) [style=app,below right=0.866em and 0.5em of app] {};
  \node (arg1) [below left=0.866em and 0.5em of app2.center] {};
  \node (arg2) [below right=0.866em and 0.5em of app2.center] {};
  \node (cont) [left=1em of arg1.center] {};
  \path
    (fn.center) edge [->-] (app)
    (app) edge [->-] (cont.center)
    (app2) edge [->-] (app);  
  \path
    (arg1.center) edge [->-] (app2)
    (arg2.center) edge [->-] (app2);    
\end{tikzpicture}}}}
\quad\qquad
\vcenter{\hbox{\scalebox{0.8}{
\begin{tikzpicture}
  \node [style=app] (lam) {};
  \node (body) [above left=0.866em and 0.5em of lam.center] {};
  \node (cont) [below left=0.866em and 0.5em of lam.center] {};
  \node [style=lam] (app) [right=2em of lam] {};
  \node (var) [above right=0.866em and 0.5em of app.center] {};
  \node (arg) [below right=0.866em and 0.5em of app.center] {};
  \draw[->-] (body.center) to (lam);
  \draw[->-] (lam) to (cont.center);
  \path (app) edge [->-] (lam);  
  \draw[->-] (var.center) to (app);
  \draw[->-] (app) to (arg.center);  
\end{tikzpicture}}}}
\overset{HI}\LRA
\vcenter{\hbox{\scalebox{0.8}{
\begin{tikzpicture}
  \node [style=app] (lam) {};
  \node (var) [above right=0.5em and 0.866em of lam.center] {};
  \node (body) [above left=0.5em and 0.866em of lam.center] {};
  \node [style=lam] (app) [below=2em of lam] {};
  \node (cont) [below left=0.5em and 0.866em of app.center] {};
  \node (arg) [below right=0.5em and 0.866em of app.center] {};
  \draw[->-] (body.center) to (lam);
  \draw[->-] (var.center) to (lam);
  \path (lam) edge [->-] (app);  
  \draw[->-] (app) to (cont.center);  
  \draw[->-] (app) to (arg.center);
\end{tikzpicture}}}}
$$
\end{proposition}
\begin{proposition}\label{prop:comm-imploid-moves}
The following are symmetric imploid moves:
$$
\vcenter{\hbox{\scalebox{0.8}{\begin{tikzpicture}
  \node (app) [style=app] {};
  \node (fn) [above=1em of app] {};
  \node (app2) [style=app,below left=0.866em and 0.5em of app] {};
  \node (arg2) [below right=0.866em and 0.5em of app2.center] {};
  \node (cont) [below left=0.866em and 0.5em of app2.center] {};
  \node (arg1) [right=1em of arg2.center] {};
  \path
    (fn.center) edge [->-] (app)
    (app) edge [->-] (app2)
    (app2) edge [->-] (cont.center);  
  \path
    (arg1.center) edge [->-] (app)
    (arg2.center) edge [->-] (app2);
\end{tikzpicture}}}}
\overset\chi\LRA
\vcenter{\hbox{\scalebox{0.8}{\begin{tikzpicture}
  \node (app) [style=app] {};
  \node (fn) [above=1em of app] {};
  \node (app2) [style=app,below left=0.866em and 0.5em of app] {};
  \node (arg2) [below right=0.866em and 0.5em of app2.center] {};
  \node (cont) [below left=0.866em and 0.5em of app2.center] {};
  \node (arg1) [right=1em of arg2.center] {};
  \path
    (fn.center) edge [->-] (app)
    (app) edge [->-] (app2)
    (app2) edge [->-] (cont.center);
  \path
    (arg2.center) edge [->-,bend right=60] (app)
    (arg1.center) edge [-->-] (app2);
\end{tikzpicture}}}}
\quad\qquad
\vcenter{\hbox{\scalebox{0.8}{\begin{tikzpicture}
  \node (out) {};
  \node [style=unit] (unit) [above=1em of out] {};
  \path 
    (unit) edge [->-] (out.center);
\end{tikzpicture}}}}
 \ \ \overset{\mathrm{C}}\LRA
\vcenter{\hbox{\scalebox{0.8}{\begin{tikzpicture}
\draw [decoration={markings, mark=at position 0 with {\arrow{>}}, mark=at position 0.25 with {\arrow{>}}, mark=at position 0.5 with {\arrow{>}}, mark=at position 0.7 with {\arrow{>}}, mark=at position 0.83 with {\arrow{>}}},postaction={decorate}] (0,0) circle [radius=0.5];
\draw (0,-1) node (root) {};
\draw (0,-0.5) node (lx) [style=lam] {};
\draw (-0.354,-0.354) node (ly) [style=lam] {};
\draw (-0.354,0.354) node (lz) [style=lam] {};
\draw (0.354,0.354) node (argy) [style=app] {};
\draw (0.354,-0.354) node (argz) [style=app] {};
\draw [->-] (lx) to (root.center);
\draw [-->-] (ly) to (argy);
\draw [-->-] (lz) to (argz);
\end{tikzpicture}}}}
\quad\qquad
\vcenter{\hbox{\scalebox{0.8}{\begin{tikzpicture}
  \node (cont) {};
  \node [style=app] (app) [above=1em of cont] {};
  \node (fn) [above left=0.5em and 0.866em of app] {};
  \node (arg) [above right=0.5em and 0.866em of app] {};
  \path
    (arg.center) edge [->-] (app)
    (fn.center) edge [->-] (app)
    (app) edge [->-] (cont.center);
\end{tikzpicture}}}}
\overset\gamma\LRA
\vcenter{\hbox{\scalebox{0.8}{\begin{tikzpicture}
  \node (cont) {};
  \node [style=app] (app) [above=1em of cont] {};
  \node (fn) [above left=0.5em and 0.866em of app] {};
  \node (arg) [above right=0.5em and 0.866em of app] {};
  \path
    (arg.center) edge [->-,out=180,in=130] (app)
    (fn.center) edge [->-,out=0,in=50] (app)
    (app) edge [->-] (cont.center);
\end{tikzpicture}}}}
$$
\end{proposition}
\noindent
The rules above (many interderivable) do not give a complete set of generators for imploid moves.
However, they are \emph{topologically complete} in the following sense.
\begin{proposition}\label{prop:preserve-topological-orientation}
All of the moves listed in \Cref{prop:imploid-moves,prop:comm-imploid-moves} preserve topological orientation.
\end{proposition}
\begin{theorem}\label{thm:topological-completeness}
Let $V_n$ be the \defn{$n$-spine} defined inductively by 
$V_0 \defeq
\vcenter{\hbox{\scalebox{0.8}{\begin{tikzpicture}\node (out) {};\node [style=unit] (unit) [above=1em of out] {};\path (unit) edge [->-] (out.center);\end{tikzpicture}}}}$,
$V_{n+1} \defeq
\vcenter{\hbox{\scalebox{0.8}{\begin{tikzpicture}
  \node (app) [style=app] {};
  \node (fn) [above=1em of app] {$V_n$};
  \node (cont) [below left=0.5em and 0.866em of app] {};
  \node (arg) [below right=0.5em and 0.866em of app] {};
  \path
    (fn) edge [->-] (app)
    (app) edge [->-] (cont.center)
    (arg.center) edge [->-] (app);  
\end{tikzpicture}}}}
$, and let $V'_n \defeq 
\vcenter{\hbox{\scalebox{0.8}{
\begin{tikzpicture}
  \node (in) {$V_n$};
  \node (out) [below=2em of in] {};
  \path (in) edge [->>-] node [pos=0.5,style=glue] {} (out.center);
\end{tikzpicture}}}}$.
Starting from the $V'_n$, the imploid moves in (the first three rows of) \Cref{prop:imploid-moves} generate all rooted essentially 3-valent planar maps with their topological orientation.
With the addition of (any of) the symmetric imploid moves in \Cref{prop:comm-imploid-moves}, they generate all rooted essentially 3-valent maps of arbitrary genus.
\end{theorem}
\noindent
As an immediate corollary of \Cref{thm:topological-completeness} we get another proof of \Cref{lemma:top-flow-is-global}: any flow on a topologically oriented map $T$ (with $n+1$ boundary arcs counting the root) can be pulled back to a flow on $V'_n$ while preserving the boundary, and so the global flow condition on $T$ may be read off directly from the local flow conditions on $V'_n$.

Although the link may seem astonishing at first, this topological completeness theorem is closely related to the classical result in combinatory logic that the combinators B, C, and I form a complete basis for linear lambda terms \cite[\S9F]{Hindley1997}, as well as its planar restriction stating that B and I form a complete basis for planar lambda terms.
It may be further refined by restricting to \emph{non-unital imploid moves} (that is, moves not involving 1-valent vertices), which generate all bridgeless essentially 3-valent maps with their topological orientation (i.e., unitless linear lambda terms).

Tantalizingly, these completeness results also appear connected to a basic but motivating result in the theory of \emph{knotted trivalent graphs} (KTGs), which in one formulation states that any KTG (and hence any knot) can be generated from the planar tetrahedron
$\vcenter{\hbox{\begin{tikzpicture}[scale=0.4]
\draw (0,0) circle [radius=0.5];
\draw (-0.433,-0.25) node (ly) [style=pt] {};
\draw (0,0.5) node (lz) [style=pt] {};
\draw (0.433,-0.25) node (ax) [style=pt] {};
\draw (0,0) node (ay) [style=pt] {};
\draw (ly) to (ay);
\draw (lz) to (ay);
\draw (ay) to (ax);
\end{tikzpicture}}}$
and crossed tetrahedron
$\vcenter{\hbox{\begin{tikzpicture}[scale=0.4]
\draw (0,0) circle [radius=0.5];
\draw (-0.354,-0.354) node (ly) [style=pt] {};
\draw (-0.354,0.354) node (lz) [style=pt] {};
\draw (0.354,0.354) node (argy) [style=pt] {};
\draw (0.354,-0.354) node (argz) [style=pt] {};
\draw (ly) to (argy);
\draw (lz) to ($(lz)!0.3!(argz)$);
\draw ($(lz)!0.7!(argz)$) to (argz);
\end{tikzpicture}}}$
using unzip, bubbling, connect sum, and the unknot.
(See \cite[Theorem 1]{Thurston2001ktg} and \cite[Appendix]{BarNatanDancso2013}.
Indeed, Thurston's article inspires our terminology of ``unzipping'' and ``bubbling'' for $\beta$ and $\eta$, backing an analogy made by Buliga \cite{Buliga2013glc}.)
This strong formal similarity
suggests it could be worthwhile to develop a more refined treatment of the exchange law as a \emph{braiding} on linear lambda terms (cf.~\cite{Mellies2018rtl}), moving up a dimension from 3-valent maps to KTGs.
For example, it may be interesting to relate imploid flows to \emph{qualgebra colorings} of KTGs \cite{Lebed2015}.

Late in the development of the theory of imploid-valued flows described here (and motivated by the parallel connections discussed in \cite{Z2017assoc}), I discovered with excitement that it has much in common with the ``graphic theory of associativity'' proposed by Tamari in a relatively obscure conference publication \cite{Tamari1982graphic}, which built on seeds planted thirty years earlier in his thesis \cite{Tamari1951phd}.
Tamari's approach can be seen as slanted towards monoids rather than imploids, but is in a sense more foundational, beginning with the minimalistic algebraic structure of a \emph{partial binary operation} (or ``bin'') and considering how to match different principles of associativity with different well-oriented bridgeless planar 3-valent maps.
(Since bins are partial, even the unzip move isn't available, and the result is an infinite hierarchy of independent, higher associative laws.)

\section{Polarized flows and bidirectional typing}
\label{sec:polarized-flows}

We close by briefly sketching another perspective on imploid-valued flows that makes explicit their connection to linear logic proof-nets \cite{Girard1987}, while also being implicitly tied to the important type-theoretic concept of \emph{bidirectional typing} \cite{PierceTurner2000,DunfieldKrishnaswami2013} (and related ideas such as \emph{polarized subtyping} \cite{Dolan2016}).

\begin{figure}[t]
$$
\vcenter{\hbox{\scalebox{1.5}{\begin{tikzpicture}
  \node (in) {};
  \node (edge) [style=axiom,below=1em of in] {};
  \node (le) [right=0.5em of edge] {\footnotesize$(a \le b)$};
  \node (out) [below=1em of edge] {};
  \path
    (in.center) edge [->-neg] node [right] {\footnotesize$a$} (edge)
    (edge) edge [->-pos] node [right] {\footnotesize$b$} (out.center);
\end{tikzpicture}}}}
\ 
\vcenter{\hbox{\scalebox{1.5}{\begin{tikzpicture}
  \node (in) {};
  \node (edge) [style=cut,below=1em of in] {};
  \node (le) {} ; 
  \node (out) [below=1em of edge] {};
  \path
    (in.center) edge [->-pos] node [right] {\footnotesize$a$} (edge)
    (edge) edge [->-neg] node [right] {\footnotesize$b$} (out.center);
\end{tikzpicture}}}}
\quad\vrule\quad
\vcenter{\hbox{\scalebox{1.5}{\begin{tikzpicture}
  \node (lam) [style=lam] {};
  \node (root) [below=1em of lam] {};
  \node (var) [above right=0.5em and 0.866em of lam] {};
  \node (body) [above left=0.5em and 0.866em of lam] {};
  \path
     (lam) edge [->-pos] node [left] {\footnotesize$\rimp{a}{b}$} (root.center)
     (lam) edge [->-neg] node [near start,right,yshift=-2pt] {\footnotesize$a$} (var.center)
     (body.center) edge [->-pos] node [near end,left,yshift=-2pt] {\footnotesize$b$} (lam);  
\end{tikzpicture}}}}
\quad
\vcenter{\hbox{\scalebox{1.5}{\begin{tikzpicture}
  \node (app) [style=app] {};
  \node (fn) [above=1em of app] {};
  \node (cont) [below left=0.5em and 0.866em of app] {};
  \node (arg) [below right=0.5em and 0.866em of app] {};
  \path
    (fn.center) edge [->-neg] node [right] {\footnotesize$\rimp{a}{b}$} (app)
    (app) edge [->-neg] node [near start,left,yshift=2pt] {\footnotesize$b$} (cont.center)
    (arg.center) edge [->-pos] node [near end,right,yshift=2pt] {\footnotesize$a$} (app);  
\end{tikzpicture}}}}
$$
\caption{Defining relations for polarized flows.}
\label{fig:polflow}
\end{figure}
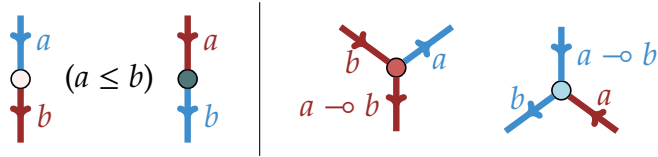

One reason the notion of imploid-valued flow is more subtle than the classical notion of abelian group-valued flow is that the defining relations (\Cref{fig:localflow}) intertwine the preorder 
with the implication operation. 
In the corresponding type system for linear lambda terms (\Cref{prop:flows-as-derivations}), \emph{subtyping} is built into the rules for typing application and abstraction, so that typing a term reduces to checking a big collection of constraints of the form $a \imp b \le c$ or $c \le a \imp b$.

In contrast, the definition of a \emph{polarized flow} (see \Cref{fig:polflow}) employs a more rigid separation between $\imp$ and $\le$, relying on the presence of both 3-valent and 2-valent vertices
(for simplicity, we leave out 1-valent vertices from this discussion: they can be dealt with similarly to 3-valent vertices).
Formally, now we are working with maps which are not merely oriented but also \emph{signed} (cf.~\cite{Kauffman1989tutte}), that is, equipped with a function $\pi : E \to \set{+1,-1}$ assigning each edge a positive (red) or negative (blue) polarity.
We assume that $\pi$ is \emph{proper} in the sense that the sum of polarities around each 3-valent vertex is either +1 or -1, and the sum around each 2-valent vertex is 0.
We likewise assume that $\pi$ is compatible with the underlying orientation in the sense that the right half of \Cref{fig:polflow} can be overlaid onto \Cref{fig:localflow}.
(This means that the orientation markers are for the most part redundant, although it is still necessary to distinguish vertices with one negative input and one positive output, which we color in white, from vertices with one positive input and one negative output, which we color in black.)
\begin{definition}
Let $T$ be a well-oriented 3-valent map, and let $\pi$ be a well-polarized, well-oriented essentially 3-valent map.
We say that $\pi$ is a \defn{polarization of $T$}, written $\pi \refs T$, if the underlying oriented map of $\pi$ is an edge subdivision of $T$.
\end{definition}
\begin{defprop}
\label{prop:minimal-pol}
Any well-oriented 3-valent map $T$ has a \defn{minimal polarization} $\pi_T \refs T$, which can be constructed by applying the two fixed patterns of polarization 
for 3-valent vertices, and then subdividing any edge whose ends have opposite polarity by a 2-valent vertex (either white or black).
\end{defprop}
\begin{proposition}
\label{prop:minimal-pol-top}
Black vertices of $\pi_T$ correspond to $\beta$-redices of $T$.
\end{proposition}
\noindent
Clearly, any polarized flow is a flow, by forgetting polarities.
Conversely, any flow on a 3-valent map can be turned into a polarized flow after sufficient edge subdivision, although the resulting polarization may not necessarily be the minimal one.

Our interest in polarization is that it gives another way of decomposing flows inductively, different from and complementary to the inductive definition of the topological orientation.
\begin{definition}
Let $\pi$ be a well-polarized, well-oriented essentially 3-valent map.
The \defn{w--b orientation} of $\pi$ is given by reversing the orientation of negative edges
(hence white vertices become \emph{sources}, black vertices become \emph{sinks}).
\end{definition}
\begin{proposition}\label{prop:topological-implies-acyclic}
If the underlying orientation of $\pi$ is topological then its w--b orientation is acyclic.
\end{proposition}
\noindent
By \Cref{prop:topological-implies-acyclic}, any topological $\pi$ can be decomposed into a forest of rooted trees, with their leaves glued together by white vertices and their roots glued together by black vertices.
(In the literature on proof-nets, these trees are called \emph{formulas}, white vertices are called \emph{axioms}, and black vertices are called \emph{cuts}.)
This makes it possible to reduce the problem of building a polarized flow on $\pi$ to the following recipe:
\begin{enumerate}
\item At each white vertex $w$, assign its negative and positive ends some values $w^-$ and $w^+$ such that $w^- \le w^+$.
\item Apply the rules on the right side of \Cref{fig:polflow} to propagate values to the remaining edges.
\item At each black vertex $\beta$ whose ends have now been assigned values $a^+$ and $b^-$, check that $a^+ \le b^-$.
\end{enumerate}
In more abstract terms following the discussion of \Cref{sec:imploid-flows:funimps}, any topological $\pi$ has a \defn{universal polarized flow} valued in the imploid freely generated from its white vertices
modulo the relations 
induced on its black vertices.

The universal polarized flow on (the minimal polarization of) a linear lambda term is analogous to its \emph{principal type-scheme} \cite{Hindley1969,Hindley1989}, but with the difference that it includes explicit subtyping constraints corresponding to $\beta$-redices.
A consequence is that the universal polarized flow can be computed very efficiently in a single traversal of the term, without performing any $\beta$-normalization either explicitly or implicitly (getting around Mairson's P-completeness result \cite{Mairson2004}).
Of course, complexity may come back into the picture if we want to \emph{instantiate} the universal polarized flow to obtain (say) a flow valued in a free imploid (or a nowhere-unit flow valued in a finite imploid), since this involves the discharging of these subtyping constraints.

An extended example (paying tribute to one of Tutte's earliest contributions to map coloring \cite{Tutte1946ohc}) may be found in \Cref{sec:appendix:polflows}.

\bibliographystyle{plain}
{\small\bibliography{imploids-arxiv}}

\begin{thebibliography}{10}

\bibitem{AppelHaken1977}
K.~Appel and W.~Haken.
\newblock Every planar map is four colorable. parts i and ii.
\newblock {\em Illinois Journal of Mathematics}, 21:429--567, 1977.

\bibitem{BarNatanDancso2013}
Dror Bar-Natan and Zsuzsanna Dancso.
\newblock Homomorphic expansions for knotted trivalent graphs.
\newblock {\em J. Knot Theory and Its Ramifications}, 22, 2013.

\bibitem{BoGaJa2013}
O.~Bodini, D.~Gardy, and A.~Jacquot.
\newblock Asymptotics and random sampling for {BCI} and {BCK} lambda terms.
\newblock {\em Theoretical Computer Science}, 502:227--238, 2013.

\bibitem{BourkeLack2017braidedskew}
John Bourke and Stephen Lack.
\newblock Braided skew monoidal categories.
\newblock December 2017.
\newblock arXiv:1712.0827.

\bibitem{Buliga2013glc}
Marius Buliga.
\newblock Graphic lambda calculus.
\newblock {\em Complex Systems}, 22(4):311--360, 2013.

\bibitem{Church1936}
Alonzo Church.
\newblock An unsolvable problem of elementary number theory.
\newblock {\em American Journal of Mathematics}, 58(2):345--363, 1936.

\bibitem{CYZ2017chordmaps}
Julien Courtiel, Karen Yeats, and Noam Zeilberger.
\newblock Connected chord diagrams and bridgeless maps.
\newblock arXiv:1611.04611, October 2017.

\bibitem{DayLaplaza1978}
B.~J. Day and M.~L. Laplaza.
\newblock On embedding closed categories.
\newblock {\em Bull. Austral. Math. Soc.}, 18:357--371, 1978.

\bibitem{DeGroote1999}
Philippe de~Groote.
\newblock An algebraic correctness criterion for intuitionistic multiplicative
  proof-nets.
\newblock {\em Theoretical Computer Science}, 224(1--2):115--134, 1999.

\bibitem{Dolan2016}
Stephen Dolan.
\newblock {\em Algebraic Subtyping}.
\newblock Phd thesis, University of Cambridge, 2016.

\bibitem{DunfieldKrishnaswami2013}
Joshua Dunfield and Neelakantan~R. Krishnaswami.
\newblock Complete and easy bidirectional typechecking for higher-rank
  polymorphism.
\newblock {\em SIGPLAN Notices}, 48(9):429--442, September 2013.

\bibitem{EilenbergKelly1966}
Samuel Eilenberg and G.~Max Kelly.
\newblock Closed categories.
\newblock In {\em Proceedings of the Conference on Categorical Algebra}, pages
  421--562. Springer-Verlag, 1966.

\bibitem{Eynard2016}
Bertrand Eynard.
\newblock {\em Counting Surfaces}.
\newblock Number~70 in Progress in Mathematical Physics. Birkh\"auser, 2016.

\bibitem{Girard1987}
Jean-Yves Girard.
\newblock Linear logic.
\newblock {\em Theoretical Computer Science}, 50:1--102, 1987.

\bibitem{Gonthier2008}
Georges Gonthier.
\newblock Formal proof---the {F}our {C}olor {T}heorem.
\newblock {\em Notices of the AMS}, 55(11):1382--1393, 2008.

\bibitem{GKRV2017tuttepolymaps}
Andrew Goodall, Thomas Krajewski, Guus Regts, and Llu\'is Vena.
\newblock A tutte polynomial for maps.
\newblock {\em Combinatorics, Probability and Computing}, page to appear, 2018.

\bibitem{Hindley1969}
J.~Roger Hindley.
\newblock The principal type-scheme of an object in combinatory logic.
\newblock {\em Transactions of the American Mathematical Society}, 146:29--60,
  1969.

\bibitem{Hindley1989}
J.~Roger Hindley.
\newblock Bck-combinators and linear lambda-terms have types.
\newblock {\em Theoretical Computer Science}, 64(1):97--105, 1989.

\bibitem{Hindley1997}
J.~Roger Hindley.
\newblock {\em Basic Simple Type Theory}.
\newblock CUP, 1997.

\bibitem{Holyer1981}
Ian Holyer.
\newblock The np-completeness of edge-coloring.
\newblock {\em SIAM Journal on Computing}, 10(4):718--720, 1981.

\bibitem{Hyland2017moderndress}
Martin Hyland.
\newblock Classical lambda calculus in modern dress.
\newblock {\em Mathematical Structures in Computer Science}, 27(5):762--781,
  2017.

\bibitem{Jaeger79}
F.~Jaeger.
\newblock Flows and generalized coloring theorems in graphs.
\newblock {\em J. Combinatorial Theory Series B}, 26:205--216, 1979.

\bibitem{JonesSingerman1978}
Gareth~A. Jones and David Singerman.
\newblock Theory of maps on orientable surfaces.
\newblock {\em Proceedings of the London Mathematical Society}, 37:273--307,
  1978.

\bibitem{JonesSingerman1994}
Gareth~A. Jones and David Singerman.
\newblock Maps, hypermaps, and triangle groups.
\newblock In L.~Schneps, editor, {\em The Grothendieck Theory of Dessins
  d'Enfants}. CUP, 1994.

\bibitem{Joyce1982}
David Joyce.
\newblock A classifying invariant of knots: the knot quandle.
\newblock {\em J. Pure Applied Algebra}, 23:37--65, 1982.

\bibitem{Kauffman1989tutte}
Louis~H. Kauffman.
\newblock A tutte polynomial for signed graphs.
\newblock {\em Discrete Applied Mathematics}, 25:105--127, 1989.

\bibitem{LZgraphs}
Sergei~K. Lando and Alexander~K. Zvonkin.
\newblock {\em Graphs on Surfaces and Their Applications}.
\newblock Number 141 in Encyclopaedia of Mathematical Sciences. Springer, 2004.

\bibitem{Lebed2015}
Victoria Lebed.
\newblock Qualgebras and knotted 3-valent graphs.
\newblock {\em Fundamenta Mathematicae}, 230(2):167--204, 2015.

\bibitem{Mairson2004}
Harry~G. Mairson.
\newblock Linear lambda calculus and ptime-completeness.
\newblock {\em J. Functional Programming}, 14(6):623--633, November 2004.

\bibitem{MarklSchniderStasheff}
Martin Markl, Steve Schnider, and Jim Stasheff.
\newblock {\em Operads in Algebra, Topology and Physics}, volume~96 of {\em
  Mathematical Surveys and Monographs}.
\newblock AMS, 2002.

\bibitem{Mellies2018rtl}
Paul-Andr\'e Melli\`es.
\newblock Ribbon tensorial logic, 2018.
\newblock This volume.

\bibitem{PierceTurner2000}
Benjamin~C. Pierce and David~N. Turner.
\newblock Local type inference.
\newblock {\em ACM Transactions on Programming Languages and Systems},
  22(1):1--44, January 2000.

\bibitem{Serre1980}
Jean-Pierre Serre.
\newblock {\em Trees}.
\newblock Springer-Verlag, 1980.
\newblock Translated from the French by John Stilwell.

\bibitem{OEIS}
N.~J.~A. Sloane.
\newblock The {O}n-{L}ine {E}ncyclopedia of {I}nteger {S}equences,
  \number\year.
\newblock Published electronically at \url{https://oeis.org}.

\bibitem{Statman1974thesis}
Richard Statman.
\newblock {\em Structural complexity of proofs}.
\newblock Phd thesis, Stanford University, 1974.

\bibitem{Street2013skew}
Ross Street.
\newblock Skew-closed categories.
\newblock {\em J. Pure Applied Algebra}, 217(6):973--988, 2013.

\bibitem{Szlachanyi2012}
Kornel Szlach\'anyi.
\newblock Skew-monoidal categories and bialgebroids.
\newblock {\em Advances in Mathematics}, 231(3--4):1694--1730, 2012.

\bibitem{Tait1880}
P.~G. Tait.
\newblock Remarks on the colouring of maps.
\newblock {\em Proc. Royal Soc. Edinburgh}, 10(4):501--503, 1880.

\bibitem{Tamari1951phd}
Dov Tamari.
\newblock {\em Mono\"ides pr\'eordonn\'es et cha\^ines de {M}alcev}.
\newblock Th\`ese, Universit\'e de Paris, 1951.
\newblock Partially published in Bull. Soc. Math. France 82 (1954), 53--96.

\bibitem{Tamari1982graphic}
Dov Tamari.
\newblock A graphic theory of associativity and word-chain patterns.
\newblock In A.~Dold and B.~Eckmann, editors, {\em Combinatorial Theory},
  volume 969 of {\em Lecture Notes in Mathematics}, pages 302--320. Springer,
  1982.

\bibitem{Thomas1998}
Robin Thomas.
\newblock An update on the four-color theorem.
\newblock {\em Notices of the American Mathematical Society}, 45(7):848--859,
  1998.

\bibitem{Thurston2001ktg}
Dylan~P. Thurston.
\newblock The algebra of knotted trivalent graphs and turaev's shadow world.
\newblock In {\em Geometry \& Topology Monographs}, volume~4 of {\em Invariants
  of knots and 3-manifolds (Kyoto 2001)}, pages 337--362. 2004.

\bibitem{Tutte1946ohc}
W.~T. Tutte.
\newblock On hamiltonian circuits.
\newblock {\em Journal of the London Mathematical Society}, 21:98--101, 1946.

\bibitem{Tutte1954}
W.~T. Tutte.
\newblock A contribution to the theory of chromatic polynomials.
\newblock {\em Can. J. Math.}, 6:80--91, 1954.

\bibitem{Tutte1962hamiltonian}
W.~T. Tutte.
\newblock A census of hamiltonian polygons.
\newblock {\em Can. J. Math.}, 14:402--417, 1962.

\bibitem{Tutte1962planartriangulations}
W.~T. Tutte.
\newblock A census of planar triangulations.
\newblock {\em Can. J. Math.}, 14:21--38, 1962.

\bibitem{Tutte1963planarmaps}
W.~T. Tutte.
\newblock A census of planar maps.
\newblock {\em Can. J. Math.}, 15:249--271, 1963.

\bibitem{Tutte1966}
W.~T. Tutte.
\newblock On the algebraic theory of graph colourings.
\newblock {\em Journal of Combinatorial Theory}, 1:15--50, 1966.

\bibitem{Tutte1984book}
W.~T. Tutte.
\newblock {\em Graph Theory}, volume~21 of {\em Encyclopedia of Mathematics and
  its Applications}.
\newblock Addison-Wesley, 1984.

\bibitem{Tutte1998}
W.~T. Tutte.
\newblock {\em Graph Theory as I Have Known it}.
\newblock Oxford, 1998.

\bibitem{Vidal2010phd}
Samuel Vidal.
\newblock {\em Groupe Modulaire et Cartes Combinatoires: G\'en\'eration et
  Comptage}.
\newblock Phd thesis, Universit\'e Lille I, France, July 2010.

\bibitem{Yetter1990}
David Yetter.
\newblock Quantales and (non-commutative) linear logic.
\newblock {\em J. Symbolic Logic}, 55:41--64, 1990.

\bibitem{Z2015counting}
Noam Zeilberger.
\newblock Counting isomorphism classes of $\beta$-normal linear lambda terms.
\newblock arXiv:1509.07596, 2015.

\bibitem{Z2016trivalent}
Noam Zeilberger.
\newblock Linear lambda terms as invariants of rooted trivalent maps.
\newblock {\em J. Functional Programming}, 26, 2016.

\bibitem{Z2017assoc}
Noam Zeilberger.
\newblock A sequent calculus for a semi-associative law.
\newblock In {\em Formal Structures for Computation and Deduction (FSCD 2017)},
  pages 33:1--33:16, 2017.

\bibitem{ZG2015corr}
Noam Zeilberger and Alain Giorgetti.
\newblock A correspondence between rooted planar maps and normal planar lambda
  terms.
\newblock {\em Logical Methods in Computer Science}, 11(3:22):1--39, 2015.

\end{thebibliography}

\ifwebversion
\pagebreak
\appendix

\section{Proofs of results}
\label{sec:appendix:proofs}

\section*{Section 2}

\begin{proof}[Proof of \Cref{prop:exchange-from-dni}.]
To derive \eqref{clopset:exch} from \eqref{clopset:dni}, we first derive an alternate form of the composition law:
\begin{equation}
 \rimp ab \le \rimp{(\rimp bc)}{(\rimp ac)} \tag{comp'} \label{clopset:comp'}
\end{equation}
Derivation of \eqref{clopset:comp'} from \eqref{clopset:dni}:
\begin{align*}
\rimp ab &\le \rimp{(\rimp{(\rimp ab)}{(\rimp ac)})}{(\rimp ac)} \tag{\ref{clopset:dni}\comment{ by $\rimp ac$}}\\
         &\le \rimp{(\rimp bc)}{(\rimp ac)}  \tag{\ref{clopset:comp} in negative pos.}
\end{align*}
Derivation of \eqref{clopset:exch} from \eqref{clopset:comp'} and \eqref{clopset:dni}:
\begin{align*}
\rimp{a}{(\rimp bc)} &\le \rimp{(\rimp{(\rimp bc)}c)}{(\rimp ac)} \tag{\ref{clopset:comp'}} \\
        &\le \rimp b{(\rimp ac)} \tag{\ref{clopset:dni} \comment{ by $c$} in negative pos.}
\end{align*}
In the other direction, we derive \eqref{clopset:dni} from \eqref{clopset:exch} under assumption of left normality:
\begin{align*}
I &\le \rimp{(\rimp ab)}{(\rimp ab)} \tag{\ref{clopset:id}} \\
 &\le \rimp a{(\rimp{(\rimp ab)}b)} \tag{\ref{clopset:exch}} \\
a &\le \rimp{(\rimp ab)}b \tag{left normality}
\end{align*}
\end{proof}

\begin{proof}[Proof of \Cref{prop:imp2mon}.]
Unwinding definitions, it is easy (but instructive) to check that the reverse inclusions
\begin{align*}
(R\bmul S)\bmul T &\supseteq R \bmul (S \bmul T) \\
I\bmul R &\supseteq R \\
R &\supseteq R\bmul I
\end{align*}
follow from the axioms \eqref{clopset:comp}, \eqref{clopset:id}, and \eqref{clopset:unit}, respectively, for any
upsets $R,S,T \refsup P$ of an imploid.
(We leave this as a fun warmup exercise for the reader!)
\end{proof}

\begin{proof}[Proof of \Cref{prop:impcommpsh}.]
\begin{enumerate}
\item
Suppose $P$ satisfies \eqref{clopset:dni}, and let $p \in S \bmul R$.
By definition, there exists $r$ such that $\rimp rp \in S$ and $r \in R$.
But then $\rimp{(\rimp rp)}p \in R$ by \eqref{clopset:dni} and upwards closure, hence $p \in R \bmul S$.
Conversely, suppose that $R \bmul S \supseteq S \bmul R$ for all $R,S \refsup P$, and consider $R = \Up{a}$, $S = \Up{(\rimp ab)}$ where $a,b \in P$ are arbitrary.
It is easy to check that $b \in S \bmul R$, but then if $b \in R\bmul S$
there must exist $p$ such that $a \le \rimp pb$ and $\rimp ab \le p$, implying $a \le \rimp{(\rimp ab)}b$.
\item
Suppose $P$ satisfies \eqref{clopset:exch}, and let $p \in (R \bmul T) \bmul S$.
By definition, there exists $s$ such that $s \imp p \in R \bmul T$ and $s \in S$, and $t$ such that $t \imp (s \imp p) \in R$ and $t \in T$.
But then $s \imp (t \imp p) \in R$ by \eqref{clopset:exch} and upwards closure, from which $p \in (R \bmul S) \bmul T$.
Conversely, suppose that $(R \bmul S)\bmul T \supseteq (R \bmul T)\bmul S$ for all $R,S,T \refsup P$, and consider $R = \Up{a \imp (b\imp c)}$, $T = \Up{b}$, $S = \Up{a}$ where $a,b,c \in P$ are arbitrary.
It is easy to check that $c \in (R \bmul T) \bmul S$, but then if $c \in (R \bmul S) \bmul T$ there must exist $p$ and $q$ such that $a \imp (b \imp c) \le p \imp (q \imp c)$ and $b \le p$ and $a \le q$, implying $a \imp (b \imp c) \le b \imp (a \imp c)$.
\end{enumerate}
\end{proof}

\begin{proof}[Proof of \Cref{prop:mon2imp}.]
Again, it is easy to check that for any downsets $J,K,L \refsdown M$ of a skew monoid, the inclusions
\begin{align*}
\rimp KL &\subseteq \rimp{(\rimp JK)}{(\rimp JL)} \\
I &\subseteq \rimp KK \\
\rimp IK &\subseteq K
\end{align*}
follow from axioms \eqref{mopset:assocr}, \eqref{mopset:lunit}, and \eqref{mopset:runit}, respectively.
\end{proof}

\begin{proof}[Proof of \Cref{prop:double-yoneda-embedding}.]
We just have to check $\UpDown{(\rimp ab)} \equiv \rimp{\UpDown{a}}{\UpDown{b}}$, which reduces to showing that
\begin{align*}
R \ni \rimp ab &\iff \forall S.\ S\ni a\ \Rightarrow\ \exists c.\ R \ni \rimp cb\ \wedge \ S \ni c.
\end{align*}
The implication from left to right is immediate taking $c = a$, while the implication from right to left is immediate taking $S = \Up{a}$.
\end{proof}

\begin{proof}[Proof of \Cref{prop:deductive-closure}.]
This is just an adaptation of the standard free monoid construction (appropriately dualized) to the skew monoidal setting: by mechanically unrolling the definition of $!R$ and applying the skew monoid laws, we can show that $!R$ is the greatest comonoid in $\check{P}$ under $R$.
\end{proof}

\begin{proof}[Proof of \Cref{prop:leRpreorder}.]
That $a \le b$ entails $I \le \rimp ab$ immediately implies $\irel{R}$ is an extension of $\le$ (and reflexive), since $R \supseteq \Up{I}$.
For transitivity, suppose that $\rimp ab \in R$ and $\rimp bc \in R$. 
Applying \eqref{clopset:comp} and upwards closure to the latter assumption we obtain $\rimp{(\rimp ab)}{(\rimp ac)} \in R$, and hence $\rimp ac \in R$ by deductive closure.
\end{proof}

\begin{proof}[Proof of \Cref{prop:dni-closure-exists}.]
The operation 
\begin{align*}
DNI &= S \mapsto S \wedge \{c \mid a \in S, \rimp{(\rimp ab)}b \le c\}
\end{align*}
defines a monotone operator $DNI : \check{P} \to \check{P}$ on the complete lattice $\check{P}$, and so by the Knaster-Tarski theorem we can compute the dni-closure of $R$ as the greatest fixed point of $DNI$ containing $R$.
(NB: since the ordering in $\check{P}$ is reverse inclusion, $\wedge = \cup$, and ``greatest'' = ``is contained in any other fixed point containing $R$''.)
\end{proof}

\begin{proof}[Proof of \Cref{prop:norm:tfae}.]
The equivalence $(1) \Leftrightarrow (2)$ is similar to the proof of \Cref{prop:impcommpsh}(1).
%
%
For $(1)\Rightarrow (3)$: \eqref{prop:norm:tfae:case3i}
 Let $a \in R$. Then $\rimp{(\rimp aa)}a \in R$ by dni-closure, hence $\rimp Ia\in R$ by \eqref{clopset:id} in negative position and upwards closure.
\eqref{prop:norm:tfae:case3ii} Let $\rimp ab \in R$, and $c \in P$. Then $\rimp{(\rimp{(\rimp ab)}{(\rimp ac)})}{(\rimp ac)} \in R$ by dni-closure, hence $\rimp{(\rimp bc)}{(\rimp ac)} \in R$ by \eqref{clopset:comp} in negative position and upwards closure.

For $(3)\Rightarrow (1)$: Let $a \in R$ and $b \in P$. Then $\rimp Ia \in R$ by \eqref{prop:norm:tfae:case3i}, and $\rimp{(\rimp ab)}{(\rimp Ib)} \in R$ by \eqref{prop:norm:tfae:case3ii}. But then $\rimp{(\rimp ab)}b \in R$ by \eqref{clopset:unit} and upwards closure.

Finally, the equivalence $(3) \Leftrightarrow (4)$ is immediate after noting that the closure condition
[$\rimp bc \in R \Rightarrow \rimp{(\rimp ab)}{(\rimp ac)} \in R$]
is always valid for any upset $R$.
\end{proof}

\begin{proof}[Proof of \Cref{prop:tensor-preserves-dni}.]
The first part follows easily from case \eqref{prop:norm:tfae:case2} of \Cref{prop:norm:tfae} and some applications of \eqref{mopset:assocr}, while the second follows from the first and the formula for the deductive closure $!R$ in \Cref{prop:deductive-closure}.
\end{proof}

\begin{proof}[Proof of \Cref{prop:quot:imploidLN}.]
We've already verified that $\quot PR$ is an imploid by \Cref{prop:leRpreorder} and \Cref{prop:norm:tfae}\eqref{prop:norm:tfae:case4}, while left normality ($I \irel{R} \rimp ab \Rightarrow a \irel{R} b$) follows from the \eqref{clopset:unit} axiom and upwards closure.
\end{proof}

\begin{proof}[Proofs of \Cref{prop:pullback,prop:Idnidedup,prop:kerclosure,prop:kerfaithful}.] Immediate.
\end{proof}

\begin{proof}[Proof of \Cref{prop:quotient-is-quotient}.]
The function $[-]$ is automatically a strong homomorphism, and the claim that $R = \ker [-]$ amounts to $a \in R$ iff $I \irel{R} a$, which follows from \Cref{prop:norm:tfae}\eqref{prop:norm:tfae:case4}.
Now suppose that $f : P \to Q$ is a (potentially lax) homomorphism into some left normal imploid, and that $R \subseteq \ker{f}$.
We define $\bar{f} : \quot PR \to Q$ to be identical to $f$ on elements of $P$, and just have to check $a \irel{R} b$ implies $f(a) \le_Q f(b)$.
Well, if $\rimp ab \in R$ then $I_Q \le f(\rimp ab)$ by assumption that $R \subseteq \ker{f}$.
But then $f(a) \le_Q f(b)$ follows from the assumptions that $f$ is a homomorphism and that $Q$ is left normal.
\end{proof}

\section*{Section 3}

\begin{proof}[Proof of \Cref{prop:flows-as-derivations}.]
This is immediate from the definition of flow and our operadic description of topological orientations in \Cref{sec:imploid-flows:canonical-orientation}, although a small amount of care should be taken in interpreting the rule of variable exchange.
Formally, a 3-valent map-with-boundary corresponds to a linear \emph{term-in-context,} that is, a lambda term $T$ equipped with a specific ordering $x_1,\dots,x_n$ of its free variables.
So even though the variable exchange rule is written (in the traditional way) with the same term $T$ appearing in both the premise and the conclusion, in fact these correspond to different maps-with-boundary when interpreted as terms-in-context.
See \cite{Z2016trivalent} for a more detailed discussion.
\end{proof}

\begin{proof}[Proof of \Cref{lemma:top-flow-is-global}.]
By \Cref{prop:flows-as-derivations}, to prove that every topological flow is global we have to show that
\begin{center}
$x_1:a_1,\dots,x_n:a_n \vdash T:b$ implies $I \le \mimp{(a_1,\dots,a_n)}{b}$.
\end{center}
When $T$ is planar, this follows by induction on exchange-free derivations, appealing in the application case
$$
\infer[(c \le a \imp b)]{\Gamma,\Delta \vdash T_1(T_2) : b}{\Gamma \vdash T_1 : c & \Delta \vdash T_2 : a}
$$
to the following lemma, proved by induction on $\Delta$ (here we take $\Gamma$ and $\Delta$ to range over lists of elements of $P$):
\begin{lemma}
If $I \le \mimp{\Gamma}{(a\imp b)}$ and $I \le \mimp{\Delta}{a}$ then $I \le \mimp{(\Gamma,\Delta)}{b}$.
\end{lemma}
\noindent
In the non-planar case we also have to deal with the exchange rule, but then we simply appeal to \eqref{clopset:exch}.
\end{proof}

\begin{proof}[Proof of \Cref{corr:nowhere-unit-bridgeless}.]
By \Cref{lemma:top-flow-is-global} and the characterization of bridgeless maps as unitless terms \cite[Proposition 7.3]{Z2016trivalent}.
\end{proof}

\begin{proof}[Proof of \Cref{lemma:nontop-flow-violation}.]
We first note that any well-oriented 3-valent map admits two different \emph{constant} $\hat{2}$-flows in which either every edge is assigned the value 1 or every edge is assigned the value $2$ (these are trivially flows since $1 \imp 1 = 1$ and $2 \imp 2 = 2$).
We now proceed by case analysis of the possible orientations of $T$, and induction on the number of vertices:
\begin{enumerate}
\item[Case 
$T =
\vcenter{\hbox{\begin{tikzpicture}[scale=1.5]
\draw[style=dotted] (0,0) circle [radius=1];
\draw (0,-1) node (root) [style=root] {};
\draw (0,-0.5) node (app) [style=app] {};
\draw (-0.433,0) node (fn) [shape=circle,draw,dotted,fill=lightgray] {$T_1$};
\draw (0.433,0) node (arg) [shape=circle,draw,dotted,fill=lightgray] {$T_2$};
\draw (-0.924,0.383) node (var1) {};
\draw (-0.383,0.924) node (var2) {};
\draw (0.383,0.924) node (var3) {};
\draw (0.924,0.383) node (var4) {};
\draw [->-] (app) to (root.center);
\draw [->-] (fn) to (app);
\draw [->-] (arg) to (app);
\draw (var1.center) to (fn);
\draw (var2.center) to (fn);
\draw (var3.center) to (arg);
\draw (var4.center) to (arg);
\end{tikzpicture}}}
$]
\mbox{}\\
By assumption, either $T_1$ or $T_2$ must be non-topologically oriented.
In the first case, by the induction hypothesis $T_1$ has a flow in which all its non-root boundary arcs are assigned 1 or $2$ and its root is assigned 0.
Then after taking the constant-$2$ (or constant-1) flow on $T_2$, we assign 0 to $T$'s root to obtain our desired flow (using that $0 = y \imp 0$ for all $y>0$):
$$
\vcenter{\hbox{\begin{tikzpicture}[scale=1.5]
\draw[style=dotted] (0,0) circle [radius=1];
\draw (0,-1) node (root) [style=root] {};
\draw (0,-0.5) node (app) [style=app] {};
\draw (-0.433,0) node (fn) [shape=circle,draw,dotted,fill=lightgray] {$T_1$};
\draw (0.433,0) node (arg) [shape=circle,draw,dotted,fill=lightgray] {$T_2$};
\draw (-0.924,0.383) node (var1) {};
\draw (-0.383,0.924) node (var2) {};
\draw (0.383,0.924) node (var3) {};
\draw (0.924,0.383) node (var4) {};
\draw [->-] (app) to node [right] {\footnotesize$0$} (root.center);
\draw [->-] (fn) to node [left] {\footnotesize$0$} (app);
\draw [->-] (arg) to node [right] {\footnotesize$y$} (app);
\draw (var1.center) to node [near start] {\footnotesize$1\vee 2$} (fn);
\draw (var2.center) to node [near start] {\footnotesize$1\vee 2$} (fn);
\draw (var3.center) to node [near start] {\footnotesize$y$} (arg);
\draw (var4.center) to node [near start] {\footnotesize$y$} (arg);
\end{tikzpicture}}}
$$
Symmetrically, in the second case we apply the induction hypothesis to $T_2$ and take the constant-$2$ (or constant-1) flow on $T_1$ (using that $y \le 0 \imp 0 = 2$):
$$
\vcenter{\hbox{\begin{tikzpicture}[scale=1.5]
\draw[style=dotted] (0,0) circle [radius=1];
\draw (0,-1) node (root) [style=root] {};
\draw (0,-0.5) node (app) [style=app] {};
\draw (-0.433,0) node (fn) [shape=circle,draw,dotted,fill=lightgray] {$T_1$};
\draw (0.433,0) node (arg) [shape=circle,draw,dotted,fill=lightgray] {$T_2$};
\draw (-0.924,0.383) node (var1) {};
\draw (-0.383,0.924) node (var2) {};
\draw (0.383,0.924) node (var3) {};
\draw (0.924,0.383) node (var4) {};
\draw [->-] (app) to node [right] {\footnotesize$0$} (root.center);
\draw [->-] (fn) to node [left] {\footnotesize$y$} (app);
\draw [->-] (arg) to node [right] {\footnotesize$0$} (app);
\draw (var1.center) to node [near start] {\footnotesize$y$} (fn);
\draw (var2.center) to node [near start] {\footnotesize$y$} (fn);
\draw (var3.center) to node [near start] {\footnotesize$1\vee 2$} (arg);
\draw (var4.center) to node [near start] {\footnotesize$1\vee 2$} (arg);
\end{tikzpicture}}}
$$
\item[Case
$T =
\vcenter{\hbox{\begin{tikzpicture}[scale=1.5]
\draw[style=dotted] (0,0) circle [radius=1];
\draw (0,-1) node (root) [style=root] {};
\draw (0,-0.5) node (app) [style=lam] {};
\draw (-0.433,0) node (fn) [shape=circle,draw,dotted,fill=lightgray] {$T_1$};
\draw (0.433,0) node (arg) [shape=circle,draw,dotted,fill=lightgray] {$T_2$};
\draw (-0.924,0.383) node (var1) {};
\draw (-0.383,0.924) node (var2) {};
\draw (0.383,0.924) node (var3) {};
\draw (0.924,0.383) node (var4) {};
\draw [->-] (app) to (root.center);
\draw [->-] (fn) to (app);
\draw [->-] (app) to (arg);
\draw (var1.center) to (fn);
\draw (var2.center) to (fn);
\draw (var3.center) to (arg);
\draw (var4.center) to (arg);
\end{tikzpicture}}}
$ or
$\vcenter{\hbox{\begin{tikzpicture}[scale=1.5]
\draw[style=dotted] (0,0) circle [radius=1];
\draw (0,-1) node (root) [style=root] {};
\draw (0,-0.5) node (app) [style=lam] {};
\draw (-0.433,0) node (fn) [shape=circle,draw,dotted,fill=lightgray] {$T_1$};
\draw (0.433,0) node (arg) [shape=circle,draw,dotted,fill=lightgray] {$T_2$};
\draw (-0.924,0.383) node (var1) {};
\draw (-0.383,0.924) node (var2) {};
\draw (0.383,0.924) node (var3) {};
\draw (0.924,0.383) node (var4) {};
\draw [->-] (app) to (root.center);
\draw [->-] (arg) to (app);
\draw [->-] (app) to (fn);
\draw (var1.center) to (fn);
\draw (var2.center) to (fn);
\draw (var3.center) to (arg);
\draw (var4.center) to (arg);
\end{tikzpicture}}}
$]
\mbox{}\\
In the first case we take the constant-1 flow on $T_1$ and the constant-$2$ flow on $T_2$ (using that $2 \imp 1 = 0$), and in the second we take the constant-$2$ flow on $T_1$ and either the constant-$2$ or constant-1 flow on $T_2$ (using that $0 \imp x = 2$ for all $x$):
$$
\vcenter{\hbox{\begin{tikzpicture}[scale=1.5]
\draw[style=dotted] (0,0) circle [radius=1];
\draw (0,-1) node (root) [style=root] {};
\draw (0,-0.5) node (app) [style=lam] {};
\draw (-0.433,0) node (fn) [shape=circle,draw,dotted,fill=lightgray] {$T_1$};
\draw (0.433,0) node (arg) [shape=circle,draw,dotted,fill=lightgray] {$T_2$};
\draw (-0.924,0.383) node (var1) {};
\draw (-0.383,0.924) node (var2) {};
\draw (0.383,0.924) node (var3) {};
\draw (0.924,0.383) node (var4) {};
\draw [->-] (app) to node [right] {\footnotesize$0$} (root.center);
\draw [->-] (fn) to node [left] {\footnotesize$1$} (app);
\draw [->-] (app) to node [right] {\footnotesize$2$} (arg);
\draw (var1.center) to node [near start] {\footnotesize$1$} (fn);
\draw (var2.center) to node [near start] {\footnotesize$1$} (fn);
\draw (var3.center) to node [near start] {\footnotesize$2$} (arg);
\draw (var4.center) to node [near start] {\footnotesize$2$} (arg);
\end{tikzpicture}}}
\quad\text{or}\quad
\vcenter{\hbox{\begin{tikzpicture}[scale=1.5]
\draw[style=dotted] (0,0) circle [radius=1];
\draw (0,-1) node (root) [style=root] {};
\draw (0,-0.5) node (app) [style=lam] {};
\draw (-0.433,0) node (fn) [shape=circle,draw,dotted,fill=lightgray] {$T_1$};
\draw (0.433,0) node (arg) [shape=circle,draw,dotted,fill=lightgray] {$T_2$};
\draw (-0.924,0.383) node (var1) {};
\draw (-0.383,0.924) node (var2) {};
\draw (0.383,0.924) node (var3) {};
\draw (0.924,0.383) node (var4) {};
\draw [->-] (app) to node [right] {\footnotesize$0$} (root.center);
\draw [->-] (arg) to node [right,xshift=-2pt,yshift=-2pt] {\footnotesize$1\vee2$} (app);
\draw [->-] (app) to node [left] {\footnotesize$2$} (fn);
\draw (var1.center) to node [near start] {\footnotesize$2$} (fn);
\draw (var2.center) to node [near start] {\footnotesize$2$} (fn);
\draw (var3.center) to node [near start] {\footnotesize$1\vee2$} (arg);
\draw (var4.center) to node [near start] {\footnotesize$1\vee2$} (arg);
\end{tikzpicture}}}
$$

\item[Case
$
T = \vcenter{\hbox{\begin{tikzpicture}[scale=1.5]
\draw[style=dotted] (0,0) circle [radius=1];
\draw (0,-1) node (root) [style=root] {};
\draw (0,-0.5) node (lam) [style=lam] {};
\draw (-0.433,0) node (body) [shape=circle,draw,dotted,fill=lightgray] {$T_1$};
\draw (-0.924,0.383) node (var1) {};
\draw (0.283,0.524) node (var3) {};
\draw (0.924,0.383) node (var4) {};
\draw [->-] (lam) to (root.center);
\draw [->-] (body) to (lam);
\draw (var1.center) to (body);
\draw [-->-] (lam) to [bend right=60] (var3) to [bend right=60] (body);
\draw (var4.center) to (body);
\end{tikzpicture}}}
\quad\text{or}\quad
\vcenter{\hbox{\begin{tikzpicture}[scale=1.5]
\draw[style=dotted] (0,0) circle [radius=1];
\draw (0,-1) node (root) [style=root] {};
\draw (0,-0.5) node (lam) [style=lam] {};
\draw (-0.433,0) node (body) [shape=circle,draw,dotted,fill=lightgray] {$T_1$};
\draw (-0.924,0.383) node (var1) {};
\draw (0.283,0.524) node (var3) {};
\draw (0.924,0.383) node (var4) {};
\draw [->-] (lam) to (root.center);
\draw [-<-] (body) to (lam);
\draw (var1.center) to (body);
\draw [-<--] (lam) to [bend right=60] (var3) to [bend right=60] (body);
\draw (var4.center) to (body);
\end{tikzpicture}}}
$]
\mbox{}\\
In the first case, by assumption, $T_1$ must be non-topologically oriented.
We apply the induction hypothesis (again using that $y \imp 0 = 0$ for all $y > 0$):
$$
\vcenter{\hbox{\begin{tikzpicture}[scale=1.5]
\draw[style=dotted] (0,0) circle [radius=1];
\draw (0,-1) node (root) [style=root] {};
\draw (0,-0.5) node (lam) [style=lam] {};
\draw (-0.433,0) node (body) [shape=circle,draw,dotted,fill=lightgray] {$T_1$};
\draw (-0.924,0.383) node (var1) {};
\draw (0.283,0.524) node (var3) {};
\draw (0.924,0.383) node (var4) {};
\draw [->-] (lam) to node [right] {\footnotesize$0$} (root.center);
\draw [->-] (body) to node [left] {\footnotesize$0$} (lam);
\draw (var1.center) to node [near start] {\footnotesize$1\vee 2$} (body);
\draw [-->-] (lam) to [bend right=60] node [near start] {\footnotesize$1 \vee 2$} (var3) to [bend right=60] (body);
\draw (var4.center) to node [near end] {\footnotesize$1\vee 2$} (body);
\end{tikzpicture}}}
$$
In the second case we simply take the constant-2 flow on $T_2$ and assign 0 to the root (using that $0 \imp 2 = 2$):
$$
\vcenter{\hbox{\begin{tikzpicture}[scale=1.5]
\draw[style=dotted] (0,0) circle [radius=1];
\draw (0,-1) node (root) [style=root] {};
\draw (0,-0.5) node (lam) [style=lam] {};
\draw (-0.433,0) node (body) [shape=circle,draw,dotted,fill=lightgray] {$T_1$};
\draw (-0.924,0.383) node (var1) {};
\draw (0.283,0.524) node (var3) {};
\draw (0.924,0.383) node (var4) {};
\draw [->-] (lam) to node [right] {\footnotesize$0$} (root.center);
\draw [-<-] (body) to node [left] {\footnotesize$2$} (lam);
\draw (var1.center) to node [near start] {\footnotesize$2$} (body);
\draw [-<--] (lam) to [bend right=60] node [near start,right] {\footnotesize$2$} (var3) to [bend right=60] (body);
\draw (var4.center) to node [near end] {\footnotesize$2$} (body);
\end{tikzpicture}}}
$$
\item[Case 
$
T = \vcenter{\hbox{\begin{tikzpicture}[scale=1.5]
\draw[style=dotted] (0,0) circle [radius=1];
\draw (0,-1) node (root) [style=root] {};
\draw (0,-0.5) node (lam) [style=app] {};
\draw (-0.433,0) node (body) [shape=circle,draw,dotted,fill=lightgray] {$T_1$};
\draw (-0.924,0.383) node (var1) {};
\draw (0.283,0.524) node (var3) {};
\draw (0.924,0.383) node (var4) {};
\draw [->-] (lam) to (root.center);
\draw [->-] (body) to (lam);
\draw (var1.center) to (body);
\draw [-<--] (lam) to [bend right=60] (var3) to [bend right=60] (body);
\draw (var4.center) to (body);
\end{tikzpicture}}}
$]
\mbox{}\\
Since $T_1$ has two outputs, it is not topological.
We apply the induction hypothesis taking the root of $T_1$ to be the arc counterclockwise from the root of $T$ (again using that $y \le 0 \imp 0 = 2$):
$$
\vcenter{\hbox{\begin{tikzpicture}[scale=1.5]
\draw[style=dotted] (0,0) circle [radius=1];
\draw (0,-1) node (root) [style=root] {};
\draw (0,-0.5) node (lam) [style=app] {};
\draw (-0.433,0) node (body) [shape=circle,draw,dotted,fill=lightgray] {$T_1$};
\draw (-0.924,0.383) node (var1) {};
\draw (0.283,0.524) node (var3) {};
\draw (0.924,0.383) node (var4) {};
\draw [->-] (lam) to node [right] {\footnotesize$0$} (root.center);
\draw [->-] (body) to node [left,xshift=1pt,yshift=-2pt] {\footnotesize$1\vee 2$} (lam);
\draw (var1.center) to node [near start] {\footnotesize$1\vee 2$} (body);
\draw [-<--] (lam) to [bend right=60] node [near start,right] {\footnotesize$0$} (var3) to [bend right=60] (body);
\draw (var4.center) to node [near end] {\footnotesize$1\vee 2$} (body);
\end{tikzpicture}}}
$$
(Notice this case accounts for why we can't restrict to globally well-oriented maps in the induction.)
\item[Case 
$
T = \vcenter{\hbox{\begin{tikzpicture}[scale=0.5]
\draw[style=dotted] (0,0) circle [radius=1];
\draw (0,-1) node (root) [style=root] {};
\draw (0,1) node (var) {};
\draw[->-] (var.center) to (root.center);
\end{tikzpicture}}}
$]
\mbox{}\\
Impossible by assumption that $T$ is equipped with a non-topological orientation.
\end{enumerate}
\end{proof}

\begin{proof}[Proof of \Cref{thm:nontop-flow-is-nonglobal}.]
Implied by \Cref{lemma:top-flow-is-global,lemma:nontop-flow-violation}.
\end{proof}

\section*{Section 4}

\begin{proof}[Proof of \Cref{prop:imploid-moves}.]
The existence of boundary-preserving homomorphisms corresponding to the first four moves reduces to the definition of a preorder (i.e., reflexivity + transitivity) and of the local flow relations for 3-valent vertices (\Cref{fig:localflow}) combined with monotonicity of implication \eqref{clopset:imp}.
As already mentioned, the justification of $\beta$ and $\eta$ amounts to totality and uniqueness, respectively, of the implication operation, combined with \eqref{clopset:imp}.
Similarly, the $\compose$, $\init$, and $\unit$ transformations may be justified by appeal to the imploid axioms \eqref{clopset:comp}, \eqref{clopset:id}, and \eqref{clopset:unit}, respectively, 
with the right-to-left direction of $\init$ corresponding to left normality.
Finally, all of the remaining moves may be derived from the above.
(For example, $\rho$ can be derived using $\beta$ and $\bar\compose$, which can in turn be derived from $\compose$ using $\beta$ and $\eta$.)
We leave the details as an exercise for the reader.
\end{proof}

\begin{proof}[Proof of \Cref{prop:comm-imploid-moves}.]
The $\chi$ move reduces directly to the \eqref{clopset:exch} axiom, while C can be easily derived from $\chi$ in combination with $\iota$ and $\eta$.
The $\gamma$ move reduces directly to \eqref{clopset:dni}, which is valid in any symmetric left normal imploid (\Cref{prop:exchange-from-dni}).
\end{proof}

\begin{proof}[Proof of \Cref{prop:preserve-topological-orientation}.]
To be completely unambiguous, we should first clarify that the definition of topological orientation (\Cref{prop:topological-orientation}) extends to rooted maps with vertices of degree 2 or 1 by the addition of the following cases:
\begin{center}
\begin{inparaenum}[1)]\setcounter{enumi}{3}
\item
$
\vcenter{\hbox{\begin{tikzpicture}[scale=1]
\draw[style=dotted] (0,0) circle [radius=1];
\draw (0,-1) node (root) {};
\draw (0,0) node (unit) [style=unit] {};
\draw[->-] (unit) to (root.center);
\end{tikzpicture}}}
$ .
\end{inparaenum}
\end{center}
Since the moves listed in \Cref{prop:imploid-moves,prop:comm-imploid-moves} only include vertices of degree $<3$ with their unique possible topological orientation, all that needs to be checked is the treatment of trivalent vertices.
We already explained in \Cref{sec:flow-rewriting:background} that the graphical $\beta$ and $\eta$ moves preserve topological orientation, precisely because they restrict to the standard rewriting rules from lambda calculus when interpreted on linear terms.
One way of seeing that $\compose$ preserves topological orientation is to verify that it corresponds to the following transformation, which takes linear terms to linear terms:
$$\lambda x.T \overset\compose\LRA \lambda x.\lambda y.T[x(y)/x].$$
Similarly, $\gamma$ and $\chi$ correspond to transformations
$$
T_1(T_2) \overset\gamma\LRA T_2(T_1)\quad\text{and}\quad (T_1 T_2)T_3 \overset\chi\LRA (T_1 T_3)T_2.
$$
The remaining moves can be considered analogously, or by deriving them from the above.
\end{proof}

\begin{proof}[Proof of \Cref{thm:topological-completeness}.]
We give a graphical proof by induction on topological orientations, exhibiting a sequence of moves
$
V_n' \Rightarrow T
$
to realize any possible rooted essentially 3-valent map $T$ with $n$ non-root boundary arcs.
(It is important to note that most maps can be constructed in multiple different ways, and the proof we give here only corresponds to one particular encoding.)
For simplicity we ignore the presence of 2-valent vertices (i.e., work modulo edge subdivision), which otherwise only requires a bit of extra book-keeping.
\begin{itemize}
\item[Case 
$T =
\vcenter{\hbox{\begin{tikzpicture}[scale=1.5]
\draw[style=dotted] (0,0) circle [radius=1];
\draw (0,-1) node (root) {};
\draw (0,-0.5) node (app) [style=app] {};
\draw (-0.433,0) node (fn) [shape=circle,draw,dotted,fill=lightgray] {$T_1$};
\draw (0.433,0) node (arg) [shape=circle,draw,dotted,fill=lightgray] {$T_2$};
\draw (-0.924,0.383) node (var1) {};
\draw (-0.383,0.924) node (var2) {};
\draw (0.383,0.924) node (var3) {};
\draw (0.924,0.383) node (var4) {};
\draw [->-] (app) to (root.center);
\draw [->-] (fn) to (app);
\draw [->-] (arg) to (app);
\draw [->-] (var1.center) to (fn);
\draw [->-] (var2.center) to (fn);
\draw [->-] (var3.center) to (arg);
\draw [->-] (var4.center) to (arg);
\end{tikzpicture}}}
$]
\mbox{}\\
\begin{multline*}
\vcenter{\hbox{\scalebox{0.8}{\begin{tikzpicture}
  \node (cont) {};
  \node [style=app] (app) [above=1em of cont] {};
  \node [scale=0.8,shape=circle,draw,dotted,fill=lightgray] (t1) [above left=0.8em and 1.385em of app] {$T_1$};
  \node [scale=0.8,shape=circle,draw,dotted,fill=lightgray](t2) [above right=0.8em and 1.385em of app] {$T_2$};
  \node (v1) [above left=0.5em and 0.866em of t1] {};
  \node (v2) [above right=0.5em and 0.866em of t1] {};
  \node (v3) [above left=0.5em and 0.866em of t2] {};
  \node (v4) [above right=0.5em and 0.866em of t2] {};
  \path
    (t1) edge [->-] (app)
    (t2) edge [->-] (app)
    (app) edge [->-] (cont.center);
  \path
    (v1.center) edge [->-] (t1)
    (v2.center) edge [->-] (t1)
    (v3.center) edge [->-] (t2)
    (v4.center) edge [->-] (t2);
\end{tikzpicture}}}}
\LLA
\vcenter{\hbox{\scalebox{0.8}{\begin{tikzpicture}
  \node (cont) {};
  \node [style=app] (app) [above=1em of cont] {};
  \node [style=app] (v2t1) [above left=1em and 1.5em of app] {};
  \node [style=app] (v1t1) [above left=0.5em of v2t1] {};
  \node [style=unit] (v0t1) [above left=0.5em of v1t1] {};
  \node [style=app] (v2t2) [above right=1em and 1.5em of app] {};
  \node [style=app] (v1t2) [above left=0.5em of v2t2] {};
  \node [style=unit] (v0t2) [above left=0.5em of v1t2] {};
  \node (v1) [above right=0.5em of v1t1] {};
  \node (v2) [right=0.5em of v1] {};
  \node (v3) [above right=0.5em of v1t2] {};
  \node (v4) [right=0.5em of v3] {};
  \path
    (v0t1) edge [-->-] (v1t1)
    (v1t1) edge [-->-] (v2t1)
    (v2t1) edge [->-] (app)
    (v0t2) edge [-->-] (v1t2)
    (v1t2) edge [-->-] (v2t2)
    (v2t2) edge [->-] (app)
    (app) edge [->-] (cont.center);
  \path
    (v1.center) edge [->-] (v1t1)
    (v2.center) edge [->-] (v2t1)
    (v3.center) edge [->-] (v1t2)
    (v4.center) edge [->-] (v2t2);
\end{tikzpicture}}}}
\overset{\assoc^*}\LLA
\vcenter{\hbox{\scalebox{0.8}{\begin{tikzpicture}
  \node (cont) {};
  \node [style=app] (app) [above=1em of cont] {};
  \node [style=app] (v4t1) [above left=0.5em of app] {};
  \node [style=app] (v3t1) [above left=0.5em of v4t1] {};
  \node [style=app] (v2t1) [above left=0.5em of v3t1] {};
  \node [style=app] (v1t1) [above left=0.5em of v2t1] {};
  \node [style=unit] (v0t1) [above left=0.5em of v1t1] {};
  \node (v1) [above right=0.5em of v1t1] {};
  \node (v2) [right=0.5em of v1] {};
  \node [style=unit] (v0t2) [right=0.5em of v2] {};
  \node (v3) [right=0.5em of v0t2] {};
  \node (v4) [right=0.5em of v3] {};
  \path
    (v0t2) edge [->-] (v3t1)
    (v0t1) edge [-->-] (v1t1)
    (v1t1) edge [-->-] (v2t1)
    (v2t1) edge [-->-] (v3t1)
    (v3t1) edge [-->-] (v4t1)
    (v4t1) edge [-->-] (app)
    (app) edge [->-] (cont.center);
  \path
    (v1.center) edge [->-] (v1t1)
    (v2.center) edge [->-] (v2t1)
    (v3.center) edge [->-] (v4t1)
    (v4.center) edge [->-] (app);
\end{tikzpicture}}}}
\overset{\unit}{\LLA}
\vcenter{\hbox{\scalebox{0.8}{\begin{tikzpicture}
  \node (cont) {};
  \node [style=app] (app) [above=1em of cont] {};
  \node [style=app] (v3t1) [above left=0.5em of app] {};
  \node [style=app] (v2t1) [above left=1em of v3t1] {};
  \node [style=app] (v1t1) [above left=0.5em of v2t1] {};
  \node [style=unit] (v0t1) [above left=0.5em of v1t1] {};
  \node (v1) [above right=0.5em of v1t1] {};
  \node (v2) [right=0.5em of v1] {};
  \node (v3) [right=1em of v2] {};
  \node (v4) [right=0.5em of v3] {};
  \path
    (v0t1) edge [-->-] (v1t1)
    (v1t1) edge [-->-] (v2t1)
    (v2t1) edge [-->-] (v3t1)
    (v3t1) edge [-->-] (app)
    (app) edge [->-] (cont.center);
  \path
    (v1.center) edge [->-] (v1t1)
    (v2.center) edge [->-] (v2t1)
    (v3.center) edge [->-] (v3t1)
    (v4.center) edge [->-] (app);
\end{tikzpicture}}}}
\end{multline*}
To be explicit, the first (= rightmost) step of the derivation is a $\unit$ move, the second corresponds to a sequence of $\assoc$ moves, and the last to two parallel applications of the induction hypothesis.

\item[Case
$
T = \vcenter{\hbox{\begin{tikzpicture}[scale=1.5]
\draw[style=dotted] (0,0) circle [radius=1];
\draw (0,-1) node (root) {};
\draw (0,-0.5) node (lam) [style=lam] {};
\draw (-0.433,0) node (body) [shape=circle,draw,dotted,fill=lightgray] {$T_1$};
\draw (-0.924,0.383) node (var1) {};
\draw (0.283,0.524) node (var3) {};
\draw (0.924,0.383) node (var4) {};
\draw [->-] (lam) to (root.center);
\draw [->-] (body) to (lam);
\draw [->-] (var1.center) to (body);
\draw [-->-] (lam) to [bend right=60] (var3) to [bend right=60] (body);
\draw [->-] (var4.center) to (body);
\end{tikzpicture}}}
$]
\mbox{}\\
\begin{multline*}
\vcenter{\hbox{\scalebox{0.8}{\begin{tikzpicture}
  \node (cont) {};
  \node [style=lam] (lam) [above=1em of cont] {};
  \node [scale=0.8,shape=circle,draw,dotted,fill=lightgray] (t1) [above left=0.5em and 0.866em of app] {$T_1$};
  \node (v1) [above left=0.5em and 0.866em of t1] {};
  \node (v3) [above right=0.5em and 0.866em of t1] {};
  \node (v2) [below right=0.5em of v3] {};
  \path
    (t1) edge [->-] (lam)
    (lam) edge [->-] (cont.center);
  \path
    (v1.center) edge [->-] (t1)
    (v3.center) edge [->-] (t1);
  \draw [->-] (lam) to [bend right=60] (v2) to [bend right=60] (t1.north);
\end{tikzpicture}}}}
\LLA
\vcenter{\hbox{\scalebox{0.8}{\begin{tikzpicture}
  \node (cont) {};
  \node [style=lam] (lam) [above=1em of cont] {};
  \node [style=app] (v3t1) [above left=0.5em of lam] {};
  \node [style=app] (v2t1) [above left=0.5em of v3t1] {};
  \node [style=app] (v1t1) [above left=0.5em of v2t1] {};
  \node [style=unit] (v0t1) [above left=0.5em of v1t1] {};
  \node (v1) [above right=0.5em of v1t1] {};
  \node (v3) [right=1.732em of v1] {};
  \node (v2) [below=1em of v3] {};
  \path
    (v0t1) edge [-->-] (v1t1)
    (v1t1) edge [-->-] (v2t1)
    (v2t1) edge [-->-] (v3t1)
    (v3t1) edge [-->-] (app)
    (lam) edge [->-] (cont.center);
  \path
    (v1.center) edge [->-] (v1t1)
    (v3.center) edge [->-] (v3t1);
  \draw [->-] (lam) to [bend right=60] (v2) to [bend right=60] (v2t1);
\end{tikzpicture}}}}
\overset{\chi^*}{\LLA}
\vcenter{\hbox{\scalebox{0.8}{\begin{tikzpicture}
  \node (cont) {};
  \node [style=lam] (lam) [above=1em of cont] {};
  \node [style=app] (v3t1) [above left=0.5em of lam] {};
  \node [style=app] (v2t1) [above left=0.5em of v3t1] {};
  \node [style=app] (v1t1) [above left=0.5em of v2t1] {};
  \node [style=unit] (v0t1) [above left=0.5em of v1t1] {};
  \node (v1) [above right=0.5em of v1t1] {};
  \node (v3) [right=1.732em of v1] {};
  \node (v2) [below=1em of v3] {};
  \path
    (v0t1) edge [-->-] (v1t1)
    (v1t1) edge [-->-] (v2t1)
    (v2t1) edge [-->-] (v3t1)
    (v3t1) edge [-->-] (app)
    (lam) edge [->-] (cont.center);
  \path
    (v1.center) edge [->-] (v1t1)
    (v3.center) edge [->-] (v2t1);
  \draw [->-] (lam) to [bend right=60] (v2) to [bend right=60] (v3t1);
\end{tikzpicture}}}}
\overset{\eta}\LLA
\vcenter{\hbox{\scalebox{0.8}{\begin{tikzpicture}
  \node (cont) {};
  \node [style=app] (v2t1) [above=2.5em of cont] {};
  \node [style=app] (v1t1) [above left=0.5em of v2t1] {};
  \node [style=unit] (v0t1) [above left=0.5em of v1t1] {};
  \node (v1) [above right=0.5em of v1t1] {};
  \node (v3) [right=1.732em of v1] {};
  \node (v2) [below=1em of v3] {};
  \path
    (v0t1) edge [-->-] (v1t1)
    (v1t1) edge [-->-] (v2t1)
    (v2t1) edge [-->-] (cont.center);
  \path
    (v1.center) edge [->-] (v1t1)
    (v3.center) edge [->-] (v2t1);
\end{tikzpicture}}}}
\end{multline*}
\item[Case 
$
T = \vcenter{\hbox{\begin{tikzpicture}[scale=0.5]
\draw[style=dotted] (0,0) circle [radius=1];
\draw (0,-1) node (root) {};
\draw (0,1) node (var) {};
\draw[->-] (var.center) to (root.center);
\end{tikzpicture}}}
$]
\mbox{}\\
$$
\vcenter{\hbox{\scalebox{0.8}{\begin{tikzpicture}
\draw (0,-0.866) node (root) {};
\draw (0,0) node (var) {};
\draw[->-] (var.center) to (root.center);
\end{tikzpicture}}}}
\quad \overset\init\LLA \quad
\vcenter{\hbox{\scalebox{0.8}{\begin{tikzpicture}
  \node (cont) {};
  \node [style=app] (app) [above=1em of cont] {};
  \node [style=unit] (v0) [above left=0.5em and 0.866em of app] {};
  \node (v1) [above right=0.5em and 0.866em of app] {};
  \path
    (v0) edge [->-] (app)
    (v1.center) edge [->-] (app)
    (app) edge [->-] (cont.center);
\end{tikzpicture}}}}
$$
\item[Case
$
T = \vcenter{\hbox{\begin{tikzpicture}[scale=0.5]
\draw[style=dotted] (0,0) circle [radius=1];
\draw (0,-1) node (root) {};
\draw (0,0) node (unit) [style=unit] {};
\draw[->-] (unit) to (root.center);
\end{tikzpicture}}}
$]
\mbox{}\\
Immediate since $T = V_0$.
\end{itemize}
\end{proof}

\section*{Section 5}

\begin{proof}[Proof of \Cref{prop:minimal-pol-top}.]
By inspection of \Cref{fig:polflow}, black vertices can only arise in the minimal polarization in configurations of the form
$$
\vcenter{\scalebox{0.8}{\hbox{
\begin{tikzpicture}
  \node [style=lam] (lam) {};
  \node (var) [above right=0.5em and 0.866em of lam] {};
  \node (body) [above left=0.5em and 0.866em of lam] {};
  \node [style=cut] (cut) [below=1.5em of lam] {};
  \node [style=app] (app) [below=1.5em of cut] {};
  \node (cont) [below left=0.5em and 0.866em of app] {};
  \node (arg) [below right=0.5em and 0.866em of app] {};
  \draw[->-pos] (body.center) to node [left] {\hide{\footnotesize$b_1$}} (lam);
  \draw[->-neg] (lam) to node [right] {\hide{\footnotesize$a_1$}} (var.center);

  \path (lam) edge [->-pos] node [right] {\hide{\footnotesize$a_1 \imp b_1$}} (cut);
  \path (cut) edge [->-neg] node [right] {\hide{\footnotesize$a_2 \imp b_2$}} (app); 
  \draw[->-neg] (app) to node [left] {\hide{\footnotesize$b_2$}} (cont.center);  
  \draw[->-pos] (arg.center) to node [right] {\hide{\footnotesize$a_2$}} (app);
\end{tikzpicture}}}}
$$
which correspond to $\beta$-redices of the unpolarized map.
\end{proof}

\begin{proof}[Proof of \Cref{prop:topological-implies-acyclic}.]
Immediate by induction on topological orientations.
(The converse is easily seen to be false: for instance consider the minimal polarizations of the non-topological maps in \Cref{ex:nonglobal-flow}, which do not have cycles in their w--b orientations.)
\end{proof}

\onecolumn

\section{An extended example}
\label{sec:appendix:polflows}

\subsection*{A bridgeless planar 3-valent map (the ``Tutte graph'')}
\small

\begin{center}
\includegraphics[width=0.45\textwidth]{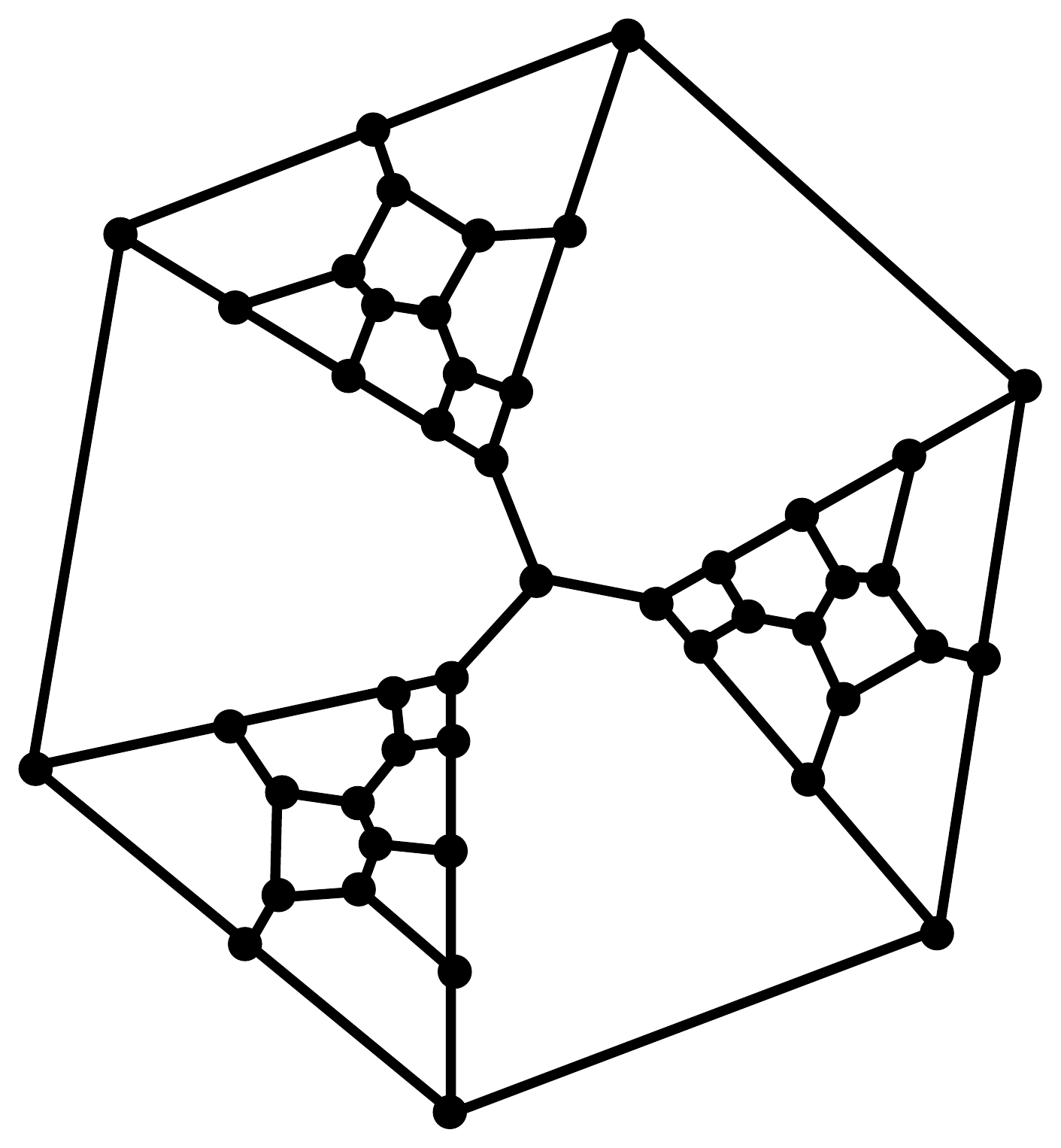}
\end{center}
(From W.~T.~Tutte, ``On Hamiltonian Circuits'', {\it Journal of the London Mathematical Society} 21 (1946), 98--101.)

\subsection*{A rooting of the above together with its topological orientation}

\begin{center}
\includegraphics[width=0.45\textwidth]{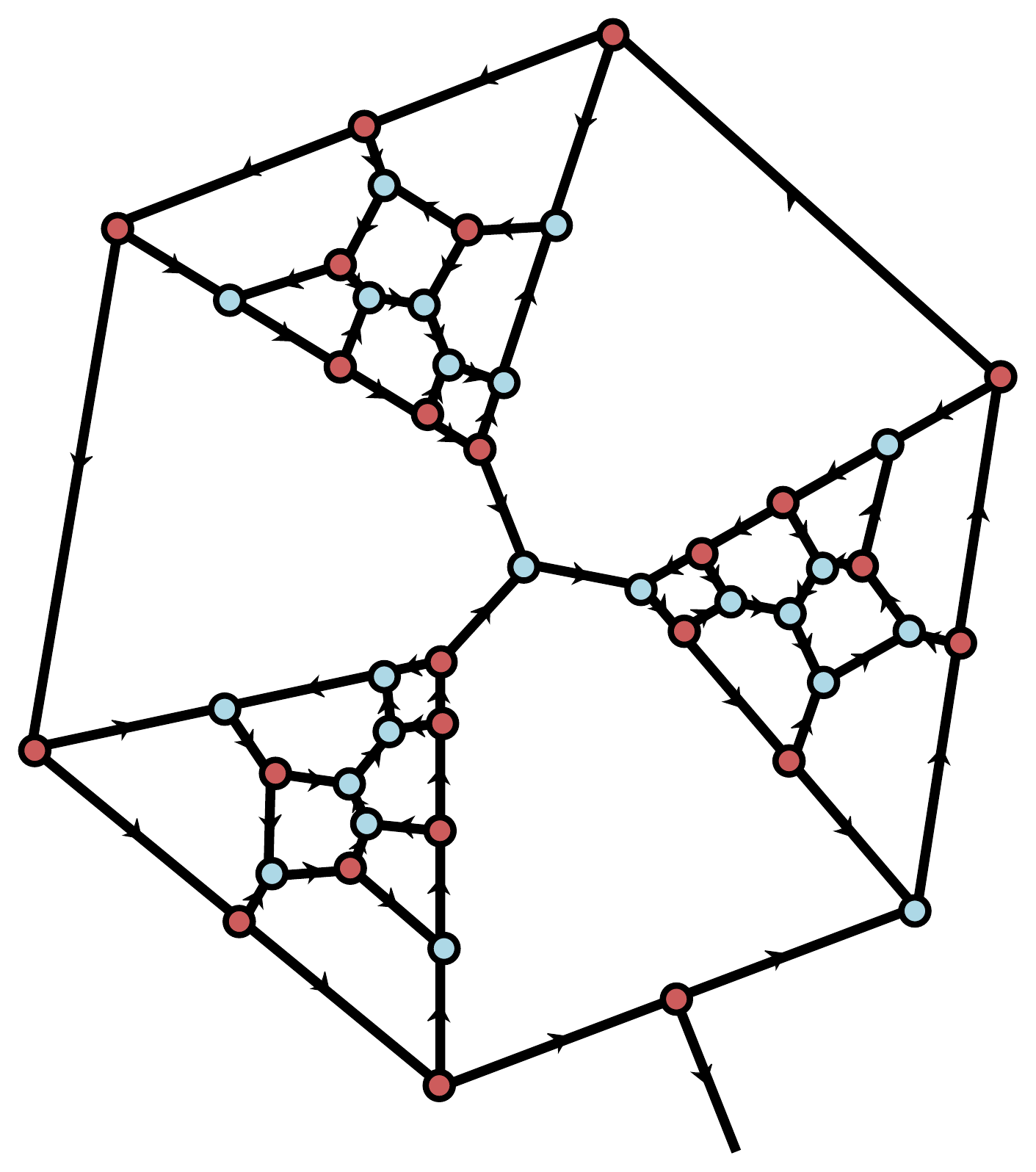}
\end{center}

\subsection*{The corresponding linear lambda term (with variables and unique $\beta$-redex indicated)}
\footnotesize

\begin{center}
\includegraphics[width=0.45\textwidth]{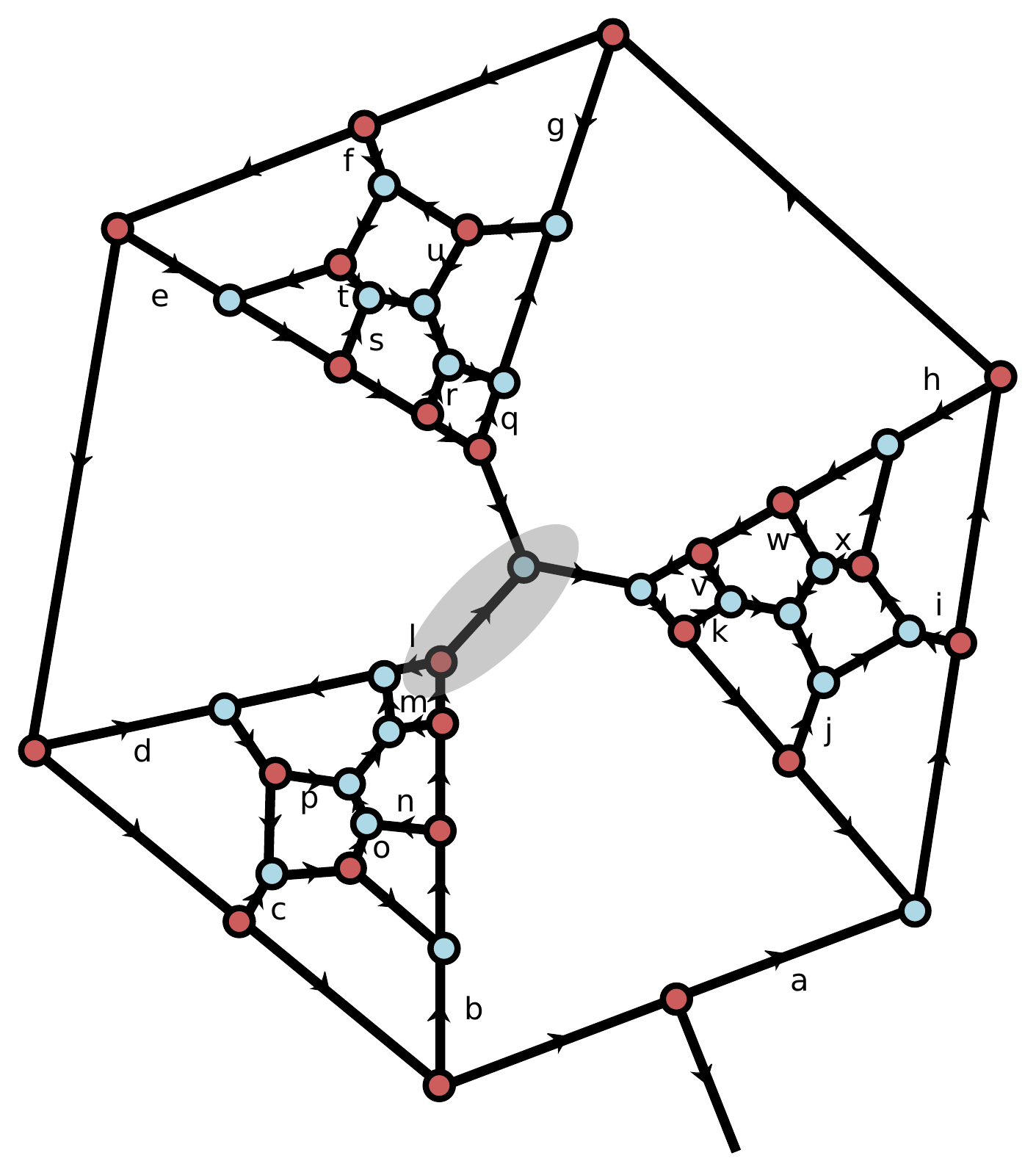}
\end{center}

$\lambda a\lambda b\lambda c\lambda d\lambda e\lambda f\lambda g\lambda h\lambda i.a(\lambda j\lambda k.\uline{((\lambda l\lambda m\lambda n.b(\lambda o.c(\lambda p.d(l(m((no)p))))))(\lambda q\lambda r\lambda s.e(\lambda t.f(\lambda u.g(q(r((st)u)))))))}(\lambda v\lambda w.h(\lambda x.i(j((kv)(wx))))))$


\subsection*{The minimal polarization of the above}

\begin{center}
\includegraphics[width=0.45\textwidth]{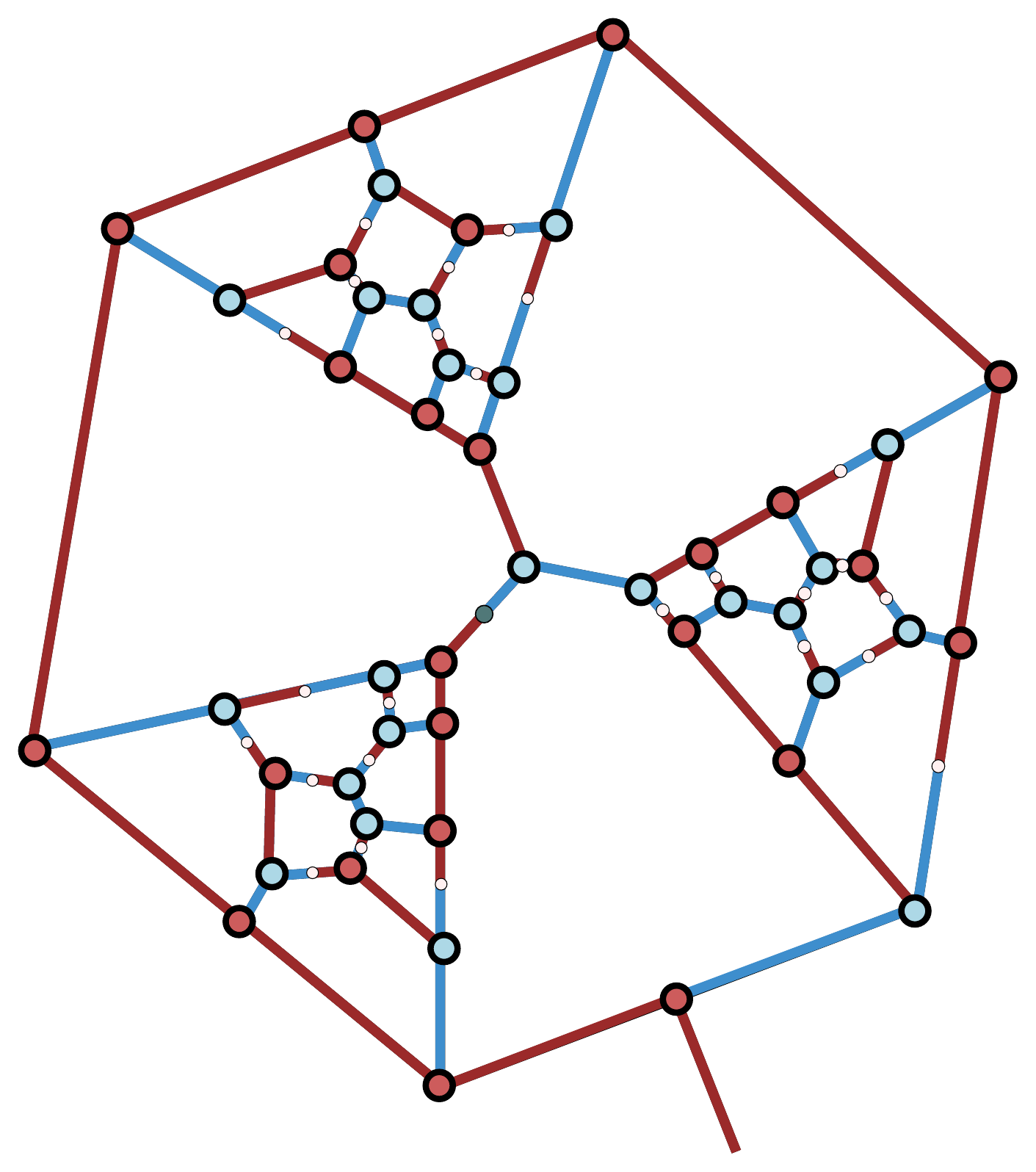}
\end{center}

\subsection*{The corresponding universal polarized flow}
\small

\begin{center}
\includegraphics[width=0.45\textwidth]{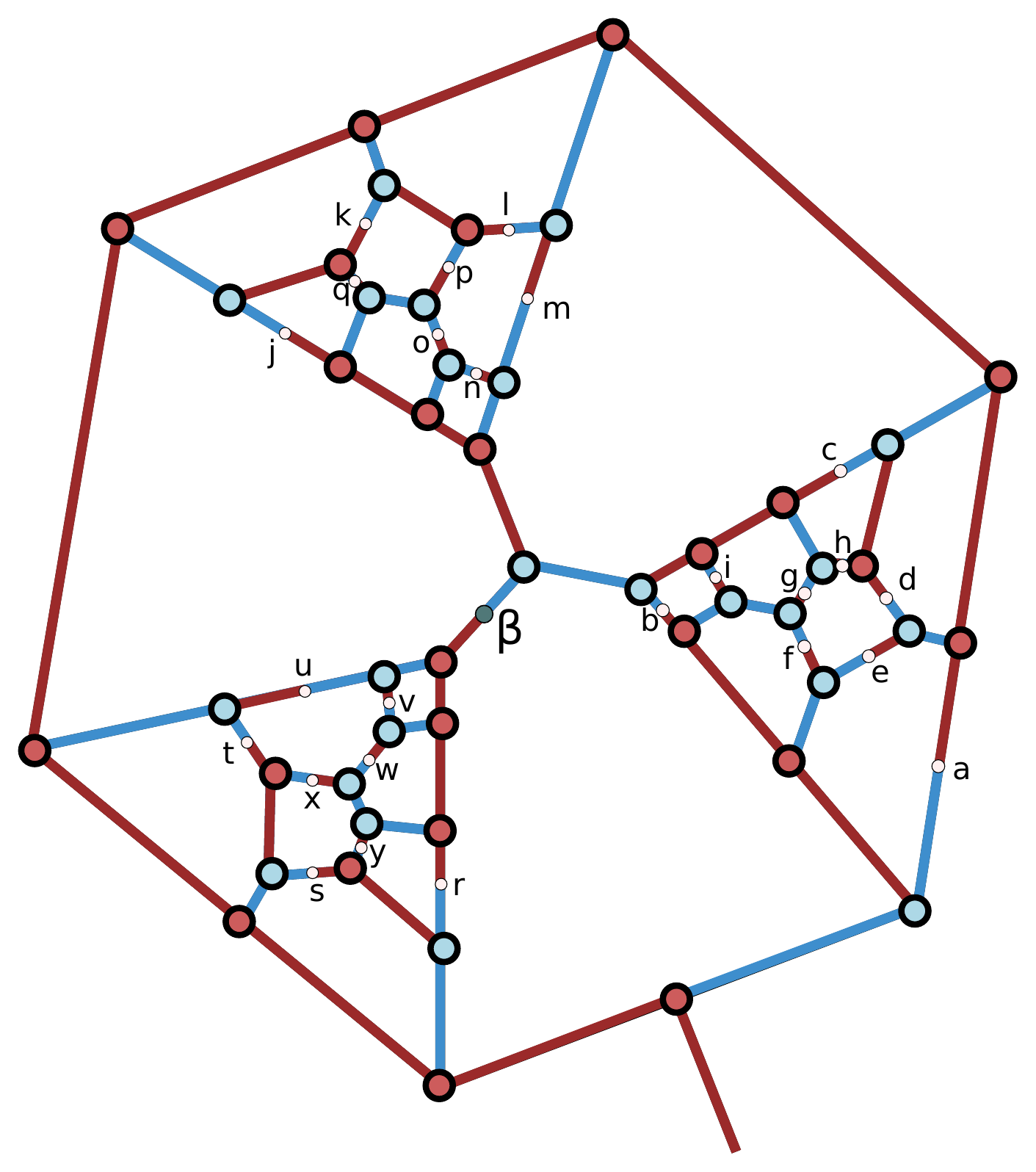}
\end{center}

The universal polarized flow is valued in the imploid freely generated over the intervals $a^- \le a^+,\dots,y^-\le y^+$ modulo the following relation (writing $[\sigma\tau]$ for $\sigma \imp \tau$):
$$
\beta : [[v^+  u^-]  [[w^+  v^-]  [[y^+  [x^+  w^-]]  r^+]]] \le [[[n^+  m^-]  [[o^+  n^-]  [[q^+  [p^+  o^-]]  j^+]]]  [[i^-  [[h^+  g^-]  c^+]]  b^-]]
$$
The type assigned to the root is:
$$
[[[[f^+  e^-]  [[i^+  [g^+  f^-]]  b^+]]  a^-] 
  [[[y^-  s^+]  r^-]  [[[x^-  t^+]  s^-]  [[u^+  t^-]  [[[q^-  k^+]  j^-]  [[[p^-  l^+]  k^-]  [[m^+  l^-]  [[[h^-  d^+]  c^-]  [[e^+  d^-]  a^+]]]]]]]]]
$$

\subsection*{A $\V$-flow realized as an instance of the universal polarized flow}
\medskip

\begin{minipage}{0.5\textwidth}
\qquad\includegraphics[width=\textwidth]{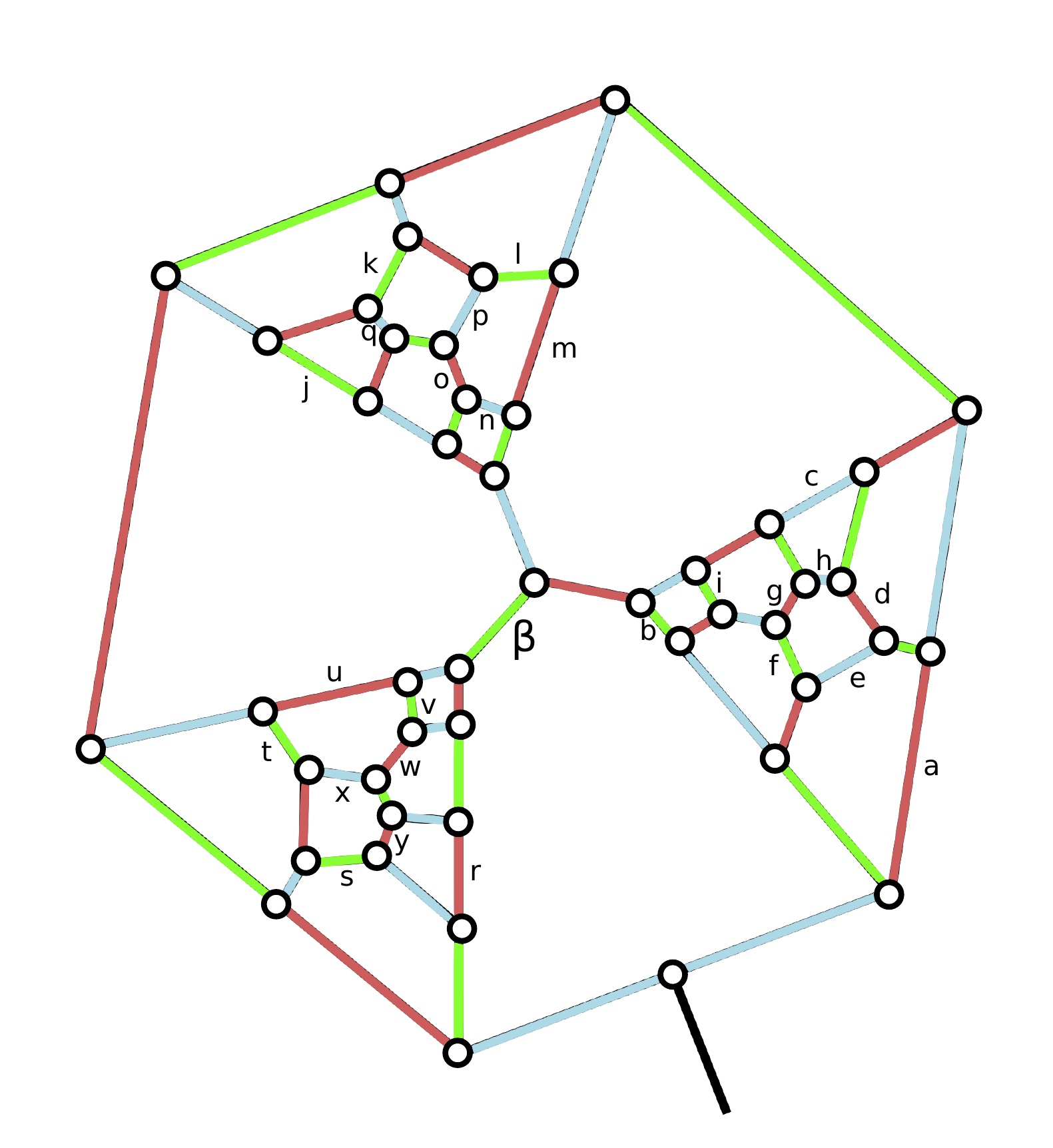}
\end{minipage}
\begin{minipage}{0.5\textwidth}
\begin{align*}
a = d = g = m = o = r = u = w = y & = R  \\
b = f = i = j = k = l = s = t = v & = G \\
c = e = h = n = p = q = x & = B \\
 \beta : G & = G 
\end{align*}
\end{minipage}



\fi
\end{document}